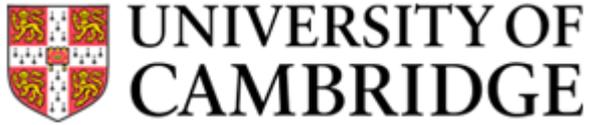

# CDBB West Cambridge Digital Twin: Lessons Learned

Justas Brazauskas, Matt Danish, Vadim Safronov, Rohit Verma, Richard Mortier, Ian Lewis
The Computer Laboratory

# I. Introduction

This work was carried out from September 2019 to September 2022 by two post-doctoral researchers supported by 10% each of two co-PIs funded by the Centre for Digital Built Britain within the Construction Innovation Hub, with additional contributions made by Ph.D. students funded through other sources. While looking broadly at opportunities for technical innovation in construction and the built environment, we anticipated the following future developments:

- **Dramatic increases in the spatial density of deployed sensors**. This is generally referred to as the long-anticipated 'Internet of Things' where we focused particularly on the context of the built environment.
- **Increased desirability of real-time processing of incoming data**. This enables support of a significantly larger set of applications, including those that simply cannot be supported without real-time support being built into the data processing platform from inception, which includes support for sensor readings, sensors, sensor types, real-time readings, building objects and people.
- **Integration of more intelligence into sensors**. Examples include those incorporating imaging and machine learning, as well as production of increasingly intelligent and complex derived analyses.
- **Acknowledgement that people dominate the built environment**. This means that it is inevitable that people's presence and actions drive many day-to-day behaviours as well as triggering exceptions, and the privacy implications of future environments and buildings having knowledge of the occupant presence, location, and behaviour.
- **Need for all occupants to be first first-class participants in in-building systems**. With pervasive deployment and detailed inferencing, it is no longer realistic to treat such systems as limited only to interaction with a trained Building Manager.

To explore this space and understand the real challenges and opportunities therein, we designed and built a complete working instance of the Adaptive City Platform (ACP). Without doing so it would not have been possible to explore the many practical and nuanced issues arising in such a complex system, and it would not have been possible to evaluate how techniques addressing challenges and issues performed. The 'digital architecture' of the ACP is illustrated below.

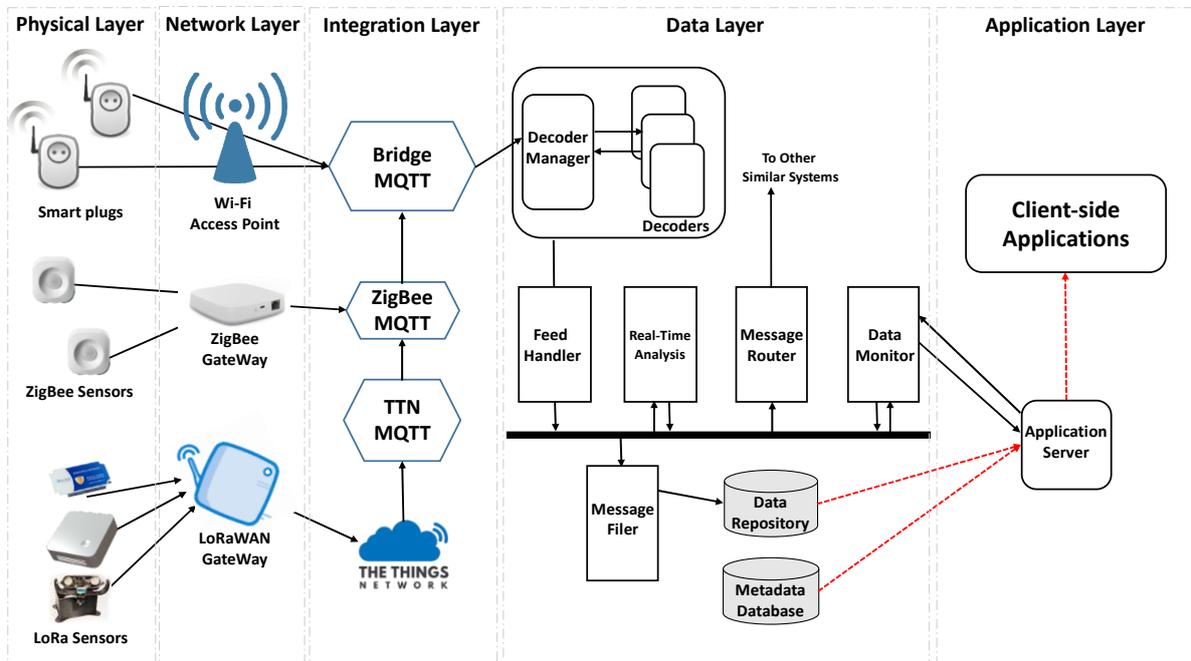

The ACP is a substantial and complex system, comprising many distinct parts. The remainder of this report is organised around the "lessons learned" from building and deploying each of those parts but it should be remembered that it is the whole platform that was the goal. We conclude with a brief reflection on the process of running a project under unrealistically stringent financial controls in the face of the fundamental uncertainties that are the key feature of research work, exacerbated by a global pandemic.

Note that in what follows, where a simple report that 'this is what we implemented' actually covers significant work debating, prototyping and testing a variety of approaches before settling on what we believe is the approach that worked best given the constraints and circumstances.

## II. Infrastructure and Sensing

### II.1. Providing LPWAN networking for built-environment sensor deployments

Sensors deployed 'in-building' or 'on site' will typically use a wired technology, often Modbus or similarly dated industrial-use technology, to communicate sensor data back to a central Building Management System (BMS). This has the advantages that sensors can be powered and have a guaranteed communication channel back to the BMS. However, wired connectivity does not scale to deployment of much larger numbers of sensors, nor when sensors need to be deployed both inside and outside the building, due to the installation efforts involved and legacy constraints on the data types and rates that can be communicated. Similar issues arise in the installation of additional sensors after the building is completed. Low-power wireless technologies targeting IoT are emerging, presenting opportunities for low-cost, high-volume, extensible deployment of sensors with multi-year battery life.

In this project we therefore wanted to evaluate use of such Low-power Wide Area Networking (LPWAN) technologies in a substantial in-building deployment of sensors. We did so, building on the group's prior experimental work with LPWAN technologies, selecting LoRaWAN due to its low cost, low power, wide support, and flexibility. Nominally, in clear air, a LoRaWAN device is expected to deliver 10km range with a 10-year battery life at an incremental per-device build cost of under £10. It does so by providing very low (although still adequate for simple sensor data) data rates. No other currently prevailing wireless technology (Wi-Fi, Bluetooth, mobile network LTE) can provide anything close to this energy efficiency: battery-powered devices using those technologies would need frequent recharging, at least daily, and so are unsuitable for such deployments unless devices are directly connected to an electricity supply. Questions remained over the use of LoRaWAN in an in-building deployment however: in-building transmission ranges would obviously be significantly lower than those in free air but how much lower? and what would be the impact on coverage, battery-life, and cost?

To answer these questions, we deployed LPWAN gateways with high-gain antennas on the roofs of both our building (the William Gates Building) and the Institute for Manufacturing building. We found that these provided remarkably good coverage inside both buildings, counter-intuitively including the cone immediately beneath the antenna. Around 95% of our building can transmit/receive data via these antennas with the good coverage due to the LPWAN technology working well with reflections from adjacent buildings. The central antenna in the left-hand image below is the WGB installation.

However, we also confirmed that building layout and proximity to electrical interference are both significantly more likely to affect in-building LPWAN data transmission than a naive assessment of either direct or indirect transmission distance to a common antenna. Although we found that the 'gateway on the roof' was highly effective in practice, there were still isolated spaces within the building where such effects needed to be mitigated. We mitigated connectivity issues in such isolated spaces by introducing compact gateways suitable for use in-building. This was sufficiently successful that, given the broad interest in the University in environmental monitoring and data collection, the University Information Services (UIS, the University's IT Department) was happy to provide a new Wi-Fi layer (the Cam-IoT network) across the entire estate for use by our devices.

This provided just enough infrastructure that we could design, build and deploy a standalone mains-powered LoRaWAN gateway, able simply to be plugged in anywhere in the University to immediately provide LoRaWAN back-haul for our sensor deployments without requiring any further configuration. Costing just £150 each, these gateways (shown below, right-hand images) allowed us to fix any "dead zones" in our network coverage and, due to the straightforward plug-and-play deployment model, they could easily be moved around or replaced with negligible effort and without adverse effect on the platform itself.

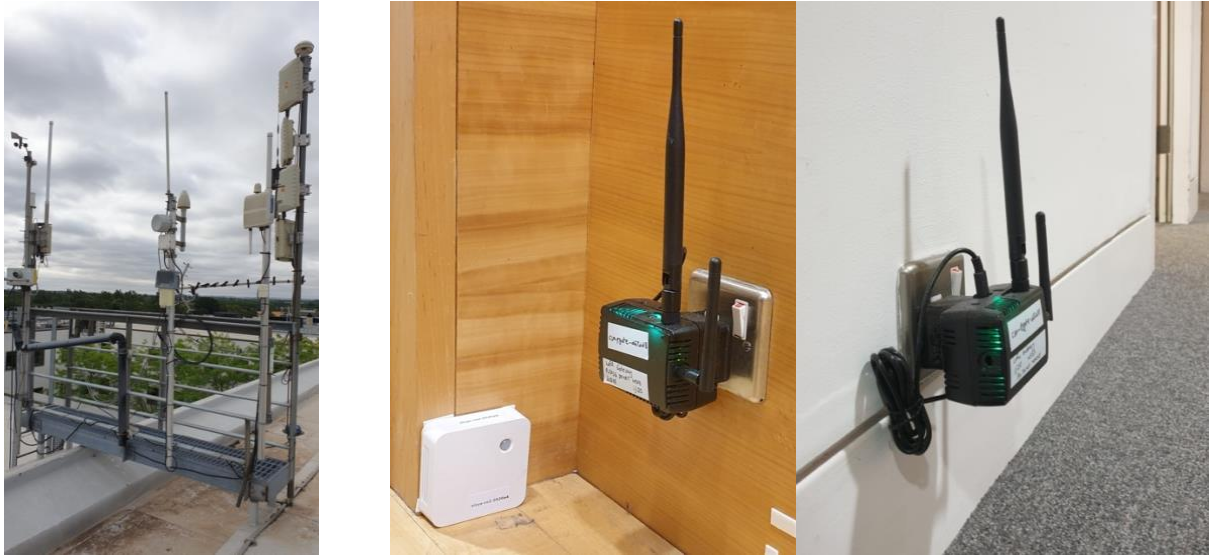

In conclusion, the scale and extended nature of this project allowed us to demonstrate that the use of external antennas augmented by selective in-building powered gateways supported by an infrastructure Wi-Fi network provides an excellent network infrastructure on which to build a dense, large-scale sensor deployment: it is cost-effective, simple to administer, highly flexible, and extremely robust.

## II.2.    The real-time nature of data

A particular focus of our work has been on issues of *time* and *timeliness* as we anticipate that many decisions in the adaptive built environments of the future will need to be taken promptly, in contrast to today's state of the art which makes a hard distinction between the relatively slow changing control-loops built into the infrastructure, e.g., turning air conditioning on and off, and the potentially medium-to-high-frequency building data generated and collected independently of those control loops. Unfortunately, as the expectation seems to be that building metadata as might be derived from plans can often be years out of date, the same lax attitude to timeliness transfers to the handling of sensor data. This results in the current industry best practice being to direct sensor data into a 'data warehouse' or 'data lake', in passing building in assumptions of ad hoc analyses, not known in advance, and likely to be executed manually at some unknown future point. The current direction of travel in the construction industry appears to be a slow evolution from the current manual timescales involved in, e.g., construction drawing updates, rather than a paradigm shift to take on board modern real-time data processing platforms as used in, e.g., Investment Banking.

Our Adaptive City Platform follows design patterns from real-time data processing systems, 'pushing' changed data throughout the system, including to the user workstation if appropriate. This enables real-time analytics to be carried out as soon as data is received, while also allowing those data to be stored, potentially in many different places according to current and anticipated future analytical requirements for such historical data. The converse, where data is initially stored and only then made available for processing introduces unavoidable latencies in performing analytics and, through the inherent properties of the system's design, prevents real-time applications from being built. The historical analyses that can be performed on accumulated stored data are widely understood and straightforward to support, so we focused instead on the challenges and opportunities of processing the information in real-time.

## II.3. Smarter sensors for real-time data

Our platform has an end-to-end 'latency' – that is, the time it takes a new sensor data reading to propagate to some analysis a user is viewing – of around one second. Most of that is the LoRaWAN network transmission time to move the data from the sensor to the gateway. This low latency means that the typical periodic reporting cycle of most sensors is rather painful to watch! For example, a naïve 5-minute reporting cycle means you will wait up to five minutes for the system to detect a room changing from unoccupied to occupied – a risible behaviour in a system that users *who may be the people who entered the room and caused its state to change* can view directly. The apparently obvious solution of reducing the reporting cycle does reduce the latency but at the cost of dramatically increased data volumes and reduced sensor battery lifetimes for very little additional information and still without achieving close to optimal latencies.

One way to do better is to provide greater intelligence nearer the sensors, the sources of data, themselves. We thus designed, built, and deployed 'smart' sensors, able to provide information in a timely fashion (e.g., with sub-second latencies) without blindly sending simple readings at a higher rate. As a practical example, consider the Coffee Pot Sensor that we developed and used to provide information about the status of the group's coffee pot, shown below.

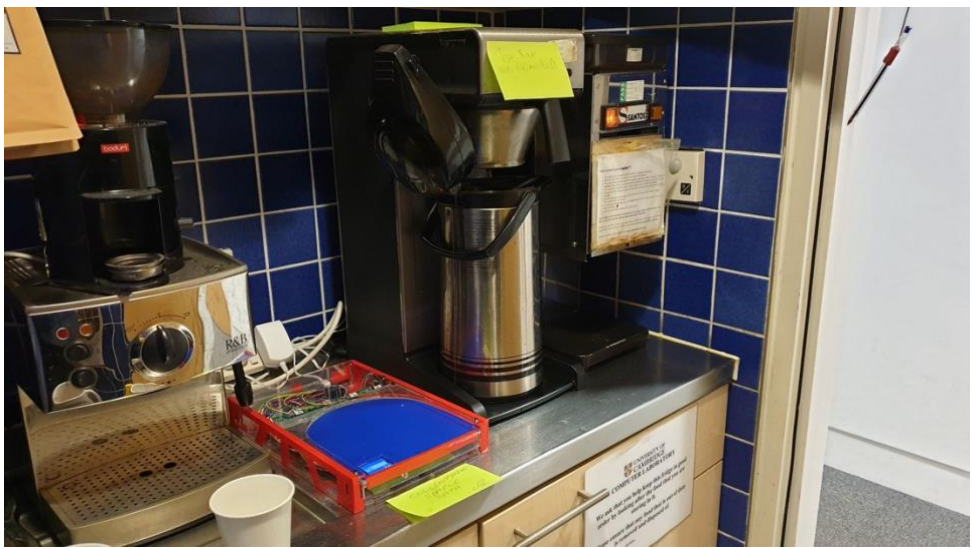

As shown in the photograph, the Coffee Pot Sensor comprises a weight sensor for weighing the coffee pot when it is placed on the node that also provides a bridge to the LoRaWAN network for two energy sensors that measure the power consumption of the coffee grinder and the coffee percolator, plus a motion sensor that detects when people are moving near the kitchen area. The node processes the raw data from these various sensors to report when cups of coffee are poured and when the pot becomes full or empty but is also able to predict when a new pot of coffee is likely to appear.

By building and deploying such a "smart" sensor we were able to observe that it achieved sub-second latencies while reducing the message count to our platform compared to the raw sensor data by a factor of 10,000,000 (ten million). Taken in isolation this sensor might perhaps be considered rather over-engineered – but it does demonstrate how transformative low-latency reporting of sensor data can be achieved without swamping a network with low value data and generating huge data sets. Similar benefits can reasonably be expected in more widely applicable deployments such as intelligent traffic systems enabling intelligent sensors to send important information immediately rather than relying on simplistic periodic reporting and the latencies it introduces, without swamping the network and data storage infrastructure.

We also found that this technique could be adapted to very generic use cases by modifying the software provided in off-the-shelf smartplugs, illustrated below. Useful for monitoring the energy use of almost any plugged-in device, with suitable (non-standard) programming we enabled the smartplug to send notable data points immediately they occurred rather than having them swamp the network with low-value repeating messages indicating that, e.g., the downstream device is still turned off.

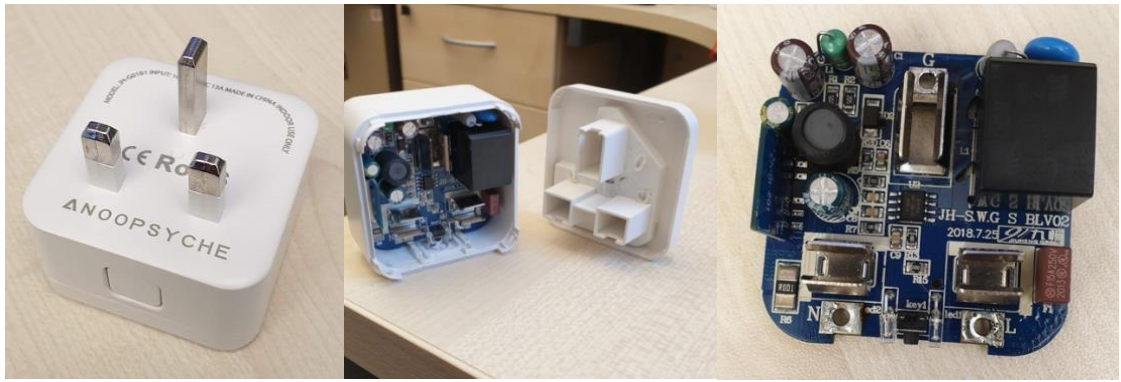

Indeed, there some devices on the market that provide this capability without requiring software modification. After extensive evaluation of the LoRaWAN sensor market, including testing a considerable number of devices, we identified suppliers providing devices at least partially aligned to our objectives. E.g., Elsys provide several LoRaWAN sensors including a Passive Infra-Red (PIR) sensor (shown below), as commonly used in burglar alarms, that will immediately wake up and send their data payload if the IR sensor triggers.

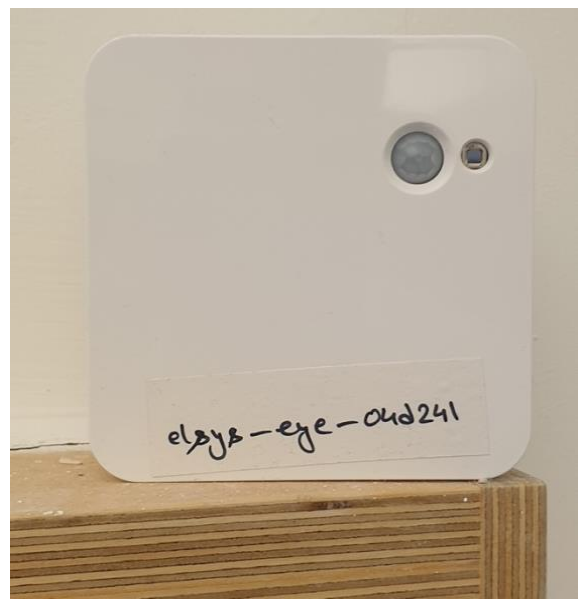

This also demonstrates a factor that contributed to the logistical difficulties of this project: the provision of LoRaWAN sensors is a nascent industry and a considerable effort of trial and error is needed when selecting products best suited to an application. The majority of LoRaWAN sensors provide only naïve periodic wake-up and transmission of data, as part of the low-duty-cycle necessary for battery-powered sensors. Even sensors that provide a 'transmit on alert' function

7 / 24

rarely differentiate this 'alert triggered' message from the routine messages that are sent periodically making handling of such 'alert' messages awkward.

Overall, it became clear that current practices where large quantities of sensor data are stored in a 'data warehouse' before making them available for processing can give a sense of satisfaction due to the apparently detailed and voluminous data, while obscuring the serious problems resulting from the high latencies experienced by individual data points. If you are processing data 24 hours after collection, then it does not seem important that it took an individual data point 5 minutes to arrive – but it will present serious (digital) future architectural handicaps both by limiting the analyses that are practical as well as the cumulative nature of latencies as they stack up from multiple sensors.

## II.4. Intelligent sensors

Taking the above approach of smarter sensors to a further extreme, we followed the edge-computing paradigm and developed an intelligent camera (*DeepDish*, shown below) by coupling the low-power Raspberry Pi computing platform with the Google Coral EdgeTPU accelerator. This provides a customised machine-learning computer vision platform that transforms video footage into anonymised data about the movement of people and vehicles in public spaces in real-time, without exposing the raw video footage off the device.

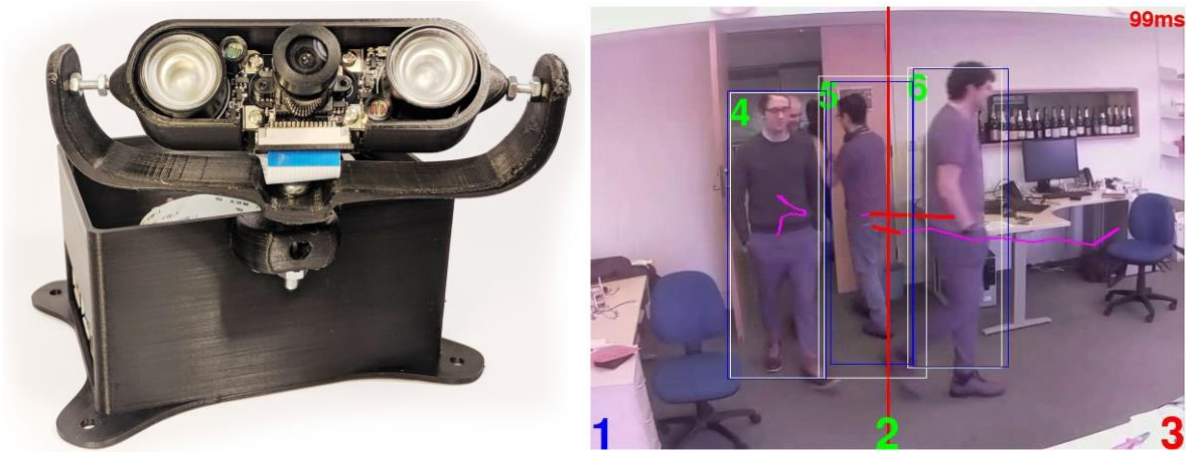

The image above shows an example of the processing that DeepDish carries out: a live video feed of one of the research group's shared offices is being processed in near real-time (the current frame having taken 99ms to process, giving an approximate frame rate of around 10Hz) to determine the number of unique individuals in frame and, by assigning each a unique randomly-generated ID (4, 5, 6 in the frame below), to detect when those individuals cross an in-frame boundary (the vertical red line). This allows the DeepDish camera to count the number of people entering and leaving a space, and thus the occupation of that space.

Overall, DeepDish demonstrated that such approaches to tracking are entirely achievable with relatively low-power (6—7W) edge hardware albeit with lower accuracy than a full-power platform could provide.

*References*
1. Vadim Safronov, Justas Brazauskas, Matthew Danish, Rohit Verma, Ian Lewis, and Richard Mortier. 2021. Do we want the New Old Internet? Towards Seamless and Protocol-Independent IoT Application Interoperability. In Proceedings of the Twentieth ACM Workshop on Hot Topics in Networks (HotNets '21). Association for Computing Machinery, New York, NY, USA, 185–191. https://doi.org/10.1145/3484266.3487374

# III. Data Handling and Analysis

## III.1. The ACP common real-time data pipeline

The real-time data pipeline is a five-layer architecture comprising Physical, Network, Integration, Data, and Application Layers.

**Physical Layer**. This refers to devices, including sensors, actuators, and building meters, that report data and exchange messages using a range of supported protocols and formats. Our deployment uses a wide range of sensors, as follows:

| Sensor | Measures | Channel | Cost |
|---|---|---|---|
| Smart plugs | power | Wi-Fi | $45 |
| Infrared Motion | motion | ZigBee | $17 |
| Door/Window | open/close events | ZigBee | $9 |
| $CO_2$ | $CO_2$, humidity, temperature, light, motion | LoRaWAN | $205 |
| Temperature | temperature | LoRaWAN | $129 |
| Tilt | tilt angle | LoRaWAN | $99 |
| Door/Window | open/close events | LoRaWAN | $ 99 |
| Water Leak | presence of water | LoRaWAN | $119 |
| Occupancy | occupancy, temperature, humidity, light, motion | LoRaWAN | $123 |
| DeepDish | people count | Wi-Fi | $100 |

**Network Layer**. This includes the devices that support message transfer from sensors, e.g., Wi-Fi access points, ZigBee and LoRaWAN gateways, Modbus components. Our deployment mixes Zigbee and Wi-Fi networks for short-range connectivity within areas, and a LoRaWAN network to provide backhaul interconnection within and between buildings.

**Integration Layer**. Provides services to collate and homogenise data received over different network types from different hardware devices, simplifying the process of building applications that consume sensor data. For example, consider a room containing smartplugs, LoRa sensors, and a smart electric meter. The smartplugs might upload data over Wi-Fi, LoRa sensors through LoRaWAN, and the smart electric meter via Modbus. Services in the Integration Layer ensure that data received over all these protocols are straightforwardly available through a single channel. The channel used by the ACP is a publish/subscribe ("pub-sub") system built over the Message Queuing Telemetry Transport (MQTT) and MQTT bridging.

**Data Layer**. Provides services to manage data streams arriving from the Integration Layer. Decoders receive and normalise the data from multiple sensors, before republishing it to a Real-Time Server (RTS) that hosts modules handling real-time stream processing, message routing, data storing, and making data available to the external application. The ACP's RTS is asynchronous and non-blocking, built using Vert.x. This layer also houses the configuration database that stores the spatio-temporal metadata of all the sensors as well as the object-level components in a hierarchical structure.

**Application Layer**. The client-facing layer, this provides APIs and UIs for those needing to access sensor data. The ACP implements this as a a server that acts as the point-of-contact with the platform for all client-side applications trying to access sensor data.

This data pipeline is informed by the following lessons learned during the design process:

- **All data is inherently spatiotemporal**. Over a long enough period, metadata about any sensor or part of the building is bound to change. For instance, a sensor could be moved to a new room changing its location, or a room be divided into two rooms changing its boundary. Tagging every datum with a timestamp and location information, and retaining all these historical changes, enables change tracking of deployments.
- **Container objects exist in a spatial hierarchy**. Sensors are deployed in different locations having different usage and access control policies depending on the granularity of location considered (e.g., building, floor, room, desk). This defines each component as a container that can hold one or more other objects, each of which could have its own set of sensors. This ensures changes to the fabric of the building are handled with minimal changes, e.g., moving a desk between rooms involves changing only its parent. An example of our in-building coordinate system is shown below; X and Y coordinates are distance from origin in metres and Z coordinate is the combination of floor number and relative height on the floor.

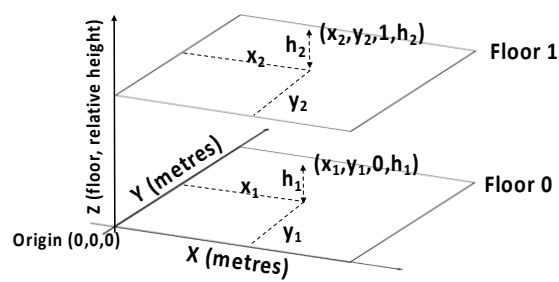

- **Stream processing of data**. Using a publish/subscribe model for data exchange ensures no polling of real-time data streams is required, reducing delivery latency, and so enabling real-time data analysis.
- **Asynchronous message transfer in a non-blocking framework**. Asynchronous, non-blocking message transfer ensures that any message from any sensor or system module could be analysed in real-time, and multiple modules can work on the same data concurrently.

### III.2. Real-time data-decoding architecture

For our experimental platform we assumed the need to support a diverse range of sensors transmitting their readings over a variety of network methods, with very little in common between them. Consequently, we designed an architectural layer on the platform to 'normalise' the data coming from the sensors making it more generally consumable further down the pipeline without constraining the healthy variety of features available across the industry. Some building management platforms require consistent formatting of data from the sensors that they support, while at the same time the vendors make efforts to incorporate sensors from multiple suppliers requiring some degree of incremental programming to accommodate those sensors.

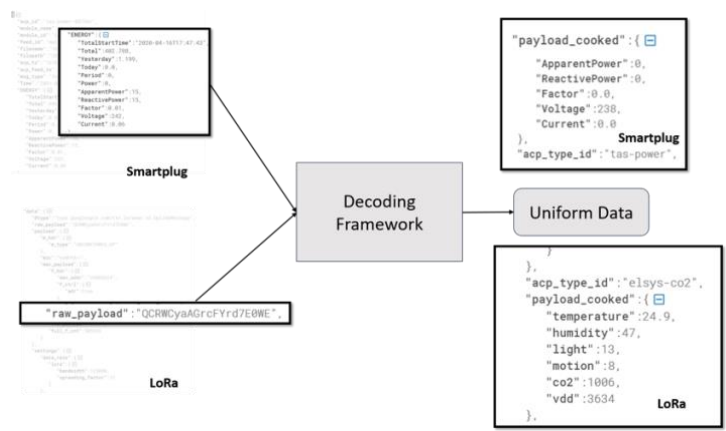

The messages transmitted by the sensors vary based on factors including the sensor type, vendor, design, and network type (e.g., Wi-Fi, ZigBee, LoRa). For example, the smartplugs we used put their unique device number in the MQTT topic rather than the message, in contrast to most LoRa sensors which include this information in the message itself. Moreover, although the platform strives to generate the data point's timestamp as far upstream as possible, and ideally at the sensor itself, it is not always possible to do so. For example, many real-world sensors simply cannot as they do not have a real-time clock. Simply receiving and storing the message without assigning time to it would make further processing a complex task, especially in a real-time system. The decoders take care of these problems, normalising messages from each sensor to include common features such as a timestamp if none were provided by the sensor. We implemented a set of Decoders for the different classes of messages received from the MQTT broker, all hosted by the Decoder Manager which supports 'plug-in' decoders, determines which decoder to use based on the message, and automatically registering new decoders added to the decoder set.

This component is extremely valuable and adds a great deal of useful flexibility to the ACP. Before it was added, original incoming data would contain unique data per sensor type as well as arriving in a plethora of original formats including 'raw binary' which made the rest of the processing pipeline complex and brittle. By decoding incoming data into the JSON data format, with any binary data retained as an encoded string within that JSON payload, these challenges were avoided.

Also useful was the design choice that the decoders would always *add* to the message rather than replace it. As illustrated in the image above the general behaviour of a decoder would include adding a *payload_cooked* JSON property to the sensor reading with standardised 'feature' values, for example any sensor defined to provide a CO2 reading would be decoded to include a "co2" value in parts-per-million in this *payload_cooked* field while retaining the original message in its entirety. This allows downstream processing such as a webpage with a CO2 heatmap to simply subscribe to all sensor data containing this "co2" value while ignoring the rest of the structure; other analytics that includes custom programming for a particular type of CO2 sensor can still access all of the original data in the message.

### III.3. A stream-processing data platform

As we have stated above, we assume that *timeliness* of reporting and analysis of all information flowing through the system is a key requirement for the real-time management of a building or multi-building site. This is in stark contrast to much prevailing Digital Twin work that takes a "data warehouse approach" for the sensor data, captured and held for retrospective historical analysis. While the ACP platform can and does record multiple copies of a given datum as does a data warehouse, the core platform simply *streams* data end-to-end. As the diagram below shows, different *actors* are responsible for storing data in accordance with the requirements of particular (sets of) applications, while other actors perform data manipulation and analytics, and deliver real-time information to end-users.

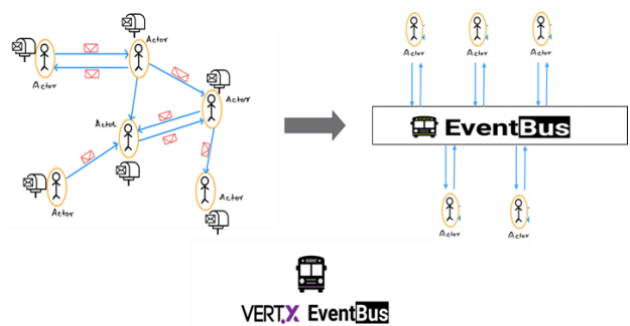

Using this real-time streaming model when building Digital Twins avoids the temporal pause that occurs when a data warehouse is considered the final destination for the real-time information flow. The Data Layer's Real-Time Server (RTS) ensures that the platform follows the stream processing approach of handling data.

The Real-Time Server (RTS) supports minimal latency processing of real-time data and asynchronous & non-blocking data management. This is guaranteed by following the Actor Model using Vert.x to receive and support analysis of data in real-time by multiple Vert.x modules called *verticles*. The actor model supports concurrency by enforcing that each actor (here implemented as a verticle with Vert.x) only interacts with other actors (verticles) through messages posted to its message-box. Aspects of this model that we found relevant to the implementation of our Digital Twin included:

- **Asynchronous message passing**. The actor model guarantees that the RTS adheres to an asynchronous message-passing paradigm, providing real-time data handling from whichever source it is received. The *FeedHandler* receives a message arriving at the RTS and publishes on the *EventBus* to be used by other verticles. Unlike the standard actor model, where each actor has its own message-box, in our implementation, the Vert.x *EventBus* acts as the common message-box for all verticles. *EventBus* removes the possibility of deadlocks that could occur because of some delay in data processing at some actors ensure non-blocking modules.
- **Non-blocking modules**. Using the Vert.x library with the actor model provides support to build a non-blocking framework for the RTS. The message exchange is performed using a publish/subscribe approach. Any verticle accesses data by subscribing to the *EventBus* and sends messages by publishing it to the *EventBus*. The model also ensures that any communication between two verticles also happens only through the *EventBus*.
- **Modular server**. Each verticle is an independent actor ensuring that the RTS is modular. This guarantees that any number of modules could be added or removed as needed without affecting any existing verticles. As a result, a standard production implementation could have thousands of verticles accessing data concurrently.

As well as following the actor model for concurrency, all verticles on the platform act as stream processors. Unlike most systems, which store the data in a storage unit and then query or perform computation over it, stream processing differs in two key ways:

- **Events substitute messages**. Verticles on the platform react to the incoming stream of events instead of a message or a batch of messages. Many sensors will send periodic updates reporting the status quo – these are typically not of interest to verticles, which are concerned rather with events indicating some change of state. For instance, a verticle controlling lights in a room might only be interested in the events indicating a change from unoccupied to occupied, or vice versa, and not in processing periodic messages of current occupancy which the sensor sends. Working with events also ensures that the platform handles timeliness of the event being processed by a verticle.
- **Reversing the norm**. Unlike the store-then-analyse approach, stream processing focuses first on enabling real-time reactive processing of data (events). Upon receiving a relevant event, a stream processing application (a verticle) reacts to the event by updating some information, creating another event, or simply storing it. The result is that data can still be archived for historical processing, but this does not negatively affect the performance of real-time processing. When coupled with MQTT, the RTS isn't restricted with the number of topics it can support. Moreover, IoT features like keep-alive and last will and testament are supported by MQTT and thus supported by the platform.

For our Digital Twin, the verticles (implementing the actors in the introduction to this section) were found to fall within the following classes:

- **Data ingestion verticles**, which receive data from the MQTT broker and publish the same on the *EventBus* (e.g.,*FeedHandler*),
- **Data storage verticles**, which subscribe to the *EventBus* for any new data and store it for future usage (e.g., *MessageFiler*),
- **Real-time analysis verticles**, which analyse the stream of data in real-time performing tasks like identifying events such as the measured CO2 level crossing a threshold or a power outage and broadcast the results of such analysis as derived events on the *EventBus*.
- **Outbound verticles**, which push the data to the outside world. Outbound verticles include the *MessageRouter* verticle, used to share data to other similar systems; the *RTMonitor*, used to interact with client-side applications.

### III.4. Real-time simple and complex event recognition

The increasing penetration of sensors able to feed into intelligent applications coupled with the real-time processing capability of the ACP makes smart environments even smarter by enabling applications that detect critical events in terms of complex sequences of simple data reported from one or more sensors. Complex Event Processing (CEP) is a popular strategy for detecting such events, enabling flexible identification of complex events based on identification of simple events from multiple sensors in appropriate relation to each other. However, existing approaches have several crucial limitations.

First, the existing atomic event detection approaches only consider a single sensor for an atomic event. However, data from other sensors in the vicinity of a sensor are likely to have a relationship with that sensor's reading, e.g., a high temperature reading from just one sensor in a room might be an anomaly whereas a consistently high reading from many sensors in the same room gives greater credence to the reading.

Second, any atomic event is detected using an exceeding threshold or an instantaneous divergence from the actual data distribution. However, a single instance of exceeding the threshold or a small window of diverging distribution does not capture the temporal variation in data due, for example, to noise. For example, a single instantaneous high-temperature value does not imply that the room temperature is uncomfortable unless the temporal variation in temperature reading is observed over time. Moreover, the time of the year and when the data is generated are also relevant to the interpretation of that datum: a high temperature reading during summer might be normal, but the same reading in winter could mean a broken heating system.

Finally, several CEP systems employ backtracking to detect complex events, which often go several levels deep, making the system slow and so hindering real-time detection.

In response we developed *Watcher,* a system providing robust end-to-end complex event detection while ensuring that the spatial positioning of the sensor and the temporal variation in data is considered. Key features include:

- A strategy to detect that a sensor is diverging from the underlying observed data distribution over a time period by calculating a Divergence Score.

- Detection of atomic events using the Divergence Score, correlation with nearby sensor data, and temporal information.
- A CEP engine that takes as input a set of atomic/complex events and CEP rules to detect complex events, while ensuring minimal backtracking, faster processing, and low-latency detection.

# IV. Buildings and People

## IV.1. Integrating building information with sensor data

As we suspect will be the case for most of our built environment, we found that BIM data for existing buildings at the University of Cambridge, where any was even available, was simply not fit for purpose for our Digital Twin work. Even after conversion to the standard IFC format, it was primarily 'drawing' focussed as the construction industry clearly has a much greater focus on the size and shape of construction elements than any consideration of the need for stable identifiers that will persist once the building is occupied and in use. There is also little regard for BIM data rapidly becoming out-of-date, in is due to limitations in current systems for supporting incremental updates.

We therefore had to create new BIM data for the William Gates Building that was more aligned to the needs of the Digital Twin. Doing so required that we undertake our own hands-on building survey to ensure that the BIM data produced contained all the information we required. This issue is certain to be commonly encountered, and while many researchers have proposed numerous alternative data formats, there are very few real-world implementations backed by deployment experience. Having done this tedious but necessary background work, we report the following lessons learned and recommendations based on our experiences:

- **Any digital twin work must plan to create such BIM data as will be suitable for the work**. We have seen many examples of digital twin development where this issue is glossed over resulting in fundamental weaknesses such as sensor data appearing disconnected from the building other than the recorded coordinate location of the sensor.
- **Simply exporting BIM data from drawing packages results in data structures that are too poor to support smart buildings**. The industry decades-long theme that as-built BIM data can be exported from the relevant drawing packages is naïve. While data from one computer system can be converted to another, drawing packages simply do not contain the information required by a smart building, focused as they are on traditional construction requirements.
- **A consistent per-object incremental update data strategy for *all* metadata, including sensors, sensor types, BIM objects, people, and institutions is required**. The databases containing this information have a common format such that any change to an individual data object would result in a timestamped *new* record being created and the previous entry being given the same timestamp as an *expiry* date. Again, current BIM data is typically poor in this regard, being simply composed of records that without an expiry date. Our web platform can provide the history of any object that has ever been used. In this way the records for each object form a history (somewhat as the current fad for blockchains espouses though without the ridiculous claims and problems such systems seem inevitably to suffer from).
- **Computers are fast, sufficiently so that they can easily support APIs to access metadata alongside real-time sensor data, if reasonable care is taken in implementation**. Although the data structures necessary to record BIM, sensor and other related data are considerably larger than they might previously have been, reasonable care in implementation means that querying and serving all this data with low latencies and high throughput is straightforward. For example, examining query patterns quickly shows that historical data is often queried by day, so sharding stored data by day makes such queries easy to answer efficiently. As most queries to such a system are no ad hoc interactive database-style queries, but stored procedures providing content to webpages and other human-friendly UIs, such analysis is relatively easy to do *a priori*.

- **JSON is a flexible data storage format with a rich ecosystem of tooling that makes it efficient and straightforward to manipulate**. We successfully use JSON for *all* of our data structures, storing JSON blobs in traditional relational databases to give the combined advantage of the flexibility of JSON and the robust indexing and backup tooling of DBMSs. This contrasts with, for example, XML and both traditional relational and graph database approaches, where the detailed specification of data formats required makes changes and extensions difficult to accommodate. However, while there is neither need nor benefit in standardising every single parameter, it *is* worth standardising the "envelope" format of different data objects ensuring consistent timestamps, identifiers, and object type identifiers (sensor, sensor type, building object, person, etc). This greatly simplifies consistency of tooling across object types when, e.g., viewing the history of a sensor or a building object.

- **The spatio-temporal focus of Digital Twin work makes it crucial to rigorously specify treatment of timestamps and locations as these are common to all data objects in the system**. We use timestamps with arbitrary resolution, able to record time to the second, millisecond or picosecond, and specify location in a well-defined coordinate system framework for each building while allowing different coordinate systems to be used in different buildings while ensuring each such building coordinate system has a defined transformation to global WGS84 (i.e. latitude/longitude/altitude) as well as a transformation to ($x$, $y$, $z$) meters given some arbitrary origin pertinent to that building. This is in contrast to other projects having point-cloud 3D scans for some spaces and building or site plans for the macro picture with no consideration of translation to a common coordinate system.

## IV.2. Visualising real-time built-environment data

Our platform provides authenticated web access to management and data-access pages as one would expect, alongside the real-time *RTMonitor* component referred to above (cf "A stream-processing data platform"). This streams real-time information from our platform to any page served by the web platform, e.g., a real-time heatmap for every office and space on a selected floor of the William Gates Building. As an example of the benefit of such a system, the heatmap image below depicts the default data used (temperature), but it is straightforward to feed any sensor parameter into this style of display – during the pandemic, understanding $CO_2$ concentrations around the building became unexpectedly important in managing access to and use of the building.

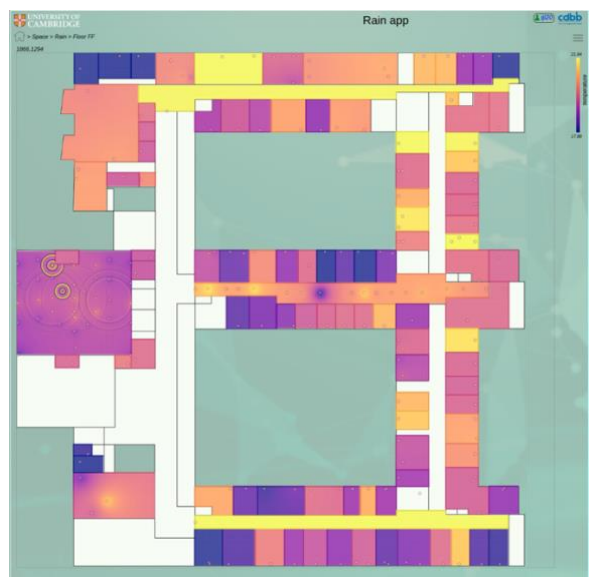

For example, the large room on the left of the screenshot is our 200-person lecture theatre LT1 within which the circles drawn are our *splash* animation used to represent each real-time update of data from an individual sensor with a sub-second latency from the real event occurring.

With this comprehensive end-to-end real-time implementation on our Digital Twin, we gained considerable practical experience summarised in the following 'lessons learned':

- Traditional 'static' web pages are well suited to illustrating spatial patterns but can conceal high latency in the data. Typically users adapt to whatever the data processing timescale was in producing the visualisation as 'normal' even when, in some cases, that might be days or even years.
- There are important temporal patterns in the data that are too easily ignored when data is shown simply as time-series graphs of individual sensors.
- Use of an animation for incoming data greatly helps in understanding temporal patterns and real-time latencies while witnessing actual events such as people entering rooms or walking down corridors.
- Given an example of a real-time heatmap, traditional techniques of rendering that heatmap are not fit for purpose for the web display of a digital twin. The usual technique is to collect all the data, calculate the colour distributions and then render the entire page, but this cannot feasibly keep up with the stream of sensor data being served to the web page. Instead, we implemented the heatmap with an incremental algorithm that recognises the spatially relevant impact of each sensor reading and only updates that part of the web page.
- The floorplan heatmap example given above exemplifies the need for an integrated data strategy for building information, sensor metadata, and sensor readings. In this example the web code uses the spatial extent of each building object (offices and other spaces) to define the area covered by each heatmap within the overall mosaic.

## IV.3. The centrality of people

Traditional building management systems are primarily concerned with physical assets in the building, reporting data such as the speed of a cooling system fan or the temperature of a physical space. Occupants of the building either do not appear in the system at all or are indirectly recorded in ancillary data such as cardkey entry logs.

Our work recognises the importance of 'people' as first-class data objects within the digital twin. This has the benefit that the dominant entity affecting the amenity of the building is included in the analysis but immediately raises important privacy concerns – though we note that existing evolving sensor deployments and computing platforms used in the built environment *already* raise privacy concerns which have generally been inadequately addressed. We assume the sensor deployments will increase in density, and the sophistication of the sensors will increase for example including the use of AI camera-based sensors so these questions are rapidly becoming more important.

Businesses already legitimately hold information on their employees, usually shared between multiple enterprise systems. Currently it is more of an implementation detail than an actual technical constraint that typical building management systems have no access to this data. We explored what personnel data is both useful and acceptable to use in a digital twin, finding the following to fit:

- Employee name and company identifier, as contained in the employee directory
- Employee departmental affiliation (in Cambridge this is multi-valued)

- The complete 'corporate entity' hierarchy (in the University, Schools and Departments, Divisions, and Institutes)
- Space allocation data such as who is allocated to which office

We enable, in principle, all building occupants to log into our platform to access data generated by and concerning them providing a degree of legibility, but also multiplying the value of the data and analysis provided as it can now support a far greater range of requirements than those with which a building manager would traditionally be concerned. However, such broad access to the system requires a privacy framework and the employee data above immediately provides parameters across which the privacy rules can be applied.

A general theme in our platform is that *all* data is spatio-temporal (weaker systems tend to consider data to be only one or the other, or neither). For example, built asset data is often spatial (recording things like floorplans) but ignores temporal change (how a given floorplan evolves over months and years). People data is treated no differently – record updates are handled using the same timestamped incremental update techniques as for the other reference data.

Supporting 'person' data as a symmetrical first-class data objects in our system helped clarify that any common metadata platform must necessarily efficiently support hierarchical data as well as simple linear searches for objects. A single-object lookup of person with identifier *ijl20* is expected to be rapid, but we also require that queries such as 'does person with identifier *ijl20* belong to the School of Technology?' are similarly rapid while recognising the organisational data objects are stored as a hierarchy. This access is not dissimilar to 'does person *ijl20* occupy the *WGB*?' where the recorded metadata for *ijl20* shows they occupy building object *SE13,* and it is necessary to traverse a hierarchy to determine that is indeed a sub-space of *WGB*.

Apart from the authentication and access to web views of the data via the privacy framework described in more detail below, we have not found it necessary to identify *individuals* within our analyses. For example, the dense array of sensors in Lecture Theatre 1 provides valuable real-time information and historical patterns for the use of the room but leaves occupants anonymous for any subsequent analysis. Similarly the DeepDish intelligent camera provides detailed count values while discarding all image data.

## IV.4. A privacy framework

As introduced above, a privacy framework is required in a future where the occupants of the building are simply peers with the Building Manager in their use of the platform, albeit with alternative appropriate views of the data. A simple example of this is to enable display screens or web pages that provide location-specific information such as this screen in Lecture Theatre 1 showing CO2 levels:

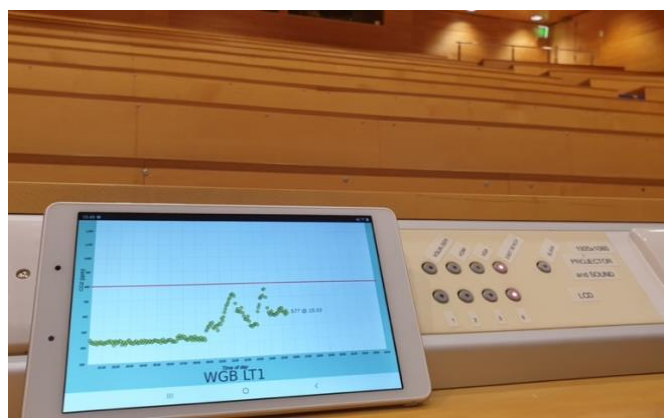

Note that the information on these screens may be derived by aggregating from multiple sensors in multiple locations, enabling privacy without requiring that the display and its sensors be co-located in the same room. The three rules we can identify as most generally applicable are:

1. Building managers can see everything about the building if not the occupants.
2. Occupants can see metadata, sensor information and derived analysis relevant to the space they occupy.
3. Data for some shared spaces within the building such as meeting rooms should be visible to all members of the department.

Hopefully it is clear from even these rules that effective linking of the data between diverse data structures, in our case including the real-time sensor data, is a pre-requisite. With these rules as an initial guide we implemented a privacy framework enabling flexible specification of access rules. Our implementation provides an abstract mechanism for defining privacy rules (in our platform called "permissions") including the three above. In fact we treat permissions as first class data objects with the same temporal historical logging as any other. Our use of JSON as our format of choice for object metadata naturally leads to the flexible use of keys between objects in declaring any relationship. JSON-LD is a format that formalises this use of relationship keys similar to the Resource Description Format of XML.

Our permission rules amount to a constraint upon the keys that can be traversed between objects as the data is accessed, and the solution chosen that permission on a 'parent' is deemed to imply that permission on the 'children' has proven to be efficient for most of the practical rules we have considered. In other words, holding 'sensor data read permission for the first floor building object' implies 'sensor read permission for sensors in individual offices on the first floor'. We exploit the fact that traversal up (i.e., through parents) in tree-structured data is typically much more efficient than searching down the tree (only ancestor are encountered, never siblings). Given the 'occupies' and 'measures' defined relationships between people, offices and sensors, essentially the "a person should be able to see sensor data from their office" becomes the declarative permissioning rule:

```
Allow 'sensor data read' permission for person P for data from sensor S
if person P has the 'occupies' relationship with building object O
and sensor S has the 'measures' relationship with building object Q
and building object Q is identical to or a child of building object O
```

The consequences of this rule are as illustrated in the web screenshots below. Anyone with access to the system can see all the sensor data from LT1 while offices are left blank. Testing access to the data to one example office (FNO5) shows it is only available to the particular user recorded as occupying that office.

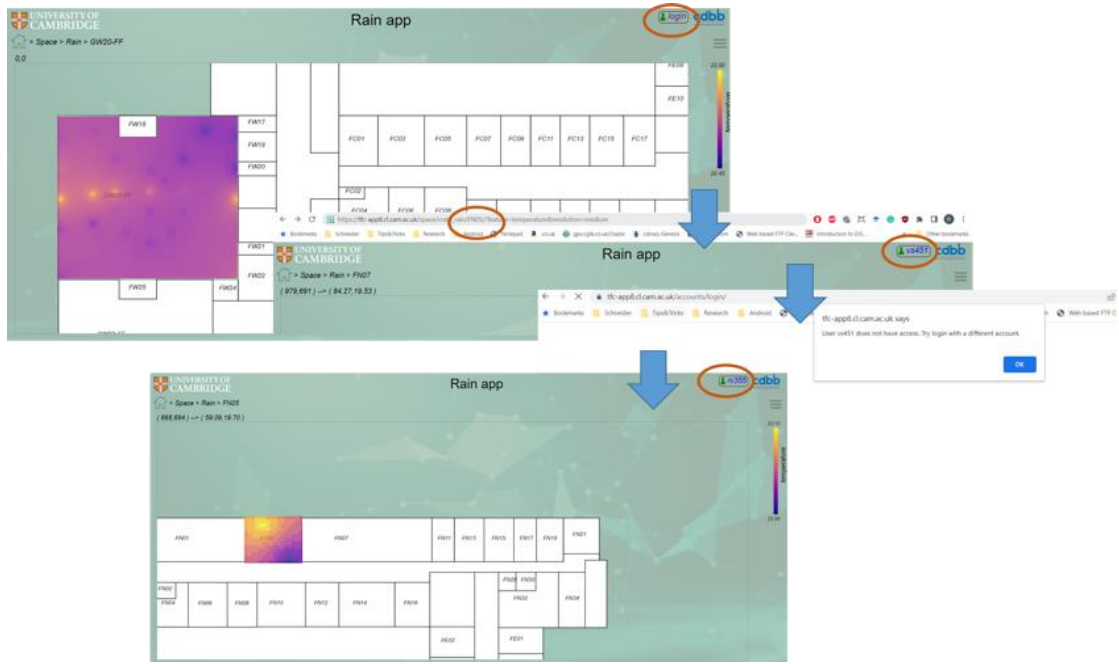

*References*

1. Justas Brazauskas, Rohit Verma, Vadim Safronov, Matthew Danish, Ian Lewis, and Richard Mortier. 2021. Real-time data visualisation on the adaptive city platform. In Proceedings of the 8th ACM International Conference on Systems for Energy-Efficient Buildings, Cities, and Transportation (BuildSys '21). Association for Computing Machinery, New York, NY, USA, 210–211. https://doi.org/10.1145/3486611.3492225

# V. Reflection on Process

As reported above, this project has produced considerable output: the ACP software itself, the data obtained through the platform, numerous papers reporting on various aspects of the project, and of course the primary goal of greater understanding of the challenges and opportunities presented by this context. This required designing and building an instance of the platform at a scale capable of collecting significant amounts of data over an extended period, which presented several inherent operational challenges that a smaller, less ambitious, lab-based deployment could not have uncovered. First, selecting, procuring, and calibrating equipment before deployment, followed by the ongoing management and operation of those deployments was achieved with extremely low staffing, just 2.3 FTEs. Second, the nature of the project required dealing with high volumes of specialised components where the individual component cost was relatively low. This meant we were often selecting and buying sensor components and pre-built sensors costing £10 to £100 each.

We were also forced to carry out this project, with its need for in-building systems development and deployment, during a period of repeated national lockdowns, building closures, and occupancy constraints due to the COVID-19 pandemic. This made an already ambitious project incredibly challenging, although the urgent demands of the pandemic very much validated some of the methods and technology we were developing in the project, providing, for example, means for the Department to assess the efficacy of different ventilation regimes as students returned.

We were successful completing the planned research work within the timescale planned and on budget, thanks in part to much appreciated and essential administrative support from CDBB. However, a few factors made this more difficult than was initially assumed and it is perhaps valuable to reflect on them to learn lessons for the future about how to carry out this kind of work efficiently:

- As the project progressed, we learned we should have spent more time establishing common ground in understanding procurement targets and patterns, and the need to phase procurement so as to best understand available product in a quickly evolving sector.
- We had the need to obtain batches of equipment in some cases, due to component, firmware, etc variation forcing process modification in calibration and deployment making it time consuming to buy piecemeal. This was not something we fully understood before starting the project.
- The period of the project was one of considerable supply chain disruption leading to high variability in availability and delivery time of components, even using a range of suppliers, and a more adaptable process would have reduced the workload. Providing accurate monthly financial reporting was extremely difficult when much of the equipment stretched to highly variable multi-month delivery times.
- Exacerbated during the Covid period, University procurement and financial processes, including the supporting tools, are in general not well suited to continuous analysis of equipment procurement and delivery. Our fall-back procedure of maintaining spreadsheets was found to have the considerable disadvantage of being unconnected to the data that might be reported by the central systems.

# VI. Appendix: Submitted and Published Papers

# Do we want the New Old Internet? Towards Seamless and Protocol-Independent IoT Application Interoperability

Vadim Safronov
University of Cambridge
vadim.safronov@cl.cam.ac.uk

Justas Brazauskas
University of Cambridge
jb2328@cam.ac.uk

Matthew Danish
University of Cambridge
mrd45@cam.ac.uk

Rohit Verma
University of Cambridge
rv355@cam.ac.uk

Ian Lewis
University of Cambridge
ijl20@cam.ac.uk

Richard Mortier
University of Cambridge
richard.mortier@cl.cam.ac.uk

## ABSTRACT
IoT is developing rapidly with frequently appearing new wireless standards and applications. However, besides a large number of IoT benefits, its further development is now being slowed down due to the repetition of old Internet development flaws while dealing with IoT heterogeneity. The current misleading trend aims to solve all IoT interoperation problems by inserting IP Addresses into those wireless protocols where the IP stack clearly slows down application performance and drains the battery, e.g. LPWANs such as LoRaWAN and SigFox. This paper tackles IoT heterogeneity from a different perspective: it is the application interoperation which matters the most. The protocols beneath the application layer shall work for smooth upper-layer service provisioning where the IP shall be just one of the many underlying integration options instead of being the essential one. Inspired by previous proposals for a more flexible inter-networking architecture, this paper applies those theoretical concepts in practice by proposing a protocol-independent distributed interoperation model for smooth service provisioning over heterogeneous IoT wireless contexts. The arguments pro the IP-agnostic IoT application interoperation are supported by the model's prototype which showed 1.6-2 times faster MQTT application operation over LoRa and WiFi compared to the legacy IP-based MQTT provisioning over that protocols.

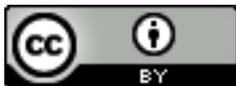

**ACM Reference Format:**
Vadim Safronov, Justas Brazauskas, Matthew Danish, Rohit Verma, Ian Lewis, and Richard Mortier. 2021. Do we want the New Old Internet? Towards Seamless and Protocol-Independent IoT Application Interoperability. In *The Twentieth ACM Workshop on Hot Topics in Networks (HotNets '21), November 10–12, 2021, Virtual Event, United Kingdom.* ACM, New York, NY, USA, 7 pages. https://doi.org/10.1145/3484266.3487374

## 1 INTRODUCTION

Smart spaces and cities have become a reality thanks to the IoT. We now have computer-controlled office climates, remote control of various smart appliances, asset automation and building management. To provide a wide range of IoT services, smart environments deploy sensors and actuators in order to react to any critical changes in a timely fashion. Diverse functional requirements and hardware constraints mean different IoT devices often use disparate communication protocols to interact with each other. E.g. LoRaWAN/Sigfox provides low-power, long-range and low-bandwidth sensors, while ZigBee/Z-Wave gives lower latency and higher bandwidth at shorter ranges.

Currently most IoT application protocols (e.g. MQTT [35]) assume they run over the commonly used Internet Protocol (IP). That assumption creates interoperation problems, especially for those IoT wireless protocols which are not meant to be layered on top of IP (e.g. LoRaWAN or other LP-WANs) as bringing IP stack to them significantly decreases their performance. Therefore, IP-agnostic IoT interoperation, with its lower memory footprint compared to IP-based IoT networks [15], can save scarce hardware resources of constrained devices.

Speaking more generally, the Internet initially has appeared as a compound of scattered local networks (e.g. MIL-NET, ARPANET [13] etc.) where the interconnection design decisions were taken on the go and thus need to be reconsidered for our current reality. Modern IoT devices with different functionalities and wireless technologies cannot be simply embraced by one generic network protocol like it was done in the early years of the Internet. Unlike classic power-plugged computers (which haven't changed so much from the network communication perspective), different IoT devices are designed for different purposes where each device performs a specific list of functions contributing to overall IoT service provisioning in a smart environment.

Inspired by general internetworking architectures [3, 12, 37], this paper argues that further IoT development shall move from current IP-dependent homogeneous interoperation towards a decentralised protocol-neutral heterogeneous model in order to seamlessly provide IoT services and applications over any set of underlying wireless technologies/protocols. It is the application interoperation which matters the most in the heterogeneous IoT reality where the IP protocol is just one of the tools for delivering upper services. Based on the concepts introduced in the Plutarch architecture [12] we propose a protocol-independent distributed model for seamless application interoperability in IoT environments. The aim of this paper is to make the initial move towards the correction of IoT interoperation paradigm on its early development stage until the technology is beyond the point of no return for escaping IP-dependency with its subsequent limitations.

Our IP-neutral IoT interoperation prototype highlights IP-free application interoperation benefits through testing seamless MQTT over WiFi and LoRa: two heterogeneous protocols where the first is high-bandwidth and works with IP while the latter comes without an IP stack and designed for low-power low-bandwidth communications. The prototype was evaluated against the standard LoRaWAN IP-based MQTT provisioning over that contexts showing 1.6-2 times faster application interoperability on the majority of links without power consumption overhead for battery-powered LoRa sensors.

## 2 RELATED WORK
### 2.1 Improving IoT interoperation

**IP stack implementation in non-IP-based wireless protocols.** Ongoing work seeks to improve IoT interoperability by providing an IP stack for non-IP protocols so that all IoT devices can be interconnected through IPv4/v6. SCHC [30] provides generic compression mechanism which can be applied to compress IPv6/UDP headers in low-power wide-area networks (LPWAN). Gimenez et al. [20] introduced an RFC draft with SCHC adaptation for LoRaWAN. The SCHC mechanism has also an open-source Python implementation called OpenSCHC [2].

Another example is ZigBeeIP [19] which is an IP-based version of a standard ZigBee specification [4]. ZigBeeIP is based on the same IEEE 802.15.4 PHY/MAC standard and integrates with 6LoWPAN [31] in order to provide IP, TCP/UDP and upper application functionality. IETF 6Lo working group [1] is focused on bringing of 6LoWPAN to 6Lo [21], i.e. implementing IPv6 in other IoT wireless low-power standards such as LoRaWAN [25] and SigFox [28].

**Porting of IP-dependent application protocols to non-IP contexts.** Stanford-Clark et al. presented MQTT-SN [36], a modified MQTT version for non-IP sensor networks which provides integration with the standard MQTT [35]. The mapping between MQTT-SN and MQTT messages is done through an aggregation gateway which receives MQTT-SN packets from all connected clients. To not overload a broker by a large number of MQTT sessions, the aggregation gateway can establish just one general MQTT connection to the broker and decide which MQTT-SN packets will go further to the server from the sensors. Although there are more similar application integration solutions such as zigbee2mqtt [23] or LoRaWAN-CoAP integration [10], they are not aimed at general application interoperation approach for neither of that protocols and work for a limited set of specific intercommunication use cases.

### 2.2 General internetworking architectures

Crowcroft et al. [12] propose Plutarch to solve the network heterogeneity problem by an explicit mapping of names, addresses and routing rules on the border of adjacent networks. The proposed concept of *contexts* and *interstitial functions* allows heterogeneous communications through the definition of query models and mapping rules between any two distinct contexts, i.e. networks with different addressing schemes, packet formats or protocol used. Inspired by the Plutarch concept and by the role-based communication architecture suggested by Braden et al. [8], Dunkels et al. propose an adaptive wireless communication infrastructure called Chameleon [14]. Chameleon architecture is protocol-neutral and provides a group of communication primitives and abstractions for seamless interconnection of heterogeneous wireless sensor networks.

Based on the ideas of layered and separated network naming [7, 11, 22] Ahglren et al. [3] propose a node identity (NID) internetworking architecture where NID is decoupled from a network protocol and can be used to bridge across heterogeneous network contexts. Unlike completely decentralised approach proposed by Plutarch, Ahglren et al. claims that a core network is still needed (either based on IP or any other addressing protocol) in order to manage the distribution of NIDs across different contexts and escape potential storing of a large number of NIDs on each communicating node.

The ideas proposed by previously mentioned internetworking architectures were covered in later works on ILNP

by Atkinson et.al [5, 6] and then by Yanagida et al. [37]. ILNP employs "Locator" and "Identifier" namespaces in order to separate a node unique identifier from transport and network protocols. The discovery and binding of NIDs is also done by ILNP. The concept is IP-neutral and can potentially be employed in other wireless sensor networks.

## 3 IOT INTEROPERATION MODEL

Of the internetworking architectures reviewed in Section 2.2, Plutarch has one of the highest levels of abstraction and applies to a wide range of intercommunication scenarios. Its flexible, IP-neutral, decentralised approach makes Plutarch a suitable basis for making distributed seamless application interoperation over heterogeneous IoT networks.

### 3.1 Main concepts

Figure 1 represents our proposed distributed and protocol-neutral interoperation architecture. The model is based on the Plutarch abstractions of *contexts* and *interstitial functions* (IF) adjusted for IoT scenarios, and on the *node identity* (NID) concept proposed by Ahlgren et al. [3]:

- *Context*: an area of homogeneity in message formats, transport protocols and naming services. For a smart environment, a context is a group of IoT devices which use same network protocols, packet formats and addressing schemes. E.g. a group of smart bulbs operating through the same WiFi network forms a single context. Similarly, several ZigBee smartplugs connected into a mesh network are considered as the same context.
- *Interstitial Function*: a software function which is implemented within a network gateway and is responsible for correct traffic transmission between different contexts. Besides being a relay which maps messages from one protocol to another, an interstitial function is also in charge of enforcing control policies on a context border as well as within a context: e.g. ensures a certain level of QoS, filters sensitive or redundant data and automates context workflows.
- *Node Identity*: a protocol-independent device unique identifier which is decoupled from the address provided by a context and its protocols. For example a NID can be derived from a public key of a public/private key pair.

IoT sensors and actuators of the same context are managed by a single *smart gateway* (SGW) with an IF which performs the following functions:

- **interconnection of adjacent IoT device groups**: provision of local IoT interoperation within a single device group, communication and traffic control between adjacent contexts.

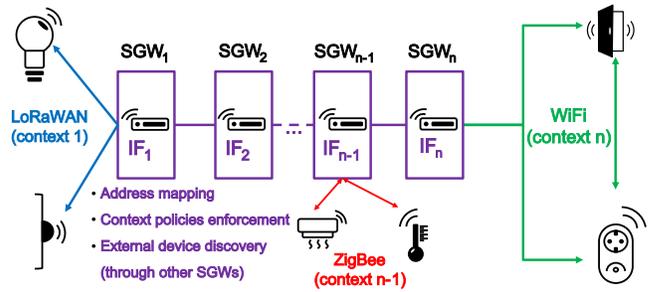

Figure 1: IoT interoperation model.

- **enforcement of context policies** defined by a user or delegated by a general control platform such as a Building Management System (BMS). Some lower-layer context policies can also be in place aiming for correct traffic transmission between the contexts and making sure that Quality of Service (QoS) remains at the same level (e.g. while switching from high-bandwidth contexts to low-power networks with much lower transmission rate and higher latency).
- **IoT device local or external discovery** through requesting other SGWs. To provide robust end-to-end (E2E) connectivity, an SGW is also responsible for device discovery in other contexts through interaction with adjacent SGWs. Each SGW has a dynamic address table that maps a device NID to its appropriate in-context address (e.g. its L2 or L3 address), or to the address of the next SGW on the path to the destination device.

### 3.2 Address resolution and routing

Interoperation between non-adjacent IoT contexts separated by several heterogeneous networks require additional address resolution and routing functions discussed in [3, 12, 37]. While the model concepts allow for support of routing functionality without sticking to IP addresses, it is crucial to mention, that in a large number of IoT applications, such as smart space automation and management, address resolution can be less tricky in fact. In smart environments the majority of IoT devices, such as sensors and actuators, stays at the same place (e.g. door open/close or temperature sensors), so the mappings can be stored statically on SGWs without requiring dynamic address resolution most of the time. Implementation and functional complexity of IP-free switching between different contexts on an SGW can be negligible, as most such IoT scenarios realistically require a few protocol types for seamless service provisioning. E.g. a smart building can employ WiFi for power-hungry and/or high-bandwidth IoT devices such as surveillance cameras and power-connected household appliances (fridges, microwaves, etc.), while using

ZigBee/Z-Wave and LoRaWAN/SigFox standards for indoor and outdoor battery-powered low-bandwidth sensors/actuators respectively.

## 3.3 Smart policies, QoS guarantees and dealing with redundant data

An IF is also responsible for enforcing administrative and user-defined policies within a controlled context. An illustrative example of such IF functionality for dealing with sensitive and redundant IoT data is given in our Smart Coffee Pot project [24]. The IoT appliance consists of a weight sensor for the coffee pot and two smart plugs that monitor power consumption from a bean grinder and a coffee brewer. All devices are connected to the same local WiFi network shared by a Raspberry Pi (RPi) model 3B+. Instead of sending raw and potentially sensitive sensor readings from a local WiFi to a remote server for data processing, the RPi side implements an IF with a local event-recognition mechanism that looks for correlations between the pot's weight and the power consumption of both the grinder and the coffee machine. Thus, nobody's sensitive nor any redundant raw sensor data is sent outside of the local WiFi. Only discrete events such as 'coffee has been made' or 'the pot is empty' are passed along to the remote web platform displaying the coffee pot's status to the public. Although the described IF implementation is WiFi-to-WiFi, the core of its smart logic can be ported to other inter-context scenarios, e.g. WiFi-LoRaWAN or ZigBee-LoRaWAN where filtering redundant traffic is crucial for keeping QoS at the same level as saves scarce resources of low-power low-bandwidth networks.

## 4 PROTOTYPING AND TESTING

### 4.1 IP-agnostic MQTT interoperation

Figure 2 depicts the main components of our IP-neutral application interoperation model's prototype, which demonstrates seamless MQTT over WiFi and LoRa wireless standards. The main prototype aims are: (1) testing the effectiveness of IP-agnostic IoT application interoperation against the legacy IP-dependent one, (2) showing the model's potential for dealing with high level of IoT heterogeneity in an IP-free manner: WiFi provides a high-bandwidth power-hungry and short-range IP network, while LoRa has no IP support and is designed for a wide propagation range at low bandwidth rates (see its official documentation [25]). Besides E2E MQTT provision over LoRa and WiFi, other main artefacts of the deployment include a LoRa MQTT client and a LoRa-WiFi IF both written in micropython [29]. The components of our interoperation model are briefly described below.

**LoRa context.** Both Sensor A and Sensor C run MQTT clients on top of a Pycom LoPy4 controller [33] connected to a Pysense shield [34]. Sensor A imitates the behaviour of a Class A LoRaWAN device (battery-powered sensor, e.g. a temperature or a humidity sensor) regularly publishing sensor readings to a given MQTT topic. Sensor A is able to receive incoming messages only for a shot bit of time just after publishing an uplink MQTT message while sleeping all other time in order to save its 500 mAh Li-Pol battery (connected to a Pysense shield). Sensor C has an external power supply and imitates the behaviour of a Class C LoRaWAN device (e.g. a smart plug or a light switch) always listening for incoming MQTT messages and regularly publishing readings to a given MQTT topic. The LoRa IoT Controller is a power-plugged LoPy4 device which is subscribed to MQTT topics of sensors A, B and C. The controller manages both sensors A and C by publishing LoRa MQTT commands to their control topics.

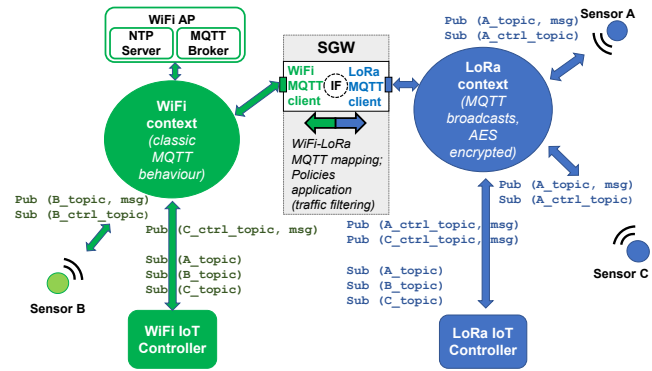

Figure 2: IP-agnostic MQTT interoperation prototype.

**Smart Gateway (SGW).** The gateway consists of a LoPy4 board with a power-plugged Pysense shield and implements an IF with WiFi and LoRa MQTT clients written in micropython. The SGW performs all necessary WiFi-LoRa mapping and control functions for seamless MQTT provision over the given contexts. The gateway functions also include context policies enforcement such as MQTT topic filtering and inter-context QoS control.

**WiFi context.** All WiFi devices run power-plugged Raspberry Pis (RPis) with the Raspberry Pi OS [18] on top. Wireless connection is provided by the WiFi Access Point which runs hostapd [26], a mosquitto MQTT broker [16] and a local NTP server [27]. Sensor B employs a paho-mqtt client [17] and imitates a WiFi MQTT sensor. The WiFi IoT Controller also employs the same paho-mqtt lib and is subscribed to all sensor A, B and C topics controlling only the local Sensor B by publishing MQTT commands to its control topic.

**LoRaWAN MQTT broadcasting.** Using a standard MQTT TCP-based broker in LoRa would result in significant bandwidth consumption greatly decreasing the application performance and, thus, resulting in disruptions of message delivery to LoRa MQTT clients. Thus, a broker-free MQTT concept (similar to MQTT/UDP approach [38]) was employed in the LoRa context. Every LoRa MQTT subscriber regularly listens to broadcast messages from other LoRa MQTT clients. If a broadcast message (an MQTT topic with some payload) is received, the subscriber accepts the message in case of a topic match. All messages are encrypted by AES-128 securing message payload from being overheard by non-authorised LoRa MQTT clients.

### 4.2 Testing the prototype

*4.2.1 Evaluation use case.* To model a general interoperation use case, all sensors were programmed to generate random values from 0 to 255 on their given MQTT topics. The SGW with IF on the WiFi-LoRa border makes MQTT messages visible for both IoT controllers subscribed to the topics of sensors A, B and C. The controllers can manage only those sensors which lie in their own context (e.g. the LoRa IoT controller can control the LoRa sensors only). Both controllers publish control MQTT messages to managed sensors when a high reading value comes from another local or external sensor (modelling sense-actuate automation activities in a smart space).

*4.2.2 The baseline.* The current well-known practice of managing WiFi and LoRa devices through MQTT assumes the employment of a full LoRaWAN stack [25]. Within the LoRaWAN paradigm sensors are not able to exchange data directly. Instead, all communications happen through a LoRaWAN Gateway (LoRaWAN GW) which makes sensors visible to an IP network by encapsulating LoRa messages into IP packets and passing them further to the LoRaWAN Server which speaks to a central MQTT broker. To implement that evaluation scenario all MQTT-specific code was removed from the LoRa sensors so they could behave like standard LoRaWAN Class A/C devices managed by the WiFi IoT Controller along with the Sensor B. To enable that legacy IP-dependent MQTT interoperation (Figure 3) the SGW was reprogrammed into a simple LoRaWAN GW [32] which transmits LoRa sensor data to the local ChirpStack LoRaWAN Server [9] with installed mosquitto MQTT broker [16].

*4.2.3 E2E latency measurements.* Figures 4 and 5 show the emprical CDFs (ECDFs) with distributions of E2E latencies of command reaction measured on Sensor B and C for both MQTT interoperation variants. The latency was calculated as the difference between the timestamps when a sensor reading was generated and when a control command came to the target sensor as a reaction to that reading. In order to minimise time offsets between the nodes and E2E latency errors, all nodes were synced with the local NTP server. The E2E latency was measured around 1000 times for both decentralised and standard approaches. The command reaction time was deliberately not measured on Sensor A due to its power-saving nature which assumes much less incoming control traffic compared to more power-hungry sensors B and C (the command reaction to the sensor A readings was still measured on both of them).

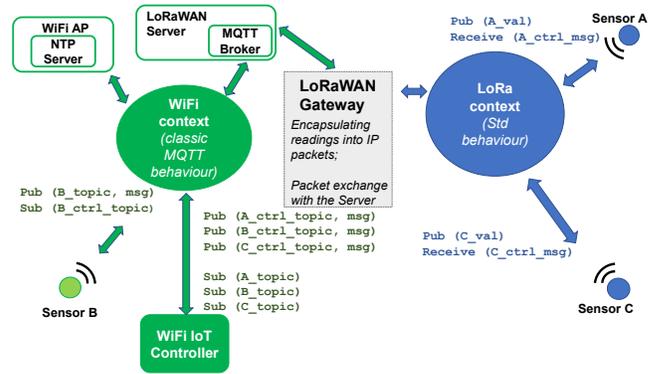

Figure 3: IP-dependent MQTT interoperation over LoRaWAN.

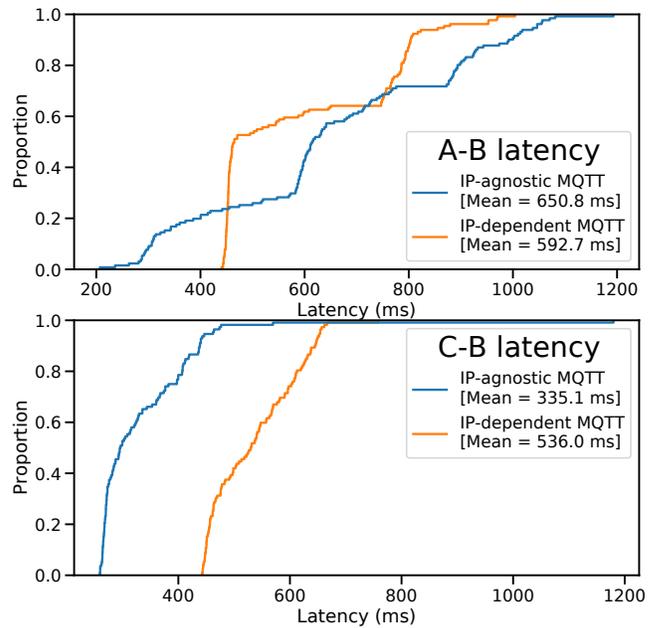

Figure 4: ECDFs with E2E latencies from sensors A and C measured on Sensor B in both evaluation scenarios.

The latency differences between the IP-agnostic and IP-dependent MQTT interoperation approaches are seen quite clearly, notwithstanding any minor shifts that might arise from NTP offset adjustments. The IP-neutral approach has lower latency on most of the communication links, especially seen for the sensor's C command reaction to the readings from the sensors A and B (Figure 5), which is 1.7-2 times faster compared to its IP-based MQTT variant. For the WiFi Sensor B (Figure 4) there are less latency differences which are around the same for A-B interoperation, however the IP-agnostic approach still has 1.6 faster reaction to the readings from the Sensor C.

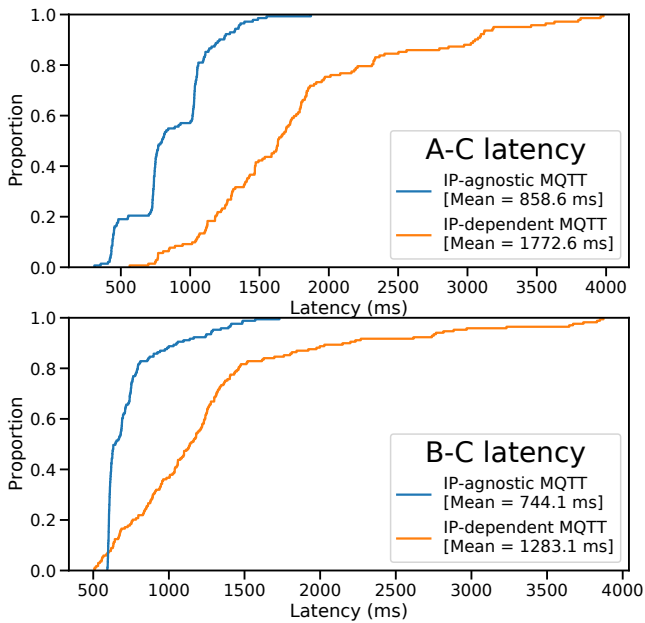

Figure 5: ECDFs with E2E latencies from sensors A and B measured on Sensor C in both evaluation scenarios.

High latency values and slow E2E feedback seen in the IP-based MQTT approach are caused by a higher number of intermediate hops between the sensors. All LoRa traffic has to pass through the LoRaWAN Server, and thus the nodes cannot talk directly to each other like they do in the IP-agnostic MQTT variant. It can be argued that the problem of IoT application interoperability lies not only in IP, but also in the MQTT implementation choices of upper-layer protocols, e.g. reliance on TCP instead of faster UDP. However, the same version of TCP-based MQTT broker was used on the WiFi side in both IP-agnostic and IP-dependent evaluation scenarios. Thus, the observed performance improvements of the IP-agnostic approach are mainly caused by the model's architectural features described in Section 4.1.

*4.2.4 Power consumption comparison.* Figure 6 shows the Sensor's A battery discharge time in both MQTT approaches. The battery voltage was measured every time when the sensor was waking up and transmitting data. To decrease the experiment duration (LoRaWAN battery can serve up to 10 years assuming that each sensor generates just a few readings per day) the sensor was programmed to produce a reading every 5 seconds within a 45-minute interval.

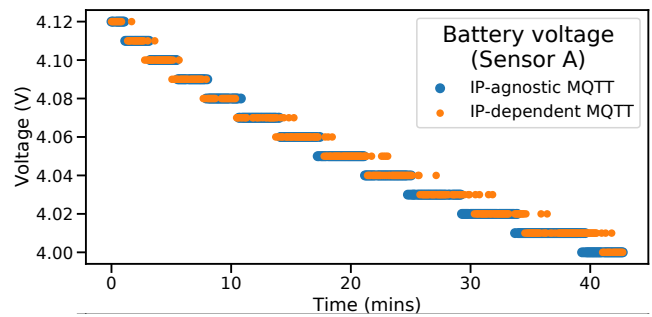

Figure 6: Sensor A battery discharge speed in both evaluation scenarios.

The results clearly show around the same battery discharge speed for both MQTT variants. Thus, no power-consumption overhead is brought by running an MQTT client on top of a LoRa sensor.

## 5 CONCLUSIONS

We presented the design and implementation of the proposed protocol-independent interoperation model for running IoT applications seamlessly. The prototype evaluation shows promising results highlighting the potential to improve IoT interoperation by moving from legacy IP-dependent communication paradigms while focusing on protocol-neutral application provision over heterogeneous IoT contexts. The benefits of escaping from the IP or any other underlying bottleneck protocols look more clear now as provide a more flexible organisation of IoT applications with decreased interoperation latency and no power-consumption overheads for battery-powered devices. This paper is a starting point and gives direction towards the development of IP-agnostic application-focused IoT architectures in order to prevent the IPv4/v6 becoming the new (old) normal with its all subsequent flaws existing since the creation of the Internet.

To better demonstrate the model's efficacy, the next step includes generalisation to a larger set of wireless IoT protocols and services. The work also opens avenues towards distributed control of smart spaces where management functions can be delegated from a central platform (e.g. a BMS) to SGWs providing more user-oriented control over IoT applications and data in a secure, resilient and privacy-preserving manner.

## ACKNOWLEDGMENTS

This research forms part of Centre for Digital Built Britain's work within the Construction Innovation Hub. The funding was provided through the Government's modern industrial strategy by Innovate UK, part of UK Research and Innovation.

# DeepDish: Multi-Object Tracking with an Off-the-Shelf Raspberry Pi

Matthew Danish
Cambridge University
mrd45@cam.ac.uk

Justas Brazauskas
Cambridge University
jb2328@cam.ac.uk

Rob Bricheno
Cambridge University
rwhb2@cam.ac.uk

Ian Lewis
Cambridge University
ijl20@cam.ac.uk

Richard Mortier
Cambridge University
rmm1002@cam.ac.uk

## ABSTRACT

When looking at in-building or urban settings, information about the number of people present and the way they move through the space is useful for helping designers to understand what they have created, fire marshals to identify potential safety hazards, planners to speculate about what is needed in the future, and the public to have real data on which to base opinions about communal choices. We propose a network of edge devices based on Raspberry Pi and TensorFlow, which will ultimately push data via LoRaWAN to a real-time data server. This network is being integrated into a Digital Twin of a local site which includes several dozen buildings spread over approximately 500,000 square metres. We share and discuss issues regarding privacy, accuracy and performance.

## CCS CONCEPTS

• **Computer systems organization** → Sensor networks; • **Hardware** → Sensor applications and deployments; **Sensor devices and platforms**.

## KEYWORDS

object detection, object tracking, edge computing

## 1 INTRODUCTION

We propose a system of edge devices based on a low-power computing board, the Raspberry Pi, that analyses the movement of people using a standard camera peripheral and can publish real-time events and anonymous statistics to a low-bandwidth and secure network. The system runs several machine-learning-based algorithms to perform multi-object tracking (MOT) on sequential image data and distils the information down to a few numbers that can be transmitted over a LoRaWAN network from sensors in the field.

The MOT problem takes as input an image sequence and a set of objects of interest. Solutions must trace their movement throughout the image sequence while maintaining the distinct identity of each object. We identify objects of interest by category (e.g., 'person') and then apply category-specific object-detection methods to automatically draw bounding boxes around all objects of interest within each frame. Solving the MOT problem then requires finding the corresponding bounding boxes in each successive frame, as well as determining if a bounding box is being drawn around a newly introduced object, and if a previously-known object has disappeared. This is known as the *tracking-by-detection* approach [4].

The object-detection problem has been studied intensively and solutions have advanced rapidly in recent years thanks to the success of supervised deep learning methods [5]. We focus in particular on the MobileNet model architecture [7] because it is optimised for use in the mobile processor context, and we use a model trained on the COCO data-set [11] that offers 91 possible object categories.

For tracking-by-detection, we employ DeepSORT [17], which extends SORT [2] with a 'feature encoder' that extracts a vector using a pre-trained convolutional neural network (CNN) on the image data within each bounding box. SORT is designed to be fast because it relies only on simple techniques of association based on scores computed using Kalman filtering and the Mahalanobis distance metric and then solved using the Hungarian method, but SORT can easily be fooled by object occlusion. By adding the feature encoder vector to the mix, DeepSORT helps avoid accidental identity switches between overlapping objects, while maintaining much of the performance required for online tracking of real-time video.

We show that this can be achieved with a low cost, stock Raspberry Pi 4B thanks to some unexpected findings we discuss in §7.

## 2 RELATED WORK

Past work by Ren et al. [14] relied on a network of edge-based servers with high-performance GPUs that could be placed close enough to gather image data from local network of cameras. Cartas et al. [3] performed object-detection by sending video frames from mobile devices to nearby small servers backed up by high-performance but more distant servers; they were only able to achieve 150ms inference time by equipping the small servers with GPUs. EdgeEye [12] similarly depends on having a GPU. Hochstetler et al. [6] bench-marked a Raspberry Pi 3B processor both with and without an Intel Movidius™ Neural Compute Stick on data from a visual recognition challenge. When setting the model input image size to 224x224, they found that the CPU alone performed at about 2.1 frames-per-second (FPS) on the object detection task, while adding the Neural Compute Stick boosted that to 17.2 FPS. The DeepX project [9] looked at ways of distributing inference tasks across heterogeneous resources such as GPUs and lower-power processing units present on certain mobile platforms, with which they were able to achieve sub-500ms object-detection inference times at significant energy savings.

DeepSORT is a popular tracking algorithm; one recent and notable work using it is by Zhang et al. [18] who considered the case of fixed-view cameras: they computed a differential filter to isolate only the portions of the view that were changing, then used the YOLOv2 [13] object-detection system with DeepSORT to perform online MOT on a high-performance server and GPU. Al-Tarawneh et al. [1] used a different style of feature vectors computed on high-performance servers to re-identify customers over the course of a day as they browsed a shop, in order to produce retail analytics of their behaviour.

## 3 EXPERIMENTAL SETUP

Our pilot project intended[1] to count the number of people entering and exiting a particular building for fire safety purposes. While person-counting is a venerable field, this is only intended as a test case for a much larger experiment that will study the movement of people in public space as part of a project to create a Digital Twin [8] of a local site, which includes several dozen buildings and covers approximately 500,000 square metres of land.

With a large enough field-of-view, we hypothesise that the counting task can be performed even if the software can only process a low number of FPS, because people would enter the view for at least several seconds. Careful placement of the camera is critical to ensure that people are within view for sufficient time on either side of the pre-determined 'counting line', while not being too far away to compromise the object detector.

### 3.1 Hardware

Our edge node is a Raspberry Pi 4B, which is a Broadcom BCM2711 system-on-chip with a quad-core Cortex-A72 (ARM v8) 64-bit processor running at 1.5GHz, with 4GB of RAM. We used a night-vision camera module recording video at 640×480 in full colour, however we also tested with the standard Pi camera, which provides a similar quality of video albeit without the infrared lighting support. We added the Fan SHIM for temperature control, and a PiJuice battery 'HAT' mounted on the GPIO interface to provide a power supply backup. The device is packaged in a 3D-printed case that allows the camera to be rotated into the desired position, as shown in Figure 1. Future work will include a LoRaWAN HAT as well.

### 3.2 Operating software

We use TensorFlow 1.15.0, including the TensorFlow Lite engine, and the controlling software is written in Python 3.7 running under Hypriot, a Debian-based operating system customised for Raspberry Pi.

---

[1]Late note: the pilot project has been cancelled due to the global COVID-19 pandemic.

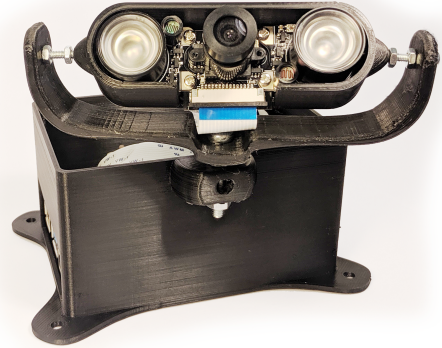

Figure 1: The counting device.

### 3.3 Algorithm

Following the tracking-by-detection concept used by DeepSORT, we break down the algorithm into three steps shown in Figure 2: object-detection, feature-extraction and tracking/association.

The object-detection model is a pre-trained quantised version of SSD MobileNet v1 [7] compiled for TensorFlow Lite, made available by Google. We feed it images of size 300×300, and from the output use only the detections that are labelled with the category 'person'. The feature-extraction step uses the DeepFLOW CNN model trained on the Motion Analysis and Re-identification Set (MARS) data-set [15] using the *cosine metric learning* technique [16]. Feature-extraction must be run separately on each person detected, therefore the algorithm scales linearly by the number of people that need tracking. Association of tracks with known history of objects is performed by the DeepSORT method of combining Mahalanobis distance computed on Kalman Filter distributions and cosine metric distance computed on extracted feature vectors. People who are new to the tracking history (based on a feature threshold) are assigned a fresh identification number, and people that fail to be found for over 30 frames are considered to have left the scene.

At each step the most recent track vector of each known person is compared against the pre-determined counting line by solving for line intersection and cross-product (to determine direction of movement). When an intersection is found, a 'count' event is generated, and the intersecting segment is highlighted in red, as shown in Figure 3. Internally, we maintain a running total of counts going in each direction, as well as the number of tracks that have been deleted after the person was not identified for 30 consecutive frames. When an HDMI monitor is hooked up to the device we display this information as an overlay on the current camera view, for debugging purposes, as well as drawing the count line, rectangles around detected people and their tracked vectors. You can also enable a web interface that shows the same debugging information.

## 4 PERFORMANCE

We used SSD MobileNet v1 because in our testing, we found it to give the best response times by far, as can be seen in Figure 4. The accuracy scores are also quite good, under the circumstances, as discussed in §6. A further speed-up can be obtained by overclocking the Raspberry Pi, as seen in Figure 5. We found that the platform remained stable up to 1,900MHz, speeding up inference by about 16.5%, with core temperatures fluctuating somewhat under 60°C with active cooling from the Fan SHIM. However, we learned that the PiJuice did not work reliably when the Pi was over-volted, and the HAT had to be removed before conducting these tests.

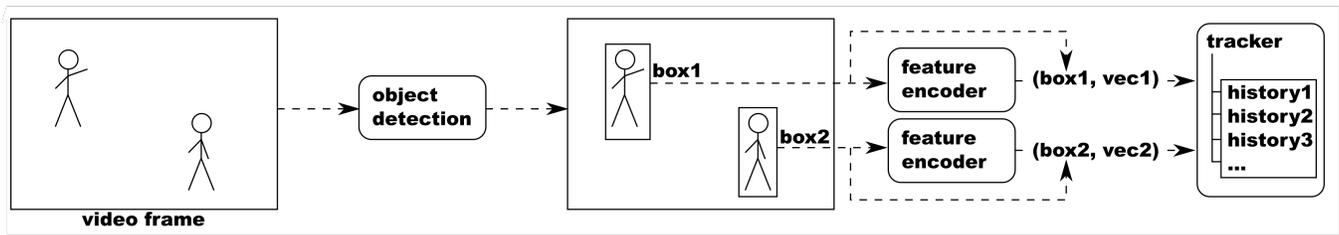

Figure 2: Tracking-by-detection pipeline.

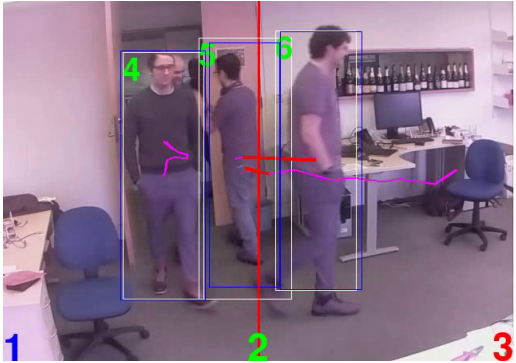

Figure 3: Screenshot from the 'Office' test video, showing two intersection events. The numbers at the bottom corners of the screen record the counts of people who have crossed the red counting line in either direction. The central number is the difference. The boxes around people are results from object-detection, and the numbers in the corners of the boxes are tracking identities. The purple lines that follow people are the tracking vectors; the intersecting segments are coloured red momentarily when an intersection event is detected.

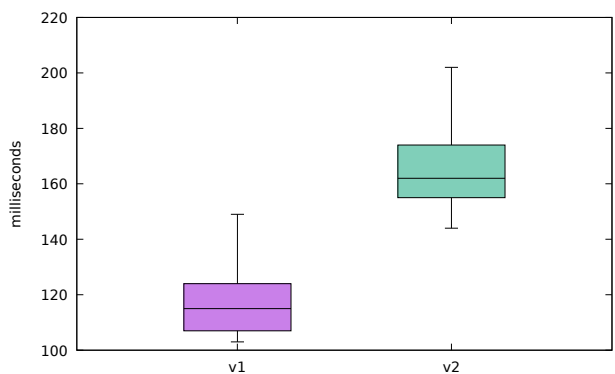

Figure 4: Inference time for object detection using different versions of SSD MobileNet.

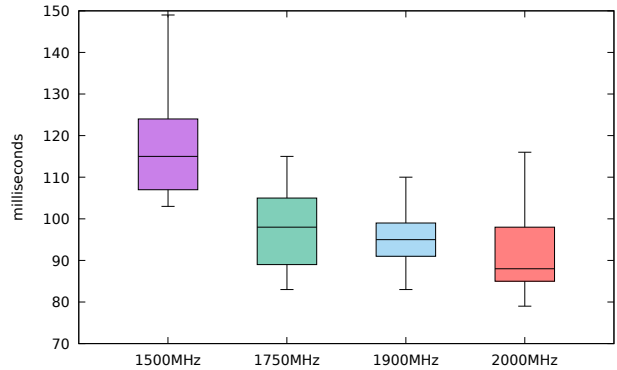

Figure 5: Object detection inference times for SSD MobileNet v1 with overclocking.

The DeepSORT feature-encoder CNN model comes trained on MARS data with an image input size of 128×64. However, we found that running it took 95ms per person per frame: too slow for real-time processing of scenes with multiple people. Therefore, we resized the image input to 64×32 and retrained the CNN model. After doing that we achieved a 35ms running time per person per frame, 63% faster than before. With overclocking the typical time for feature-encoding was reduced further to 32ms per person per frame. This is considerably more practical in our expected use case.

When overclocking to 1,900MHz, the running time can be estimated using the following numbers:

- Fixed costs per frame: approximately 130ms, composed of,
  - Object-detection: about 96ms
  - Processing (e.g. input, resizing and output): about 34ms
- Cost per person tracked: approximately 36ms, composed of,
  - Feature-encoding: about 32ms
  - Association: about 4ms

Therefore, the overall inference time per frame when tracking $n$ people at a time averages approximately $T(n) = 130 + 36n$ milliseconds. We can track up to ten people at a time while still maintaining two FPS, five people at a time at three FPS and two people at about five FPS.

**Table 1: Power draw and CPU temperature in different CPU frequencies and modes.**

| CPU Clock (MHz) | Mode | Power (W) | CPU Temp. (°C) |
|---|---|---|---|
| 1,500 | *Run* | 6.0 | 49 |
| 1,500 | *Sleep* | 3.6 | 35 |
| 1,500 | Idle Pi | 3.4 | 33 |
| 1,750 | *Run* | 7.0 | 51 |
| 1,750 | *Sleep* | 3.9 | 35 |
| 1,750 | Idle Pi | 3.6 | 34 |
| 1,900 | *Run* | 8.5 | 57 |
| 1,900 | *Sleep* | 4.2 | 36 |
| 1,900 | Idle Pi | 3.7 | 35 |

## 4.1 Power usage

Table 1 shows the power draw and CPU temperature of the Raspberry Pi under different loads. *Run* mode is the normal operation, seeking maximum performance. In *Sleep* mode the program does not invoke inference but instead periodically checks if anything has changed in the input, waking back up if necessary. Both are compared against a baseline of an idle Raspberry Pi. There are significant opportunities for power-saving during quiet periods when there is no motion in front of the camera and it is looking at a fixed scene.

## 5 PRIVACY

Putting up cameras raises privacy concerns, even in public spaces. One assurance we can offer is that since our data transmission will ultimately be carried by LoRaWAN, there is simply not enough bandwidth for it to be practical to transmit images at all. The only data transmitted are counting events and the current state of the counters. An attacker could learn the number of people within the public part of the building, but this is not considered sensitive information. For debugging purposes, the WiFi device in the Raspberry Pi has been configured with a private network protected by a pre-shared key. It is not connected to the Internet and can only be accessed by a person with the password and in close proximity to the device.

As the algorithm reads and processes each camera frame, it finds the coordinates of boxes around each object it detects, then distils the contents of each box down to a short vector of 128 numbers, and finally discards the frame. The vectors may stay in memory as long as the object is within sight but they are a very sparse encoding of pixel colours and basic shapes, with no personally identifying information associated with them.

## 6 ACCURACY

We set up a test-bed to accept pre-recorded video in place of the live camera feed into the same algorithm used by the live tracker. The test video named 'Office' comprises two minutes of filming using the Raspberry Pi camera in an office environment similar in character to the pilot project location, with seven volunteers instructed to enter and exit the office repeatedly and without any particular patterns. The test video named 'Plaza' comprises 50 seconds of video from the MOT17 challenge [10], taken from a fixed camera overlooking a pedestrian plaza with a fairly intense flow of people moving about, as seen in Figure 6.

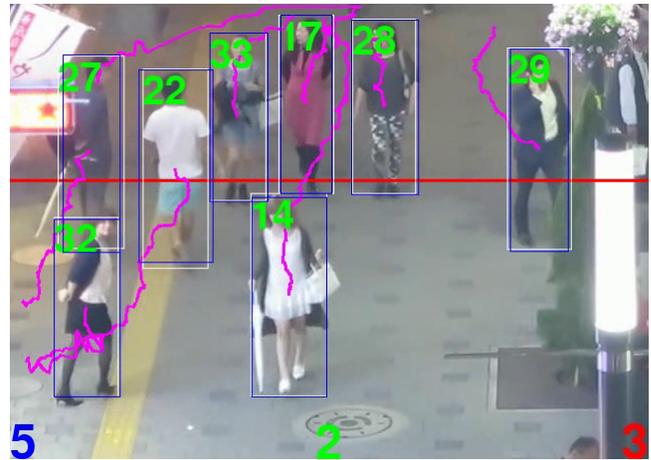

**Figure 6: Screenshot from the 'Plaza' test video.**

The code was instrumented so that it wrote the current value of several counter variables into a file every ten seconds: persons crossing the counting line in each direction (as determined by the sign of the cross-product), overall number of crossings, and number of tracking identities that have expired.

Ground truth for test videos was established using the following method: in each ten-second interval we counted the number of people crossing the counting line in each direction.

Therefore, for each ten second interval of test video, we know the number of people walking to the 'negative' side of the line (*negcount*) and the number of people walking to the 'positive' side of the line (*poscount*). These values were subtracted and then compiled into a vector covering the whole video. This allowed us to see how the algorithm varied from the ground truth over time, and penalised offsetting mistakes to some extent.

Vectors drawn from the ground truth and from test runs of the algorithm are compared using cosine-distance, subtracted from 1. The ideal score is 1, the worst score is 0. We ran the tests with a number of configurations of the test-bed, tweaking parameters to DeepSORT, and trying much slower (but more accurate) YOLO-based object detectors in addition to MobileNet. The test-bed also was able to simulate different frame-rates, for example, by dropping five frames out of every six to simulate a 200ms inference latency.

Parameters adjustable for testing include the resolution of the MARS-trained feature encoder (64×32, 128×64 or 256×128), the simulated FPS value (from 5–30), the maximum cosine distance (*max-cos-distance*) threshold for two feature vectors to be considered part of the same 'track', and the 'non-maximum suppression' (*nms-max-overlap*) threshold that eliminates spurious overlapping object detection boxes (at 1.0 the boxes must overlap completely for one to be pruned, and at 0 it would eliminate even non-overlapping boxes). Over 450 configurations were tested. A selection of scores is shown in Table 2.

Table 2: Accuracy of counting: a selection of test results.

| Score | Test name | Object-detector | Feature encoder | FPS | max-cos-distance | nms-max-overlap |
|---|---|---|---|---|---|---|
| 0.968 | Office | YOLO v3 | 128x64 | 5 | 0.3 | 0.6 |
| 0.948 | Office | MobileNet v1 | 64x32 | 5 | 0.6 | 0.6 |
| 0.938 | Office | MobileNet v2 | 64x32 | 30 | 0.9 | 0.6 |
| 0.906 | Office | MobileNet v1 | 128x64 | 5 | 0.3 | 0.3 |
| 0.878 | Office | MobileNet v2 | 64x32 | 5 | 0.9 | 1.0 |
| 0.794 | Office | MobileNet v1 | 128x64 | 30 | 0.9 | 0.6 |
| 0.656 | Office | MobileNet v2 | 256x128 | 30 | 0.6 | 0.6 |
| 0.424 | Office | MobileNet v2 | 256x128 | 15 | 0.01 | 1.0 |
| 0.986 | Plaza | YOLO v3 | 64x32 | 5 | 0.6 | 0.6 |
| 0.903 | Plaza | MobileNet v1 | 128x64 | 15 | 0.9 | 0.6 |
| 0.880 | Plaza | MobileNet v2 | 128x64 | 5 | 0.3 | 0.6 |
| 0.843 | Plaza | MobileNet v1 | 64x32 | 5 | 0.6 | 0.8 |
| 0.839 | Plaza | MobileNet v2 | 256x128 | 5 | 0.9 | 0.6 |
| 0.815 | Plaza | MobileNet v2 | 64x32 | 5 | 0.6 | 0.3 |
| 0.713 | Plaza | MobileNet v1 | 64x32 | 15 | 0.01 | 1.0 |
| 0.596 | Plaza | MobileNet v2 | 128x64 | 30 | 0.9 | 1.0 |

## 7 DISCUSSION

The most surprising finding from our experiments is that increased frame-rate was not helpful and in fact could make things worse. As seen in Table 2, using a configuration with conditions similar to our live video counter, we achieved a score of 0.948 on the 'Office' test with MobileNet v1 running at 5 FPS. This rivalled the top test score of 0.968 that was obtained using a powerful GPU running YOLO v3 (tests at both 5 and 30 FPS achieved this mark). Figure 7 shows the overall average scores at each frame-rate: the worst cases are considerably worse with 15 and 30 FPS compared to 5 FPS.

The cosine distances were often not that large between vectors encoded from the views of different people, especially if they were wearing similar colour clothing. This could result in unwanted identity swaps, but that would only affect the score in a small way if the overall count was correct. Generally, setting *max-cos-distance* to very low values tended to reduce the performance of the tracker, effectively removing any assistance from the feature encoder. This effect can be seen in Figure 8.

Lower 'non-maximum suppression' was slightly more important with MobileNet than YOLO because the latter tends to generate higher-quality object-detection boxes and get less confused when multiple people are standing together in a group. With MobileNet, it helped to suppress some spurious boxes that could be generated by clusters of people, and lower values of *nms-max-overlap* led to slightly improved performance, as can be seen in Figure 9.

Another unexpected finding is that lowering the resolution of the input to the feature encoder in order to gain performance did not affect scores overall. In Figure 10, it shows that the 64×32 feature encoder gave approximately the same scores as the slower default resolution of 128×64. We also tried experiments with an even higher resolution feature encoder, at 256×128, and that backfired, producing worse results.

Finally we compare models in Figure 11. This chart focuses on the generally good configurations: with feature encoder resolution 64×32, processing 5 FPS, having *max-cos-distance* ≥ 0.3, and

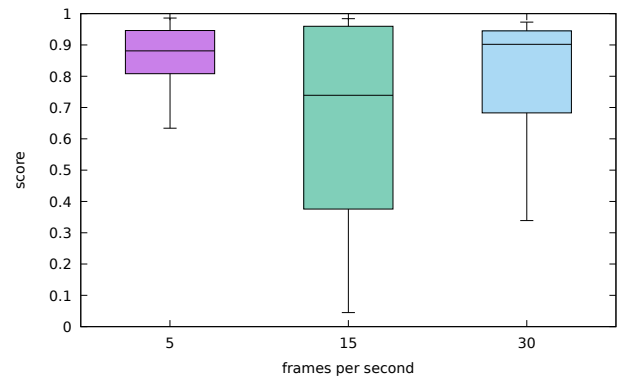

Figure 7: The effect of frame-rate on score.

*nms-max-overlap* < 1.0. Both MobileNet versions score about the same (just a slight advantage for v2) but in our judgement that is outweighed by the worse running time.

## 8 CONCLUSION

We show that practical performance can be achieved on a Raspberry Pi 4B in this application without requiring special hardware acceleration. Although our pilot project was meant to take place inside a building, we anticipate future deployments in places where network access is limited, space is at a premium and power supply may be more circumscribed. We are continuing to refine the models and parameters so this system may be used in more extensive experiments for our larger Digital Twin project.

## ACKNOWLEDGMENTS

This research forms part of Centre for Digital Built Britain's work within the Construction Innovation Hub. The funding was provided through the Government's modern industrial strategy by Innovate UK, part of UK Research and Innovation.

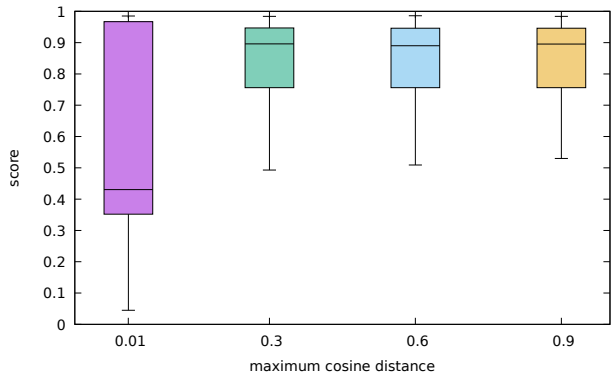

Figure 8: The effect of maximum cosine distance on score.

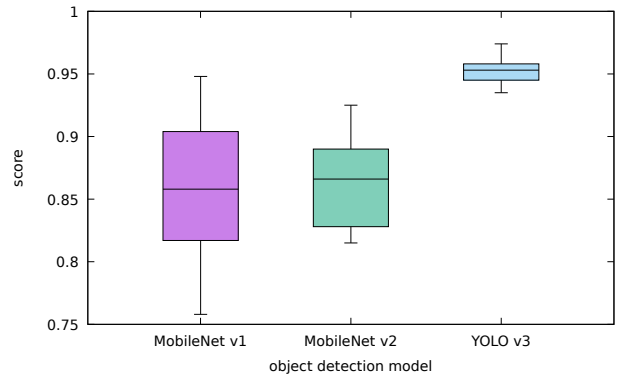

Figure 11: The effect of model type on score.

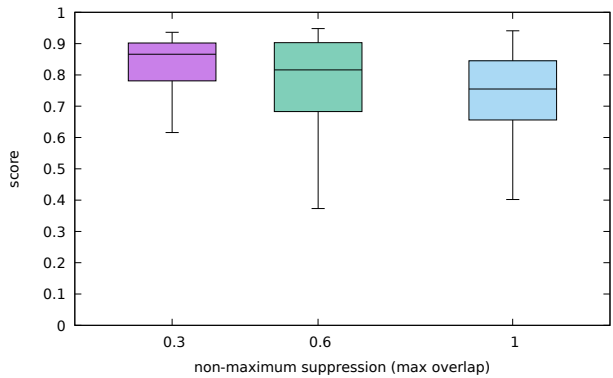

Figure 9: The effect of non-maximum suppression on score for MobileNet.

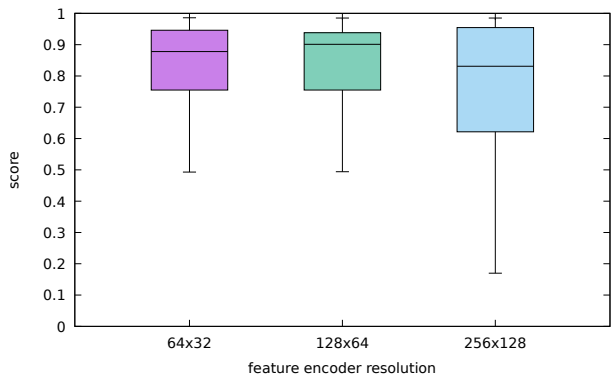

Figure 10: The effect of feature encoder resolution on score.

# DeepDish on a diet: low-latency, energy-efficient object-detection and tracking at the edge

Matthew Danish
Cambridge University
United Kingdom
mrd45@cam.ac.uk

Rohit Verma
Cambridge University
United Kingdom
rv355@cam.ac.uk

Justas Brazauskas
Cambridge University
United Kingdom
jb2328@cam.ac.uk

Ian Lewis
Cambridge University
United Kingdom
ijl20@cam.ac.uk

Richard Mortier
Cambridge University
United Kingdom
rmm1002@cam.ac.uk

## ABSTRACT

Intelligent sensors using deep learning to comprehend video streams have become commonly used to track and analyse the movement of people and vehicles in public spaces. The models and hardware become more powerful at regular and frequent intervals. However, this computational marvel has come at the expense of heavy energy usage. If intelligent sensors are to become ubiquitous, such as being installed at every junction and frequently along every street in a city, then their power draw will become non-trivial, posing a severe downside to their usage. We explore Multi-Object Tracking (MOT) solutions based on our custom system that use less power while still maintaining reasonable accuracy.

## CCS CONCEPTS

• **Computer systems organization** → **Sensor networks**; • **Computing methodologies** → **Computer vision**.

## KEYWORDS

object detection, object tracking, edge computing

## 1 INTRODUCTION

Dynamic digital twins [9] and other embedded sensor networks [7] are becoming popular ways to gather data about the built environment in real-time for immediate analysis or reaction. This requires ubiquitous installation of sensors, some of which may be naturally low-power (such as temperature sensors), but others invoke computationally intensive workloads. We have been exploring a sensor design that can solve the Multi-Object Tracking (MOT) problem using video camera input, using machine learning techniques to distil the high-bandwidth image stream down to relatively few numbers that can be easily transmitted without significant bandwidth or privacy concerns.

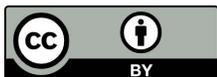

The MOT problem takes as input an image sequence and a set of objects of interest. Solutions must trace their movement throughout the image sequence while maintaining the distinct identity of each object. We approach this problem by finding bounding boxes around objects in each frame and using tracking and data association techniques to find correspondences between bounding boxes in successive frames. It is important to determine if a bounding box is being drawn around a newly introduced object, of if a previously-known object has disappeared. Altogether, this is known as the 'tracking-by-detection' approach [3].

Most approaches to solving the MOT problem are computationally and power-intensive. Feeding the video stream into software running on a high-performance workstation with a top-end GPU will produce a highly accurate result but with intensive power requirements: VoVNet [11] significantly reduces power requirements for object detection but still required 63.9W when tested on a workstation with a NVIDIA Titan X GPU. There are deployments [13] with such powerful equipment, but we believe they are using much more power than is necessary for the task, and at a large scale this raises environmental concerns. We focus on low-power solutions, ideally in the single-digit watt range.

## 2 RELATED WORK

Lujic, et al. [12] proposed InTraSafEd5G (Increasing Traffic Safety with Edge and 5G) as an application of object detection to help prevent crashes between turning drivers and people who are walking or cycling at junctions. They deployed a Raspberry Pi 4 with Coral EdgeTPU running the SSD MobileNet v2 model to detect people at a real-life installation with a camera pointing at a junction in Vienna. The presence of people was then transmitted using MQTT over 5G to drivers running a special mobile phone app that then would display a warning message in the driver's view. The average information delivery time was about 108ms in their experiment.

Fernández-Sanjuro, et al. [6] developed a real-time multi-object tracking-by-detection system on the NVIDIA Jetson TX2 embedded computing board with Pascal GPU. The object detector is based on customised YOLOv3 architecture and the tracker is a combination of an appearance-based correlation filter tracker (KCF) and a motion-based Kalman Filter tracker. To reach real-time speeds the object detector is only used on every eleventh frame. The Jetson is a more powerful computer than the Raspberry Pi and consumes approximately 15W under load, about twice as much as the Pi. SkyNet [19] is also a Jetson-based project with an object detection model developed from the bottom-up to capture hardware limitations.

DeepDish [4] explored object detection and tracking on the Raspberry Pi. It demonstrated feasible CPU-only tracking-by-detection and showed power usage could be kept to within 7W when not overclocking the CPU. One surprising result was that low framerate did not necessarily affect the accuracy of the results — in terms of a people-counting metric — in some cases the result was even improved by lower framerates. DeepDish is the basis of the current work, however the detection and tracking engine has been completely rewritten for high-performance and flexibility on various platforms, and works with either GPUs or Coral EdgeTPUs when available.

MARLIN [2] is an object detection and tracking system for augmented reality with an explicit goal of reducing energy drain on mobile phones (it was tested on the LG G6 and Google Pixel 2). They used the Tiny YOLO object detector but focused on finding ways to avoid invoking it because the detection latency is 1,200ms and it consumes an additional 1.7-1.9W when running. Object tracking is maintained using optical flow and a they trained and evaluated a 'change detector' based on random forest classifiers to determine if the object detector needed to be run again. They compared this against methods that simply skipped frames and a baseline method that ran detection on every frame. They found that MARLIN reduces power consumption by 34% on average with a typical accuracy loss of less than 10%.

## 3 EXPERIMENTAL SETUP

### 3.1 Hardware
Our edge node is a Raspberry Pi 4B, which is a Broadcom BCM2711 system-on-chip with a quad-core Cortex-A72 (ARM v8) 64-bit processor running at 1.5GHz, with 4GB of RAM. We have connected a Google Coral EdgeTPU device to one of the USB 3.0 ports to provide hardware acceleration for specially compiled models.

### 3.2 Software
We use the TensorFlow-Lite (TFLite) engine for evaluation on the Raspberry Pi, and the controlling software is written in Python 3.7 running under Hypriot, a Debian-based operating system customised for Raspberry Pi and specialised to run Docker. Our software runs within a Docker container based on the BalenaLib framework, with the necessary libraries and modules all ready to go.

### 3.3 Test harness
In order to simulate the capture of live video, we have written the program with a queue feeder thread to read video files at a constant framerate (such as 25 FPS) and offer the video frame to the main thread for a limited time period until the next one is ready. If the input queue of the main thread is not ready to accept the video frame when the time elapses, then that video frame is dropped and the feeder moves onto the next one. This allows us to simulate the effects of live video while gaining the benefit of comparison with existing well-known video sets with ground-truth data and metric-analysis scripts, such as those from MOT15 [10] including its multi-object tracking accuracy (MOTA) metric.

Power consumption is measured by a Tasmota smart plug. Whenever the power usage changes by more than 0.1W, this sensor transmits the new numbers using MQTT over Wi-Fi to our experiment-monitoring workstation. DeepDish also transmits its telemetry using MQTT to the same computer, so its data can be correlated with the power readings, and average power usage measured over the running time of a test.

## 4 ALGORITHM
We follow the basic architectural concept of DeepSORT [18], as shown in Figure 1, but with substantial modifications to object detection, feature encoding and tracking/association.

### 4.1 Object detection
We use a pluggable object detection architecture designed to support numerous models and new ones to come. At the time of this writing, here are some notable examples we support that generally fall into the following categories:

- TFLite models, and their quantised EdgeTPU versions:
  - EfficientDet Lite [1]
  - SSD MobileNet v1 and v2 [8]
- YOLOv5 [16] family of models, converted into TFLite and EdgeTPU-supporting formats:
  - In particular, the smaller and faster architectures, namely: YOLOv5n, YOLOv5s and their recent updates, YOLOv5n6 and YOLOv5s6
- Other models that fit in the TensorFlow object detection family, but are less suitable for edge computing:
  - EfficientDet family of models (D0…D7) [15]
  - CenterNet ResNet101 and CenterNet HG104 [5]

The output of the object detector is mapped into a common interface composed of bounding boxes, labels and confidence scores for each. If the background-subtraction option is enabled then we prune any bounding boxes that do not contain regions of motion. If there are no objects detected in the scene, and powersaving is enabled, then we begin an artificial slowing-down process that intentionally bottlenecks the pipeline so that the object detector runs less frequently while there are no objects in the scene. Once any object is detected, the pipeline returns to full speed.

### 4.2 Feature encoding and tracking
The DeepSORT-based feature-extraction step uses the DeepFLOW CNN model trained on the Motion Analysis and Re-identification Set (MARS) data-set [14] using the cosine metric learning technique [17]. Compared to the original work, we use smaller input sizes of 16x32 and 32x64 to reduce latency. In order to speed up processing times even further, we have also retrained a modified version of this model ('8x16-mod') with several layers removed, on a smaller input size of 8x16 pixels, and found it to run much faster with basically no loss of tracking accuracy. See Table 1 for a comparison of encoders and MOTA on the 'TUD-Stadtmitte' example from MOT15, with object-detections in this case provided by the high-quality EfficientDet D7 model running on an NVIDIA GeForce RTX 2080Ti GPU.

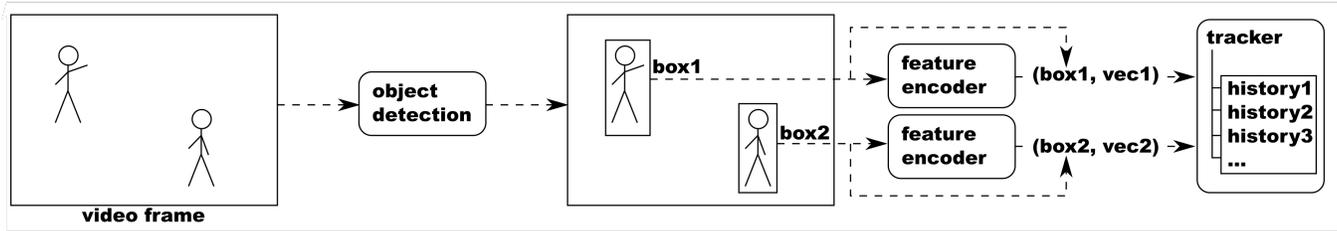

Figure 1: Tracking-by-detection pipeline.

| Feature Encoder | MOTA |
|---:|---|
| constant | 40.8% |
| 8x16-mod | 41.4% |
| 16x32 | 41.4% |
| 32x64 | 41.2% |

Table 1: Comparing the accuracy performance of four feature encoders, with object detections in all cases provided by the high-quality EfficientDet D7 model (see Section 4.2).

| Model | e2e | f2f | obj | enc | cpu | fif |
|---|---|---|---|---|---|---|
| EfficientDet-Lite0 | 459 | 175 | 162 | 78 | 231% | 2.2 |
| SSDMobileNetV1 | 409 | 104 | 65 | 99 | 296% | 3.9 |
| SSDMobileNetV2 | 384 | 99 | 64 | 92 | 296% | 3.8 |
| YOLOv5n6-320x320 | 537 | 188 | 175 | 104 | 227% | 2.4 |
| YOLOv5s6-320x320 | 455 | 194 | 183 | 50 | 183% | 2.0 |

Table 2: Latencies (in ms) for various pipeline components, comparing various object detector models (see Section 4.3).

In performing these tests, we converted the encoder model to the TFLite format and modified it to accept a fixed batch-size so that the deployed device need not run the full TensorFlow stack, but only the much smaller tflite-runtime package, while still efficiently processing batches (TFLite cannot handle variable input sizes in the same manner as the full-blown TensorFlow). As a control measure we also implemented a 'constant' image-encoder that always returns the same unit-vector for any input, which shows that a significant amount of the value of the tracking module comes from the data association portion discussed below.

Although running in batches reduces fixed overhead, feature extraction must be run separately on each person detected, therefore this portion of the algorithm scales linearly by the number of people that need tracking. Association of tracks with known history of objects is performed by the DeepSORT method of combining Mahalanobis distance computed on Kalman Filter distributions as well as the cosine metric distance computed on extracted feature vectors as discussed above. Objects that are new to the tracking history are assigned a fresh identification number, and objects that fail to be found for over 60 frames are considered to have left the scene.

### 4.3 Pipeline

One of the goals with the complete rewrite of DeepDish was to architect it in a scalable way. This has been achieved by structuring the pipeline as a series of asynchronous Python tasks connected by queues, based on the asyncio module. The frame capture loop, object detector and feature encoder all run in separate threads in order to ensure that they do not block the cooperatively-scheduled asynchronous pipeline. The result has been a success, with the pipeline capable of processing multiple 'frames-in-flight' at a time. We instrumented the code to measure performance in several ways; all of the below are averages over the course of a single run on the MOT15 'TUD-Stadtmitte' sample with the object detector running on every frame:

- **e2e** (ms) The time it takes from when a frame enters the pipeline to when the pipeline finishes with it.
- **f2f** (ms) The time between invocations of the same pipeline stage by consecutive frames, a.k.a. 1/framerate.
- **obj** (ms) The latency of the object detector per frame.
- **enc** (ms) The latency of the feature encoder per frame.
- **cpu** Average CPU utilisation.
- **fif** Average number of frames-in-flight in the pipeline.

Table 2 shows instrumentation output for several sample runs. Note that the **f2f** timings are significantly smaller than the **e2e**; this means frames are hitting the end of the pipeline much more quickly than it takes for a single frame to be processed. Some of the **e2e** latency comes from waiting in queues, however both **obj** and **enc** measure the latency of some of the significant computations that take place in the pipeline and their combination is always longer than the **f2f** time; this indicates that the pipeline is at the very least performing object detection and feature encoding in parallel on consecutive frames.

The previous generation DeepDish had frame-to-frame performance (on an overclocked Raspberry Pi) modelled as $T(n) = 130 + 36n$ milliseconds where $n$ is the number of people detected in the scene. All of the runs shown in Table 2 have comparable performance to $T(2)$, some significantly better than $T(1)$, on a recorded scene with 4-5 people on average. The differences are: use of the EdgeTPU, pipelining to take advantage of the multiple CPU cores, and a faster and lighter-weight feature-encoder.

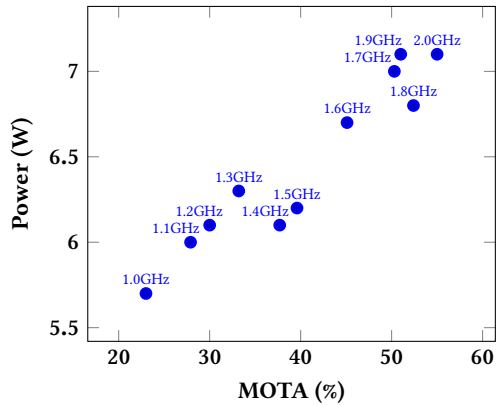

Figure 2: Power and accuracy at different CPU clock-speeds (discussed in Sections 5.1 and 6.3).

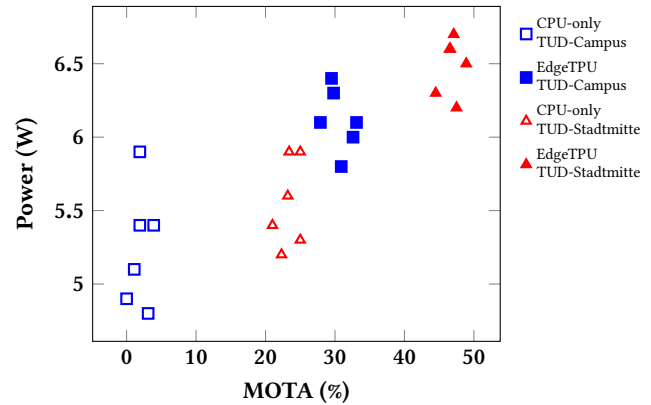

Figure 3: Two sample videos from MOT15 are tested for power usage and accuracy, first running on CPU, and then boosted with the EdgeTPU (see Section 5.2).

## 5 REDUCING POWER USAGE

### 5.1 Underclocking

We experimented with underclocking the Raspberry Pi 4B CPU from the normal 1.5GHz to see how it would affect power consumption and performance. To focus on the effect of underclocking we used the object detector model SSDMobileNetV1 and the constant feature encoder, skipping two out of every three frames. The MOT15 video 'TUD-Stadtmitte' was fed into the test harness at a steady rate of 25 FPS, whether or not the system could keep up. The scatter plot in Figure 2 shows the average MOTA and power readings from six runs at eleven different CPU frequency settings, of which the 1.5GHz is the factory setting and those below it constitute underclocking.

It should be noted there is significant variability as frame-dropping affected the results. In general, though, we can see that underclocking causes an immediate reduction in accuracy without necessarily reducing power usage, at least not until the clock-speed is drastically lowered, severely affecting accuracy.

### 5.2 EdgeTPU effect

We experimented with holding all settings constant except for swapping between the EdgeTPU and CPU-only versions of SSD-MobileNetV1. These experiments were conducted on the 'TUD-Campus' and 'TUD-Stadtmitte' videos from MOT15. Unsurprisingly, connecting and using the Coral EdgeTPU increases power consumption of the whole device, on average from 5.4W to 6.3W in these tests. However, the acceleration of the EdgeTPU also greatly increased the MOTA scores, as can be seen in Figure 3.

### 5.3 Choice of model

We looked at whether the object detector model affected the power usage of the Raspberry Pi. We tested six models, all of which use the EdgeTPU to some extent, running them six times each on five different videos (the ones listed in Table 3). The results are shown in Figure 4. The corresponding latencies are shown in Figure 5. Power consumption was effectively the same for all the models tested. Latencies varied substantially, with the MobileNet models winning handily; see Section 6.2 for further discussion.

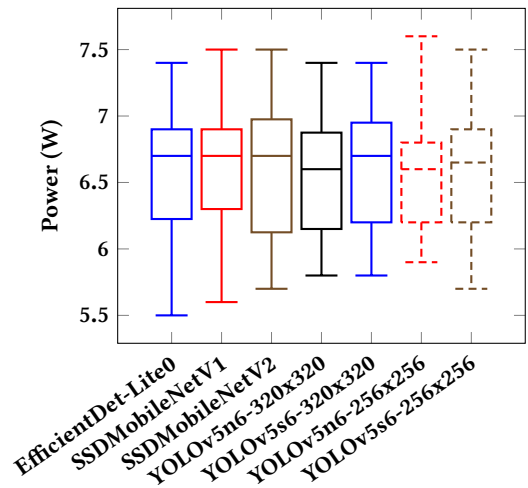

Figure 4: Comparison of power consumption by different object detector models (see Section 5.3).

## 6 ACCURACY

### 6.1 Comparison to GPU

The DeepDish software is also able to run on workstations with GPUs. We collected accuracy results of such high-power runs for comparison purposes. With a GPU available, we chose to use the high-quality EfficientDet D7 object detection model, and to run it alongside a 32x64 feature encoder generated from the original DeepSORT neural architecture. These tests were run with no time constraints on an Intel Core i9-based workstation equipped with one NVIDIA GeForce RTX 2080Ti GPU using videos from MOT15 to produce the MOTA scores in Table 3. We selected these particular videos, from the available benchmark set, because these ones largely fit the target use-case of DeepDish, which is motion-tracking people or vehicles in urban or interior scenes.

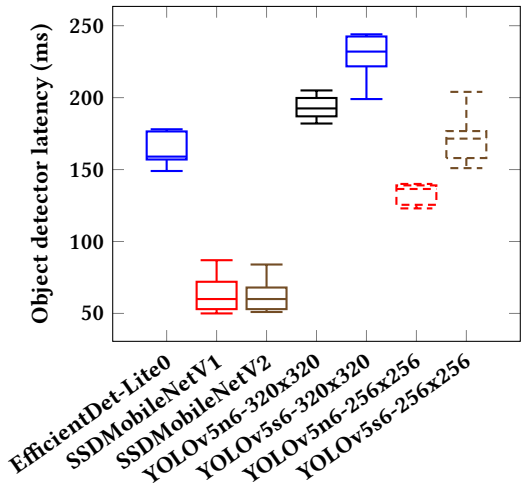

Figure 5: Comparison of latency of different object detector models (see Section 5.3).

| Video | MOTA |
|---|---|
| PETS09-S2L1 | 78.7% |
| TUD-Campus | 63.2% |
| ETH-Pedcross2 | 50.6% |
| ADL-Rundle-6 | 55.6% |
| TUD-Stadtmitte | 41.2% |

Table 3: Accuracy of DeepDish running on a high-powered workstation with GPU (see Section 6.1).

## 6.2 Accuracy and choice of model on the Pi

We examined accuracy on a model-by-model basis more closely for two of the MOT15 videos, 'TUD-Campus' (Figure 6) and 'ETH-Pedcross2' (Figure 7), first without overclocking and then at 2.0GHz and higher voltage. Both videos have a fair amount of pedestrian activity, the latter being a longer and more complex video with a moving camera. We ran them on the test harness at 25 FPS, dropping frames whenever the pipeline was no longer able to accept them in time. We note that the performance of the SSDMobileNet family continues to hold up, with its low latency (Figure 5) seeming to make a big difference in outcome. EfficientDet-Lite0 comes in third, dragged down by its higher running time, meaning that more frames must be dropped. Surprisingly, the YOLOv5 family does poorly, even using the newer 'n6' and 's6' variants; it only manages to achieve somewhat better latencies when using the 256x256 input size, but then their MOTA scores come out worse than the ones from EfficientDet-Lite0 (with its 320x320 input).

## 6.3 Overclocking

While overclocking goes against the grain of this work, because it increases power consumption, we looked at it to see whether the accuracy / power trade-off might be worthwhile sometimes. Figure 2 shows MOTA and power consumption at a range of frequencies, all the way up to 2.0GHz, in the manner described in Section 5.1. The chart shows that overclocking does increase power consumption sigificantly, but also accuracy. This is most likely because fewer frames are dropped. In this test it was possible to keep average power usage under 7W even when overclocked to 1.8GHz, while gaining most of the accuracy advantage of higher speeds.

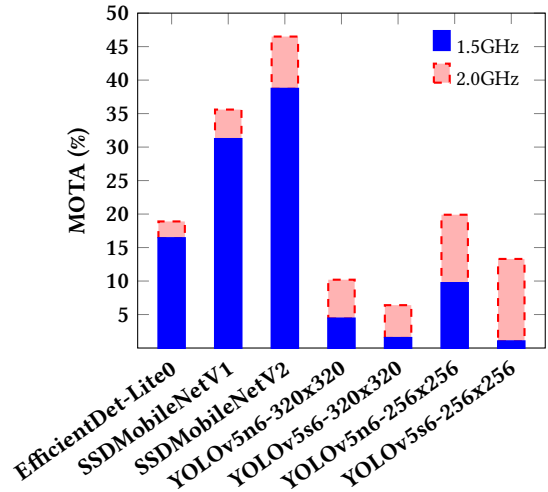

Figure 6: Video 'TUD-Campus': accuracy vs choice of model (see Section 6.2).

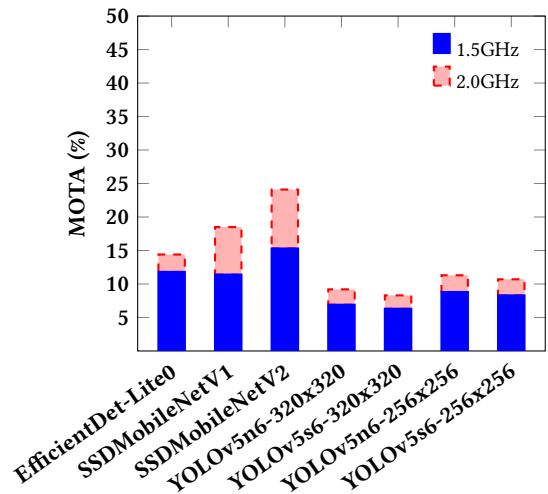

Figure 7: Video 'ETH-Pedcross2': accuracy vs choice of model (see Section 6.2).

## 7 CONCLUSION

The previous generation DeepDish relied on overclocking at least to 1.9GHz in order to bring the frame processing latency to within tolerable margins. That setup consumed 8.5W under load and brought the CPU temperature to 57C, ultimately requiring active cooling and ventilation.

In comparison, the complete redesign of the pipeline and utilisation of the EdgeTPU has resulted in a lower average power consumption with quicker frame-processing latency. The MOTA metric is affected by lower framerates in much different way that the aforementioned people-counting metric was in the older work (see Section 2 discussion on [4]). However, these lower scores can be ameliorated to some extent through overclocking at the cost of additional power consumption. In practice, the parameters and choice of model may require some tuning for each particular scene where a DeepDish is installed.

### 7.1 Source code

DeepDish is open-source and its WWW homepage is: https://github.com/AdaptiveCity/deepdish.

## ACKNOWLEDGMENTS

This research forms part of Centre for Digital Built Britain's work within the Construction Innovation Hub. The funding was provided through the Government's modern industrial strategy by Innovate UK, part of UK Research and Innovation. This work is funded in part by EPSRC grant EP/T022493/1.

# Data Management for Building Information Modelling in a Real-Time Adaptive City Platform

Justas Brazauskas
jb2328@cam.ac.uk
University of Cambridge
Cambridge, United Kingdom

Rohit Verma
rv355@cam.ac.uk
University of Cambridge
Cambridge, United Kingdom

Vadim Safronov
vs451@cam.ac.uk
University of Cambridge
Cambridge, United Kingdom

Matthew Danish
mrd45@cam.ac.uk
University of Cambridge
Cambridge, United Kingdom

Jorge Merino
jm2210@cam.ac.uk
University of Cambridge
Cambridge, United Kingdom

Xiang Xie
xx809@cam.ac.uk
University of Cambridge
Cambridge, United Kingdom

Ian Lewis
ijl20@cam.ac.uk
University of Cambridge
Cambridge, United Kingdom

Richard Mortier
rmm1002@cam.ac.uk
University of Cambridge
Cambridge, United Kingdom

## ABSTRACT
Legacy Building Information Modelling (BIM) systems are not designed to process the high-volume, high-velocity data emitted by in-building Internet-of-Things (IoT) sensors. Historical lack of consideration for the real-time nature of such data means that outputs from such BIM systems typically lack the timeliness necessary for enacting decisions as a result of patterns emerging in the sensor data. Similarly, as sensors are increasingly deployed in buildings, antiquated Building Management Systems (BMSs) struggle to maintain functionality as interoperability challenges increase. In combination these motivate the need to fill an important gap in smart buildings research, to enable faster adoption of these technologies, by combining BIM, BMS and sensor data. This paper describes the data architecture of the Adaptive City Platform, designed to address these combined requirements by enabling integrated BIM and real-time sensor data analysis across both time and space.

## CCS CONCEPTS
• **Computer systems organization** → *Real-time system architecture*; • **Information systems** → Information integration.

## KEYWORDS
smart buildings, IoT, BMS, BIM, real-time, API

## 1 INTRODUCTION
Smart infrastructure projects, e.g., building automation, are increasingly dependent on BIM and BMS for metadata and data collection, analysis and activation components. BMS' often have the ability to monitor lighting, heating, ventilation and air conditioning (HVAC) as well as electricity consumption. Typically proprietary software, most BMS vendors provide closed system products with industrial interfaces for activation, control, and basic data visualisation without the additional contextual data from BIM. Simultaneously, considerable effort is being expended on the deployment of IoT devices to increase sensor density, but unfortunately the industry remains highly fragmented [10, 30].

Additionally, legacy BMS' and much current research focuses on in-building sensor data collection, storage and presentation platforms, rarely emphasising the challenges and benefits of being able to analyse and respond to data in real-time [16, 18, 45]. BMS' have historically dealt with low-volume low-velocity data and metadata, so the adoption of IoT devices poses substantial network and system challenges in dealing with real-time data analysis, event recognition, prediction and action planning [40].

In this paper we focus on the real-time aspects of spatio-temporal data available from IoT sensing. We define a real-time platform as an asynchronous system capable of processing high-volume, high-heterogeneity data with minimal latency to collect, analyse, predict and adapt to changes in a timely manner.

A real-time data architecture is only part of the puzzle though: despite the increasing deployment of IoT devices, there are still no canonical means to join BIM and deployed sensors in a single unified system. While numerous attempts exist in the form of creating ontologies (e.g., BRICK) [8, 9] to unify static metadata management for use by building automation systems, industry recognition for metadata standards is limited [10, 11, 30]. Also, as a result augmenting BIM with IoT devices, building and facility management software must be adapted [28]. Highly siloed BMS software must become able to handle an increased amount of contextual building data in a timely manner to comply with the use of edge computing to for accident and emergency management [33] and smart home initiatives resulting in the creation of safer and more resilient smart spaces [23, 46]. New approaches that combine BIM, BMS and sensor data are thus needed.

To meet these important challenges, we propose the Adaptive City Platform (ACP), a system for collecting, processing and visualising building information and sensor data in real-time. Our contributions are: (*i*) the design of the ACP, a real-time software architecture combining BIM, BMS and IoT data into a unified low-latency, high-throughput building monitoring system; (*ii*) description of our prototype implementation of the ACP, deployed in our department building; and (*iii*) initial evaluation of our platform, in the form of a set of data management practices and detailed examples used to create a functioning front-end visualisation prototype combining heterogeneous data from BIM, BMS and IoT systems.

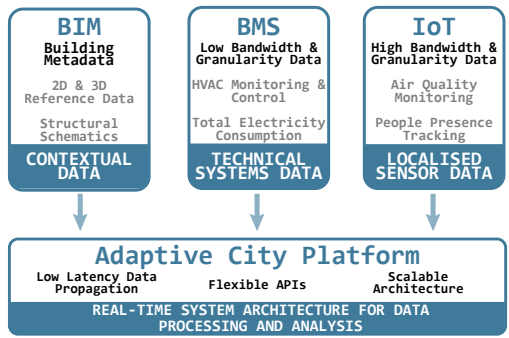

**Figure 1: BIM-IoT fusion envisioned as managed in a single real-time platform combining the contextual, building-systems and localised data provided by BIM, BMS and IoT.**

We first discuss the background and trends in BIM, BMS and IoT integration research in (§2) and describe the importance of a real-time platform in (§3), followed by our original work presented in (§4). We then discuss ACP's advantages as well as challenges it poses in areas such as privacy and sensor-derived constraints in (§5), before concluding (§6).

## 2 RELATED WORK

Building Information Modelling (BIM) first appeared in the 1970s [20], and is the design of buildings and infrastructure projects so as to enable different AEC (Architecture, Engineering, Construction) stakeholders to collaborate on a project [6]. BIM models typically comprise a database of complete structural schematics encompassing 2D plans, 3D models, as well as data about building internals, e.g., the materials and parts used in construction. They have been found to provide many benefits, including more efficient collaboration and coordination in all phases of building construction, improved planning, reduced construction costs, and a lower carbon footprint [4, 7, 13]. As a result, BIM use has increased five-fold in the last decade [34]; the most popular BIM software today, used by over 46% of architects and engineers in the industry, is Revit followed by Archicad [34]. However, despite this rise in popularity and BIM's numerous advantages, its definition remains debated [19].

As well as becoming an integral part of complete building lifecycle management including planning, construction and maintenance phases, BIM is expected to play a key role in the deployment of IoT devices in smart buildings [14]. As the number of connected devices increases, around 11 billion IoT devices are expected to be installed in buildings worldwide [1]. However, deployment of these sensors presents challenges to do with the timeliness and contextualisation of the data they produce. To create smart, automated infrastructure, these IoT devices must interface with BIM and BMS platforms to provide both visualisation and actuation capabilities within the temporal and spatial building context. This three-way BIM-BMS-IoT (abbreviated as BIM-IoT) relation forms closed loop systems that we term BIM-IoT fusion (Figure 1).

Today, building systems monitoring and actuation is managed by highly encapsulated, siloed BMS systems [10, 30]. Usually employed to monitor building-scale mechanical and electrical systems (e.g., HVAC, electricity consumption), the majority of BMS' are unfit for use with IoT as they typically lack the contextual information provided by the BIM software as well as the capacity to access and process low-latency, high-bandwidth IoT data [31, 40]. Though there are numerous benefits for further integration of these technologies, BIM-IoT fusion remains in its infancy.

Successful BIM and IoT device use for energy, health, safety and building management, and construction monitoring have been demonstrated [37, 47]. Furthermore, combined with BMS, BIM-IoT fusion allows for a new way to manage, control and ultimately automate building systems to achieve lower electricity consumption and reduce the $CO_2$ footprint [32]. While much smart infrastructure research focuses on overcoming the standardization problems and crafting ontologies for static reference data, BIM-IoT incorporation is becoming an important topic in its own right [45].

While real-time analysis of data processing is often overlooked, several papers have proposed interlinking and contextualising sensor data using BIM tools [45]. Nevertheless, most attempts remain relatively small scale and often lack good (non-blocking, asynchronous) support for real-time data. A common pitfall with data storage-centric models that rely on repeated querying is that sensor data is collected and archived while sensor metadata is made an extension of the BIM model. This results in smart building platforms that are dependent on static BIM software with no real capacity to provide a low-latency data flow, as the examples below show.

Penna et al [38] describes a system where sensor readings are linked with a BIM database using Revit and Dynamo (a visual scripting language for Revit). The proposed system integrated LoRaWAN-powered environmental sensor data such as temperature, $CO_2$, humidity with people's presence, storing it in a SQL database. The data was then loaded to Revit using a Dynamo script to visualise readings. However, the system lacked a real time component, resulting in transmission latencies of up to 15 minutes.

Rasmussen et al [26, 42] presents an integrated web system using their proposed Linked Building Data ontology with sensor and actuator readings. Although their demo featured a well described web interface and detailed API reference, it lacked a real-time component as a result of their storage-centric model, as well as also relying on Revit. Similarly, Chevallier et al [16] propose a reference architecture for Digital Twin buildings, using BIM and IoT technologies. Focused on ontology definitions, the presented system architecture and examples also feature a database-centric model reliant on queries.

Dave et al [18] proposed a platform that integrates IoT data and real-time web visualisation environment, *Otaniemi3D*, that showed the University of Aalto's campus on several scales. They describe the design criteria, system architecture, and workflow, along with examples of how their system would operate in a real-world environment. However, this also assumed a "collect, store, query" model for sensor data which fundamentally limits its real-time capability. Additionally, the static nature of their XML-based data storage makes it unsuitable for spatio-temporal data.

More widely, Tang et al [45] provides an overview of research covering both BIM and IoT, and highlighting several major challenges in research regarding smart buildings. Key overlooked aspects identified include practical implementability – many described systems were conceptual rather than usable. Moreover, prototypes were often tested under lab conditions, limiting the assessment of the ability of those systems to withstand the challenges of real-world deployments. Finally, they assert that most research they survey fails to achieve real-time information queries, and lacks actuation elements.

## 3 REAL-TIME REQUIREMENTS

As deployed IoT sensors become the standard for applications such as periodic monitoring of temperature and air quality in buildings, we propose that the next iteration of connected environments infrastructure focus around asynchronous real-time systems capable of collecting, analysing, predicting and adapting to change in a timely manner. There is a lack of attention paid in existing work to timeliness in such systems, perhaps due to the relatively limited scale of IoT deployments considered [18, 45]. We believe that the importance of event-driven real-time platform is thus insufficiently emphasised. We next clarify why this real-time component is essential, and how novel data flow and collection methods can be used to minimise response time.

### 3.1 Key Concepts and Data Types

A real-time focused system architecture is important because connected environments software need event-driven data to provide responsiveness without introducing arbitrary programmer-determined delays. Instead of being solely reliant on data querying using APIs, our proposed system architecture also enables instantaneous data flow, allowing for rapid event detection. By implementing this architecture we can overcome some of the hard limits related to IoT sensor deployment and BMS, such as the dependence on periodic data querying and static BIM systems.

In this section we outline key concepts related to an effective real-time data management platform for buildings. We distinguish different data types flowing through the platform, whilst simultaneously declaring that all data be treated as spatio-temporal data. We consider building data as spatial data, while all time-series sensor readings are temporal data, split into periodic and event-based data.

Many BIM systems assume spatial data is static by ignoring building metadata changes once construction is complete. We contend that this is false: buildings change over their life-cycle and so does their spatial data, either in the form of changes to floor plans or movement of building contents. A proper real-time platform integrating sensors, BMS and BIM must have the capacity to update spatial data properties as they change. This is particularly important in cases where building equipment often changes location, e.g., hospital beds, as coordinate data becomes (perhaps low-rate) real-time data subject to stream processing.

### 3.2 Status Reading vs Events

One of the fundamental challenges in real-time software platforms for sensor data is the lack of distinction between event data and periodic readings. This issue arises with the off-the-shelf sensor software which is often based on periodic status reporting rather than publish-subscribe and stream processing architectures.

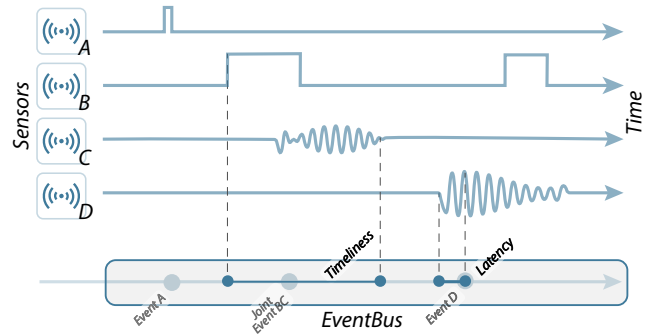

Figure 2: Concepts of Timeliness and Latency in events from spatio-temporal data.

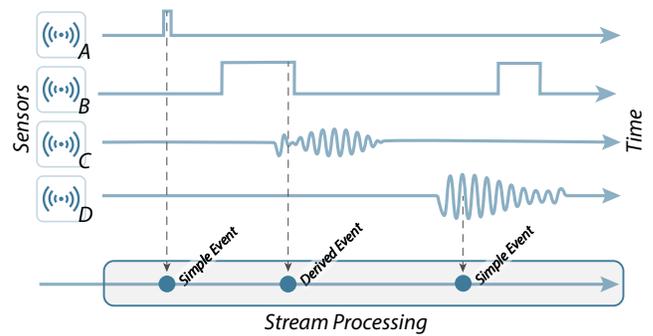

Figure 3: Simple and Derived events from spatio-temporal data, illustrated by different types of sensor readings.

While periodic readings are useful for sensor status reporting or sensing data that changes very gradually (e.g., air humidity), a platform solely based on periodic queries cannot be defined as real-time due to arbitrary blackout time-periods where no data is reported. For this reason, we believe a proper real-time architecture must focus on the concepts of *timeliness* and *events*, where messages are sent instantaneously after readings indicate a significant change in state.

We define timeliness as a characteristic of an event, in which it is appropriate to act on changed readings, e.g. the duration from the beginning of the end of an event. In other words, it is a time-frame during which our event is at its most relevant. While for some events, e.g. interrupt-triggered sensor readings, latency and timeliness are almost identical time-wise, however, for more sophisticated events that follow a sequence of multiple sensor reading, minimising latency becomes difficult due to uncertainty. As a result, we propose that every incoming event comes in with a probability value, determining the how likely the event-recognition is to be true.

Furthermore, the focus on events rather than raw data reporting allows for a flexible and easily scalable approach to data collection by having the ability to group raw sensor events into more informative, less frequent events. As the number of sensors increases, events can occur due to a single sensor being triggered, or as a result of a sequence of triggers passed down from multiple sensors. We distinguish two such categories of events: *Simple* and *Derived*.

**Event types**. Depicted in Figure 3, we derive events based on their readings, e.g., when sensor readings reach a specific threshold or an interrupt occurs, for Sensors A and D. After a trigger occurs, the sensor sends an event to the real-time platform. We define such incoming data as *Simple Events*. Sensors B and C are programmed to follow event detection based on a sequence of readings, e.g., sensor B's activation being followed by sensor C's. In such cases, events are detected as a combination of sensor readings and we consider these *Derived Events*.

Derived events are only possible when the timeliness of individual sensor readings overlap (Figure 2). Our platform permits us to look for such sequences of individual asynchronous events that result in more complex events being tracked, thus creating a richer building information environment.

Our intent is not to collect and solely rely on periodic readings and then send the data back to the platform, but rather to use stream processing to detect derived events as they happen. Real-time stream processing allows for data to be received and analysed instantly, therefore we are able not only to recognize events from a single sensor but also from a combination of multiple incoming data sources. This event-focused approach to real-time system architecture allows for non-blocking data processing with minimal additional latency and minimal network load.

## 3.3 Programmatic Access to BMS

In addition to event detection, another critical aspect of smart building infrastructure is the BMS-in-the-loop integration for building technical systems monitoring and actuation. While IoT sensor deployment is instrumental in smart buildings, programmatic access to BMS software for real-time usage suffers from the same lack of ontological standards that hinders deeper BIM-BMS integration.

The combination of a real-time platform that has access to both BMS and in-building sensors then allows for timely decision making by collecting, analysing, predicting and adapting to the sensed environment. As it stands, BMS software is limited to low granularity data (e.g., electricity consumption on a per-floor basis) and closed control loop mechanisms that would benefit from richer IoT data sources. While there are products on the market that claim to augment BMS capabilities by implementing MQTT drivers and deeper IoT integration [44], the benefits of such platforms remain to be seen. In addition to well-known building automation benefits (e.g., decreased energy usage or HVAC optimisations), deeper BMS integration with the BIM software and IoT could be particularly useful in (*i*) detecting accidents and emergencies, (*ii*) microgrid energy management, (*iii*) domain-specific applications.

(*i*) *Accident and emergency management*. The use of a real time platform for accident and emergency response can form the basis for effective immediate risk management. Broader IoT integration within BMS would play a critical role in recognising such anomalies in two ways. First, deployed sensors would be capable of detecting changes and anomalies in the behaviour of critical assets in real-time, avoiding further damage e.g., by monitoring pump vibrations and sensing any deviation from what is observed to be normal.

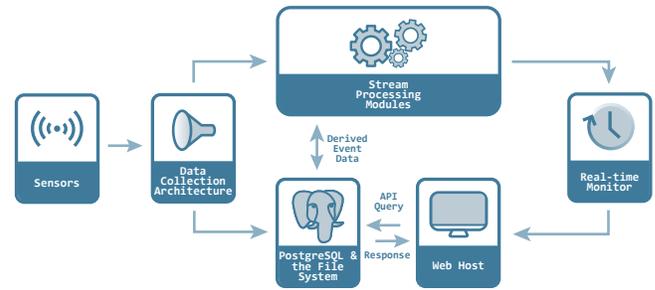

**Figure 4: Adaptive City Platform system architecture.**

Furthermore, the use of real time event recognition could be crucial in recognising gas or oil leaks in industrial settings where such accidents could be handled before becoming serious [33]. Second, it would permit a general system-wide tracking across numerous subsystems deployed in buildings e.g., HVAC. A real-time BMS with access to highly localised sensor data would be especially useful in large scale infrastructure projects e.g., airports where numerous subsystems create a complex whole. Therefore, the resulting outcome becomes not only the provision of time-critical response systems but also a generally lowered complexity that makes large infrastructure projects easier to manage.

(*ii*) *Microgrid energy management*. Event-based occupancy tracking from a multitude of sensors deployed in the built environment allows for more efficient energy usage. Having the ability to combine sensor readings coming from environmental factors, like $CO_2$ concentration [22] and presence-detection indoors [52] presents numerous possibilities for heating optimisation and electricity consumption on a finer granularity a per room basis [39, 51].

(*iii*) *Domain-specific use cases*. Real time event detection can benefit built environments where building inhabitants are at an increased risk, e.g. in hospitals, for patient tracking and fall detection in wards [21, 23]. Similarly, in a smart home context, real time accident recognition could help the elderly population receive timely help if unable to live independently [3, 46]. Lastly, real time-based people density estimation to avoid overcrowding and violent behaviour detection in crowds can be useful for managing people flow in busy areas like stations or airports and stopping crowd crushes [24, 35].

## 4 DESIGN

The core of our contribution is our real-time system, the Adaptive City Platform (ACP), as well as the accompanying example application showing our data pipeline in a real-world scenario. We connected our platform to sensors deployed in a non-BIM native building. The platform uses an ontology that we created to illustrate how real-time mechanisms allow for more versatility in BIM-IoT fusion.

Figure 4 depicts the overall system. The ACP accepts real-time data flow from in-building sensors over the MQTT protocol. The back-end consists of a non-blocking real-time platform with a set of APIs accessing the file system and a PostgreSQL database. Any incoming data is passed through our real-time platform where it is restructured, stored, and ultimately propagated to our front-end visualisation. The collected data is saved in a PostgreSQL database and the file system along with the BIM metadata.

We next describe our real-time platform architecture, APIs, and the front-end interface.

## 4.1 Storage Structure

In contrast to traditional XML based schemata used in BIM and other open data models like IndoorGML [48] or Brick [9], the data inside the Adaptive City Platform is propagated using JSON objects. There were several reasons for this. First, it enables the data to be easily readable to humans and provides numerous benefits when working on both front-end and back-end languages [29]. Secondly, we envision the data flowing through the platform to utilise a Linked Data system, and we found JSON-LD to be the most compliant with our needs.

We annotate our data with auxiliary metadata properties, such as unique name identifiers, building descriptions and location data. This is done by appending metadata properties to the propagated data rather than replacing them, resulting in self-contained data structures that do not require additional API queries to be contextualised in use. For example, a sensor data object does not rely on querying the BIM to obtain information on the sensor's location, because that data is already embedded within the sensor metadata.

Our data structures can generally be separated into three distinct types: (*i*) building data, (*ii*) sensor metadata and (*iii*) sensor readings. This separation provides flexibility in data processing, as well as making it more usable for visualisation. Finally, sensor readings are stored in a data repository, where the data is sorted by date, as well as saved individually for each sensor and then sorted by date. While the data is duplicated this way, it does become useful when working with API queries that ask for data over a time period rather than from a specific sensor. Since we focused on a single building deployment, we found that saving our data in the file system worked sufficiently well, however, with a bigger deployment the use of a different data repository, e.g. a SQL database, would be a viable option instead.

## 4.2 Platform Modules

The ACP real-time platform is based on the underlying architecture of the SmartCambridge framework [43] which is implemented using Eclipse Vert.x [17], to allow for real-time event-based data processing. The platform consists of multiple Vert.x modules termed *verticles* responsible for asynchronous, non-blocking data traversal. Figure 4 illustrates these modules as Data Collection Architecture, Stream Processing, and Real-Time Monitor (RTMonitor) blocks.

**Data Collection Architecture.** Incoming sensor data is captured by our Data Collection verticles that listen for incoming messages over MQTT. After a message is received, the modules timestamp binary sensor data, archive it in the file system and further propagate sensor readings down the platform as JSON files. Using the SmartCambridge framework allowed us to reach an average latency of 160ms from sensors sending the data to its use in the sample application.

**Stream Processing Modules.** Stream Processing is comprised of modules that further parse the received JSON files. These modules decode and write the data to the file system as well as inserting relevant metadata in the database and passing data to the RTMonitor for low-latency data visualisation. Stream Processing modules are versatile because they are able to sort the incoming data and separate it into different directories per event basis for future API use.

| /bim | returns building metadata |
|---|---|
| /space | returns rendered SVG objects based on the BIM data |
| /sensors | returns sensor metadata |
| /readings | returns sensor readings |

Table 1: API description.

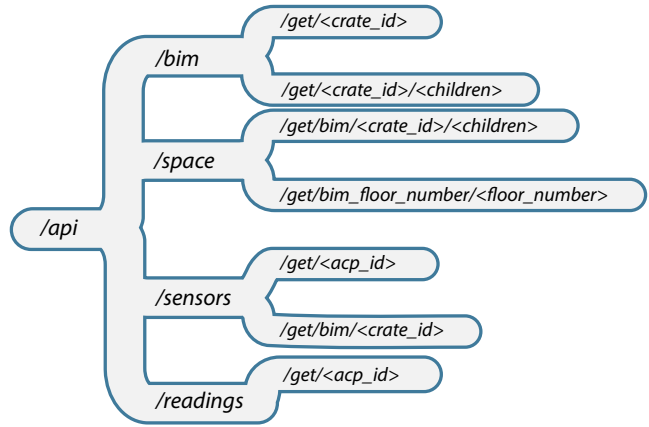

Figure 5: API structure.

**RTMonitor.** The RTMonitor is a module allowing client web pages to issue subscriptions to the Stream Processing verticles and receive updates via websockets on the specified URI the moment sensor readings are passed down the pipeline. The RTMonitor then manages these subscriptions using a token-based system, where users can issue subscriptions by specifying the sensor data they would like to receive.

However, while the RTMonitor is an integral part of the real-time platform, we also use regular API queries to fetch historical data and BIM-related metadata from our database.

## 4.3 APIs

We created four APIs to fetch the sensor metadata, sensor readings, BIM metadata and rendered SVG BIM data. These API endpoints are used in our example application to provide the scaffolding that permits the non-blocking real-time data flow straight to the front-end visualisation.

The front-end is maintained using the API endpoints that interface with the PostgreSQL database and the file system to query the BIM and sensor data. Three of the four API endpoints return JSON-formatted historical data that is easily manipulated by JavaScript or Python. The */space* API endpoint is unique in that it returns nested XML SVG objects that are rendered on screen for the building visualisation. The diagram shown in Figure 5 and the Table 1 describes the API in more detail.

## 4.4 Building Reference Data Platform

In order to evaluate the Adaptive City Platform, we deployed and tested IoT sensors in a non-BIM native building, that did not have any Revit files, other than 2D floor plans. Initially we considered adopting a BRICK ontology for the building, however, after testing it we decided to craft a JSON-LD compliant proprietary ontology with the ability to be converted to BRICK. We did this for the following reasons.

First, BRICK's focus is to capture information on building operations and improve application portability – not to create another building design model [9], like the IFC standard. This was problematic as our system was deployed in a non-BIM native building that lacked any IFC files for contextual visualisation. Thus, the ability to embed precise location coordinates and boundary data, in addition to relational information, was important for us. While achieving this using the BRICK ontology was technically feasible by saving the additional properties as literals, we found it to be impractical. The impracticality was caused by our metadata properties existing as dictionaries, and hence saving them as strings would *(a)* reduce human readability and *(b)* slower data retrieval time, as string literals would have to be converted back to objects.

Second, we found that including additional metadata properties like timestamps for recognising patterns were key. For our application area, spatio-temporal data is highly relevant, but BRICK mostly ignores such information by disregarding time, making it unsuitable for true real time system applications.

Third, we found that our existing JSON (and further JSON-LD) data structures, whilst providing more flexibility than RDFa used by BRICK, could be easily converted to BRICK ontology, providing the possibility of portability when needed. Therefore, we crafted a sample ontology to reference our building and sensor metadata in the platform.

For our building reference data we used a universal naming convention for buildings, floors and rooms called *crates*. A *crate* is an object (however, usually a building or part of a building) with a defined enclosed boundary that denotes its perimeter. *Crates* can also be nested by defining the outer *crate* as a *parent crate*. For example, we can define a *crate* that denotes a building, and within that building we have other *crates* that are its children – e.g., floors, that in return also have their own children – e.g., *room crates*. This parent-child relation in our building data allows for hierarchical data management that is useful for efficient API queries.

We use the nested tree-like structure in our BIM database to retrieve data associated with our building. For this reason, the API call */bim/get/<crate_id>/<children>* has the optional *<children>* property that allows us to specify if any children should be included for that particular crate.

Additionally, every *crate* has a *crate_type* property to select specific object types, e.g., *floor*, *building*, etc., as shown in Figure 6. Further metadata associated with *crates* includes location, boundary and crate type, as well as a UNIX timestamp that indicates the date any information had been updated. Overall, using this metadata structure permits us to index our objects in the PostgreSQL database using any of the properties defined below to allow for fast and flexible metadata retrieval and data management. A complete example of the BIM metadata is given in Table 2, and a sample response to the BIM API query is shown in Listing 1.

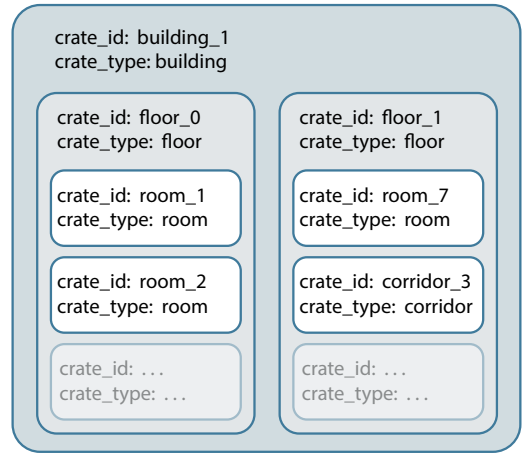

**Figure 6:** *Crate* **structure illustrating the parent-child relation and object types.**

```
{
  "crate_id": "FE11",
  "crate_type": "building",
  "acp_ts": "1589469825.165538",
  "long-name": "Computer Science Department",
  "description": "Crate Description",
  "acp_boundary": "[
  [0,0]  ,[0,78],
  [73,78],[73,0]
  ]",
  "parent_crate_id": "West Campus",
  "acp_location":{
    "f": 1,
    "x": 22.06,
    "y": 34.67,
    "z": 0,
    "system": "WGB"
  }
}
```

**Listing 1: Crate object metadata, received after querying the API endpoint */bim/get/FE11*.**

## 4.5 Sensor Reference Data Platform

Similarly to *crate_id*, sensors also have unique identifiers, the *acp_id* property, defined using the manufacturer's name and a six character-long unique identifier. Since different types of sensors often have varying attributes, our sensor metadata table is defined by two columns containing: (*i*) the *acp_id* identifiers and (*ii*) another containing the other auxiliary properties. Such auxiliary metadata includes information on sensor properties like the location and type. Complete metadata description can be found in Table 3, as well as a queried sample sensor metadata object from */sensors/get/<acp_id>* in Listing 2.

Data Management for Building Information Modelling
in a Real-Time Adaptive City Platform| crate_id | parent_crate_id | crate_type | location | boundary |
|---|---|---|---|---|
| WGB | - | "building" | { "system":"GPS", "acp_lat": -27.116667, "acp_lng":-109.366667, "acp_alt":0.0 } | { "system":"WGB", "boundary":[ [0,0],[..,...],[..,...],[245,56] ] } |
| GF | WGB | "floor" | { "system":"WGB", "x":36.5, "y":39, "f":0, "zf":0 } | { "system":"WGB", "boundary":[ [0,0],[..,...],[..,...],[245,56] ] } |
| FE11 | FF | "room" | { "system":"WGB", "x":22.06, "y":34.67, "f":1, "zf":0 } | { "system":"WGB", "boundary": [ [0,0],[0,78],[73,78],[73,0] ] } |

**Table 2: BIM metadata example, extracted from our database, showing how data can be indexed based on range of properties.**

| key | definition |
|---|---|
| acp_id | sensor identifier, globally unique e.g., elsys-eye-049876. |
| acp_ts | epoch timestamp most relevant to data reading or event, e.g., "1586461606.465372". |
| acp_type | sensor type, determines data format e.g., elsys-eye. |
| acp_event_value | qualifier or data reading for event, e.g., open. |
| acp_event | event type, for time-stamped events, e.g., openclose. |
| acp_location | location using a custom coordinate system e.g., { "system": "WGB", "x": 12, "y": 45, "f": 1 }. |
| acp_confidence | a value 0-1 indicating the reliability of the sensor reading. |

**Table 3: Sensor metadata properties.**

```
{
  "acp_id": "elsys-co2-041ba9",
  "acp_ts": "1589469979.861816",
  "type": "co2",
  "owner": "ijl20",
  "source": "mqtt_ttn",
  "features": "co2, humidity,
light, motion,
temperature, vdd",
  "acp_location": {
    "system":"GPS",
    "acp_alt": 10,
    "acp_lat": -27.116667,
    "acp_lng": -109.366667,
    "parent_crate_id": "FE11"
  }
}
```

**Listing 2: Sensor object metadata, received after querying the API endpoint */sensors/get/elsys-co2-041ba9*.**

```
{
  "acp_id": "elsys-co2-041ba9",
  "acp_ts": "1589469979.861816",
  "features": {
    "co2": 415,
    "device": "elsys_co2",
    "humidity": 36,
    "light": 0,
    "motion": 2,
    "temperature": 15.3,
    "vdd": 3659
  }
}
```

**Listing 3: Sensor reading data received after querying the API endpoint */readings/get/elsys-co2-041ba9*.**

Finally, we expanded the sensor API to be capable of retrieving sensor metadata by their physical deployment location, specifying the parent crate in the API call */sensors/bim/get/<crate_id>*.

### 4.6 Historical Sensor Data Platform

We attach UNIX timestamps to all metadata and incoming sensor readings as soon as they enter the ACP. Depending on the type of data received, JSON packets further propagate to the PostgreSQL database or the file system, where they become historical data. Historical sensor readings can be fetched by querying the API with a sensor's *acp_id* property. The API call then returns the following JSON file from the most recent entry in the database, as shown in Listing 3.

### 4.7 Spatial Coordinates Data Platform

In both building and sensor metadata we refer to our spatial coordinate system as *acp_location*. As the ACP needs to provide data for both relative (in-building) and global (WGS84) positioning references for sensors and *crates* alike, we introduce three parallel location reference systems.

**Global**. The definitive common reference system constituting of latitude, longitude and altitude. The global system is necessary for outdoor sensors as latitude and longitude coordinate system is used while interacting with the *site template* view in the sample application (Figure 7A). We define this in our BIM model by setting the *acp_location* parameter as *GPS*, shown in Table 2.

**In-building coordinates**. We use a spatial coordinate system unique to each building, typically when interacting with in-building floor plan or 3D views of sensors or data. Sensors that transmit their position in the building (particularly relevant for sensors that move around) may use this system in their sensor data. We set in-building coordinates by specifying the *acp_location*'s parameter as the building's name, as illustrated in Table 2 and Listing 1.

**Building object hierarchy**. Hierarchies are often used in BIM software, and in the ACP the hierarchy is defined with the *parent_crate* parameter. It is reasonable for a sensor (or other monitored device) to be recorded as being in location based on a *crate* i.e. a room/office, which relates to the BIM data structured as *site→building→floor→room→window*, etc. This hierarchy is often natively used when collating or browsing in-building information, such as electricity consumption in specific rooms or floors.

By annotating our metadata with a combination of these three location systems, we achieve high flexibility to manage and visualise the BIM and sensor data in multiple ways. This is exemplified in the following section where we showcase how the real-time platform can be effectively used in a real-world scenario.

## 4.8 Visualisation

To best illustrate how the ACP can perform in the wild, we conceived a data visualisation application to monitor and visualise the BIM-IoT fusion data. The visualisation acts as an example of how we can develop software based around our platform and APIs to facilitate the BMS and IoT integration.

The front-end consists of five templates showing the BIM-IoT fusion at different levels. We use multiple templates for our data visualisation, allowing us to select the granularity at which the information is displayed. The templates are loaded in a hierarchical order, where the first template (Figure 7A) represents the site view, while the last one (Figure 7D) shows the *crate-level* view, followed by a sensor readings template (Figure 8) with time-series plots.

*4.8.1 Site Template.* The site-level visualisation (Figure 7A) displays buildings and sensors in aggregated groups on the campus. The map is rendered using Leaflet and OpenStreetMap [36], an open-source JavaScript library for interactive maps.

We query our database to return the boundary data for the buildings we would like to render on the screen. After receiving the boundary coordinates in latitude and longitude we then render polygons over the original building locations on the map. Finally, users are able to navigate to the other templates by clicking on individual buildings.

*4.8.2 Building Template.* In order to render the building-level template we query the API to return individual floor and room boundaries based on the information we have in the BIM database. After fetching the boundary data, we then render the building in 3D using the Three.js library [49].

```
<g>
  <polygon
    id='FE11'
    data-crate_type='room'
    data-parent_crate='FF'
    data-floor_number='1'
    points='362.8,0 362.8,40.3 482.1,40.3 482.1,0'>
    <title>
      FE11
    </title>
  </polygon>
</g>
```

**Listing 4: A generated *crate* SVG sample on the floor-level visualisation. We embed additional metadata to SVG objects using the HTML *data-\** attribute.**

The building-level template (Figure 7B) allows for a general building view to inspect individual crates, or proceed to any of the floor plan templates by clicking on the floor icon.

*4.8.3 Floor Plan Template.* The floor-level template (Figure 7C) has several important features. Primarily, it has the ability to display heat-maps of sensor readings, as well as the most recent sensor data. Users can inspect sensor readings by hovering over the sensor icon on the floor plan. Upon doing so, an API call is executed fetching the most recent data for the particular sensor in question.

We query the */space/get_bim_floor_number/<floor>* API endpoint to acquire all crates that are on the queried floor. The API returns an SVG object along with the metadata necessary to render the floors and individual rooms. We then use D3.js [12] to render the SVG on screen and make it interactive.

With this approach we eliminate the need to possess static SVG files in the file system, and instead rely on our predefined BIM data structures that can be easily updated and generated into SVG files on demand.

Users proceed to enter the floorspace template by clicking on a crate on the floor-level template.

*4.8.4 Floorspace Template.* This final BIM-generated template (Figure 7D) is used to display the finest granularity spatial data, including the *crate*'s BIM description and metadata for all the sensors deployed in that crate. The selected *crate* of type *room* is zoomed in on, allowing to more accurately display the sensor location.

*4.8.5 Sensor Template.* As users travel down the spatial hierarchy of the web application, the data becomes increasingly more granular, allowing detailed time-series data analysis. The sensor-level template is accessed by clicking on a sensor on either the floor plan or floor space views. The user is then redirected to the sensor template where detailed spatio-temporal time-series data is shown, as illustrated in Figure 8. This template allows to visualise the data provided by every in-building sensor. Users are able to select the timeframe they wish to inspect. The visualisation changes in real-time as new data readings propagate through the platform.

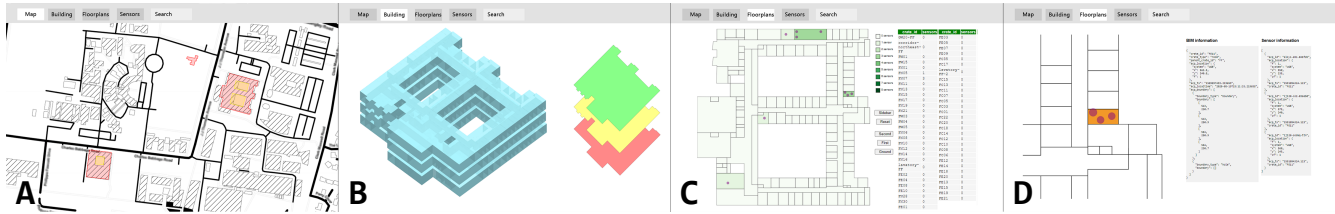

Figure 7: The 4 templates used to show BIM data on different scales: (A) Site, (B) Building, (C) Floor, (D) Floorspace.

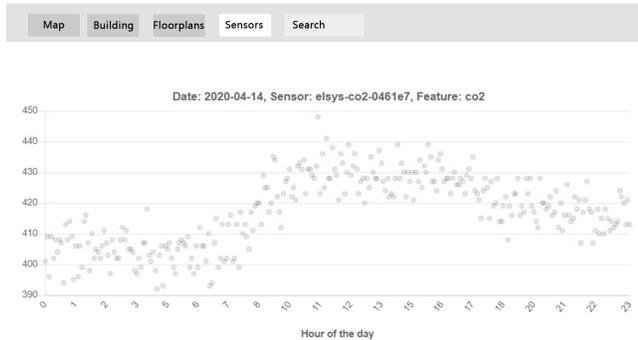

Figure 8: A template used to visualise a day's worth of historic sensor data for one of the deployed $CO_2$ sensors.

## 5 DISCUSSION

During the development of the ACP and the accompanying visualisations, we identified four major themes associated with BIM-IoT fusion: (*i*) scalability, (*ii*) sensor-derived constraints, (*iii*) BMS actuation and (*iv*) privacy.

### 5.1 Scalability

While the ACP does provide real-time sensor information, it remains a single building deployment with a limited scope. We anticipate using our modular architecture to add additional buildings simply by appending the BIM files to the current PostgreSQL database.

One challenge in doing so is to acquire the BIM data in the first place. In our example we used a non BIM-native building and as a result had no access to 3D building models and other key BIM data. Instead we parsed the SVG floor plans for our building to extract the boundary data for individual crates. While such CAD file parsing could also be adapted to other 2D-friendly formats such as DXF or DWG, the differences between the way building data is stored remains an issue. As use of BIM in the design phase of buildings increases, a potential solution would be to write a Revit plugin that would automatically generate and export the requested data in a suitable ontological standard [41], or utilise IFC-BRICK converter [9].

In terms of the back-end system performance, we have no reason to suspect that any further scalability efforts would affect the real time platform itself. Since our real-time architecture is based on the Vert.x system used in the SmartCambridge project [43] that had ∼1000 sensors sending data every 20 seconds, we have encountered no slowdown with asynchronous modules processing data with low-latency.

### 5.2 Sensor-produced constraints

Event-driven data is crucial for responsiveness but the majority of off-the-shelf sensors, with the exception of interrupt-based devices, are only capable of sending periodic readings, e.g., $CO_2$ every 5 minutes. Furthermore, such sensors are rarely reconfigurable to allow for dynamic data transmission based on user-defined threshold values, making it impossible to implement timely event-driven behaviour to track sensor readings in real-time. Finally, many sensors lack control over where data is delivered – while we chose to rewrite factory sensor firmware in some cases, the complexity and the required man-hour efforts may not necessarily be worth the desired outcome.

Overall, sensor choice and their detailed capabilities plays a key role in determining the timeliness factor of any real-time platform, therefore considerable attention should be paid to what sensors are integrated within the IoT ecosystem.

### 5.3 BMS Implementation and Actuation

As research progresses towards the adaption of a universal ontological standard to incorporate BIM and BMS with IoT technologies, machine learning-based approaches to building automation and actuation will become more important. According to Tang et al, many current BIM-IoT fusion attempts fail to achieve the enclosed loop system capability by not enacting on actuation and only being capable of sensing [45].

*5.3.1 BMS in-the-loop.* BMS implementation is often a non-trivial task due to siloed software systems, as well as some buildings not having actuation capabilities implemented and accessible by their BMS [10]. While beyond the scope of this paper, such attempts to normalise BMS data are made through the creation of building ontologies, providing unified metadata schemas.

However, a substantial number buildings have neither BMSs nor BIM-enabled facilities management systems, and therefore lack basic programmatic access. While deploying IoT sensor would make such buildings 'smarter', the lack of actuation equipment means that we would not be able to reach the full potential of modern, BIM-native buildings.

While implementing actuation components in such buildings would likely be the next step after the deployment of sensors and machine learning tools, it would also mean a substantial financial commitment to AEC stakeholders. Nevertheless, it has been reported that building automation does indicate a substantial energy savings [15, 53], thus potentially making expenses cost-effective, even ignoring the benefits that such systems could bring in risk/disaster management and anomaly detection.

*5.3.2 Data Analysis, Prediction and Actuation.* Even though HVAC systems in buildings have been around for a considerable time and have shown positive results in increasing energy efficiency [53], the mass deployment of IoT sensors in the built environment will provide much more granular data feeds. This increase in the amount and detail of data is what drives the need for fast and efficient real-time platforms like the ACP, as they permit the analysis and decision making to be done on the spot, once the right machine learning tools are implemented in the data pipeline.

The ongoing COVID-19 pandemic prevented us progressing our deployment to the point where we had dense enough deployment of sensors and rich enough data from those sensors, building automation and data analytics remains a topic of great importance for future research.

## 5.4 Privacy

Privacy in IoT-backed smart buildings has grown in importance over the last decade [2, 15, 27], but more work is needed as privacy issues remain a minefield and may hinder further technology adoption [25]. It is important to note that while sensing in the presence of people in itself causes privacy concerns, real-time tracking further amplifies these concerns, as data integrity may not always be guaranteed [45]. Even though none of the sensors are used to track individuals directly, latent information may still cause ethical issues, as it may be possible to, e.g., infer information about office workers by analysing $CO_2$ data and electricity consumption [5].

The primary concern about deploying sensors in the built environment has to do with the ownership of and the access to the collected data. Sensors are relatively easily locatable objects in the workplace and a possible source of anxiety due to the collection of data. Currently there are few IoT deployment guidelines covering office spaces and data ownership, other than the GDPR, focused on data collection and storage [50]. We envision a privacy-oriented system provided by a capability-based platform where people are given access to data based on a combination of factors such as their day-to-day proximity to sensors (you should have means to examine data collected about your behaviour), and role in the company (a building manager may need to see aggregate data for the whole building).

For example, people with sensors deployed in their offices would have the access and control of the high granularity (or frequency) time series data that they produce. However, such access would be limited to the rest of the building, providing a decreasing level of granularity (or frequency) across time and space domains. An example of this in practice could be the building facility manager having the access to entire building data on a weekly scale without specific information on individual offices.

In addition to personal privacy in the workplace, in a case study presented by Cascone et al, security of data was the second most common concern after privacy [15]. The deployment of similar web applications to ours comes with a risk of data being leaked to third parties. Therefore, considerable care should be taken to secure both the BIM data, and the sensor data so that it would only be accessible to authorised personnel. The same capability-based platform could be employed in cases where third parties are given access to some of the low frequency, low resolution spatial data.

## 6 CONCLUSION

The lack of standardisation in BIM-IoT fusion means that the AEC industry is unprepared for mass IoT sensor deployment in the built environment. Our contribution to research on BIM-IoT fusion has been to describe a high-level real-time software architecture for processing data, and to show how it can be used to visualise in-building sensor data with useful features for facility managers and building inhabitants alike.

As it stands, our real-time Adaptive City Platform is currently unique, playing a key role in the way the IoT sensors are integrated with BIM. By using the concept of events in capturing time-sensitive data in the ACP we were able to achieve low-latency stream processing capable of instantaneous updates to our in-house visualisation.

We have further shown how hierarchical data visualisation can be used to display the collected data at different levels of granularity. Furthermore, we have described how metadata is described and our data is stored, both buildings (*crates*) and sensors.

While we have demonstrated that the ACP works, there are still substantial challenges in how to tackle mass sensor deployment and further data analytics. As workplaces and buildings eventually reopen as the current phase of the COVID-19 pandemic subsides, we anticipate being able to increase the density and spread of our deployment, enabling us to collect and process more and richer data to explore how the ACP can support prediction and actuation.

# *SenseRT*: A Streaming Architecture for Smart Building Sensors

Rohit Verma
rv355@cam.ac.uk
University of Cambridge
Cambridge, United Kingdom

Justas Brazauskas
jb2328@cam.ac.uk
University of Cambridge
Cambridge, United Kingdom

Vadim Safronov
vs451@cam.ac.uk
University of Cambridge
Cambridge, United Kingdom

Matthew Danish
mrd45@cam.ac.uk
University of Cambridge
Cambridge, United Kingdom

Jorge Merino
jm2210@cam.ac.uk
University of Cambridge
Cambridge, United Kingdom

Xiang Xie
xx809@cam.ac.uk
University of Cambridge
Cambridge, United Kingdom

Ian Lewis
ijl20@cam.ac.uk
University of Cambridge
Cambridge, United Kingdom

Richard Mortier
rmm1002@cam.ac.uk
University of Cambridge
Cambridge, United Kingdom

## ABSTRACT
Building Management Systems (BMSs) have evolved in recent years, in ways that require changes to existing network architectures that follow the store-then-analyse approach. The primary cause is the increasing deployment of a diverse range of cost-effective sensors and actuators in smart buildings that generate real-time streaming data. Any in-building system with a large number of sensors needs a framework for real-time data collection and concurrent stream processing from sensors connected using a range of networks.

We present *SenseRT*, a system for managing and analysing in-building real-time streams of sensor data. *SenseRT* collects streams of real-time data from sensors connected using a range of network protocols. It supports concurrent modules simultaneously performing stream processing over real-time data, asynchronously and non-blocking, with results made available with minimal latency. We describe a prototype implementation deployed in two University department buildings, demonstrating its effectiveness.

## CCS CONCEPTS

• **Networks** → *Sensor networks*; • **Hardware** → **Sensors and actuators**; • **Computer systems organization** → **Real-time system architecture**.

## KEYWORDS

real-time data, stream processing, smart buildings, sensors, spatio-temporal data

## 1 INTRODUCTION

Increased penetration of the Internet of Things (IoT) in our lives [22], is making it increasingly easy to deploy energy-efficient sensors and actuators to perform tasks like occupancy detection [26] or indoor-environment control [17]. The number of deployed IoT devices by 2021 just for smart buildings would be around 10.8 billion (2.8 billion for residential buildings) [4]. To put that in perspective, an average home in 2020 would generate approximately 4.7 terabytes of data annually [4, 8]. Handling such amounts of data is a challenging problem, exacerbated by most such sensors generating real-time streams of data.

The data generated has both spatial and temporal aspects. For example, the humidity level falling below a threshold causing discomfort to occupants will be observed by a sensor deployed in a specific place in the building (spatial aspect) at a particular date and time (temporal aspect). The spatial aspect is relatively straight-forward to handle, but there are several subtleties to the temporal aspect: what time should be assigned as the *time of reading*, and associated with a particular sensor reading? When the data was sensed, when the sensor transmitted the reading, or when it arrived at the receiving platform?

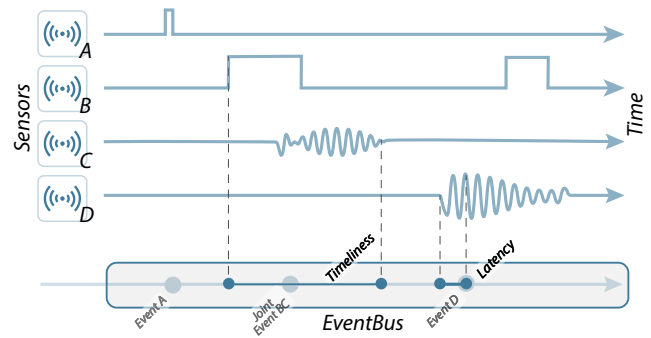

**Figure 1: Timeliness vs latency.**

This brings to the fore concepts of *latency* and *timeliness*, depicted in Figure 1, often incorrectly considered synonymous. We use *latency* to refer to the delay introduced between the time of recording the reading at the sensor and that reading arriving at the intended destination. In contrast, *timeliness* refers to the inherent characteristic of an event, the timescale on which it is appropriate to determine a reading changed. In the humidity example above, *latency* would be the delay between the time the sensor recorded the humidity and when the system determining the degree of occupant comfort received the reading; this will vary between sensors and over time. However, *timeliness* is the time period over which readings from the sensor lead to a situation where the humidity

Rohit Verma, Justas Brazauskas, Vadim Safronov, Matthew Danish, Jorge Merino, Xiang Xie, Ian Lewis, and Richard Mortier

| Protocol | Frequency | Data Rate | Range | Power | Security |
|---|---|---|---|---|---|
| **Bluetooth [52]** | 2.4 GHz | 1 Mb/s | 50 m | Low | 64 and 128 bit encryption |
| **Wi-Fi [13]** | 2.5 GHz, 5 GHz | 150-200 Mb/s (typical) | 50–100 m | High | 256 bit key encryption |
| **ZigBee [14]** | 2.4 GHz | 250 kb/s | 10–100 m | Low | 128 AES layer security |
| **LoRa [16]** | Region specific | 0.3–50 kb/s | 2–5 km | Low | 128 bit AES encryption key |
| **Modbus [6]** | 2.4 GHz | 9600 b/s | Layout dependent | High | 128 bit AES encryption (if Modbus/TLS) |

Table 1: Popular sensor communication protocols used in smart buildings

threshold overshoots. A reading from the sensor that arrives after this time period won't help in computing the event of discomfort. Any framework striving for efficient real-time data management and analysis needs to treat these concepts with care.

Furthermore, the concept of *timeliness* also explains the need of moving towards a stream processing based approach from the prevalent store-then-analyse approach [25, 49]. These existing approaches fail to provide real-time insights when the data never stops and usefulness of the streaming data is ephemeral but has to adhere to *timeliness* requirements [62]. Nowadays sensors are used to manage and control several crucial aspects of a building like identifying time-critical events of gas leaks [31] or fires [42]. In such scenarios, any delay between generation of data at the sensor and its analysis should be minimised as much as possible, wherein stream processing [60] comes in.

Building a system to handle devices in a building and the large volume of real-time data streams they generate while keeping these key spatio-temporal concepts in mind has multiple challenges. First, the heterogeneous nature of sensors and consequently the data they generate. The deployed sensors in a building could connect using channels including Bluetooth, Wi-Fi, ZigBee, LoRaWAN, or the in-building Modbus. Moreover, the definition of time of reading for sensors using these different networks could be different. The system should accept data from any such channels and normalise aspects such as time of reading. Second, analysis of the stream of data generated from all the in-building sensors in real-time should introduce minimal latency between data generation, data analysis, and publication of a result, especially for crucial events such as fire or power outage in critical areas. Moreover, processing of this stream of real-time data should be done for all sensors concurrently. Third, all data flow must be asynchronous and non-blocking while supporting concurrency.

To tackle these challenges, we develop *SenseRT*, a system for real-time data flow in smart buildings. Key contributions include:

- *SenseRT* defines a set of design principles ensuring that the architecture is easy to mutate based on the needs of the building at any time of building design.
- *SenseRT* provides custom decoders to retrieve data from sensors on different channels, accumulated via bridging framework of Message Queuing Telemetry Transport (MQTT) [59], and normalise it for easy use throughout the system.
- *SenseRT* provides a suite of modules which follow the *stream processing* [60] framework to ensure that the streams of real-time data from a large number of sensors is managed and analysed with minimal latency and adhere to timeliness bounds.

- All data handling and stream processing modules are designed to follow the *actor model* [34], which guarantees asynchronous and concurrent processing. We modify the actor model such that the concurrent modules could work in a non-blocking fashion.
- *SenseRT* provides support for client-side applications to analyse and process the data as they need.

We continue by discussing related work in smart buildings (§2), before describing the design principles and high-level system architecture (§3). We implemented a prototype of *SenseRT* in two buildings in our university and ran experiments to observe the architecture's effectiveness for seven months. We describe this implementation (§4) and a case study of an end-to-end application (§5), before evaluating the implemented prototype (§6). We conclude with a discussion of limitations and future work (§8).

## 2 RELATED WORK

The IoT market is expected to increase to 5.8 billion endpoints in 2020, a 21% increase from 2019, with utilities (electricity, water) the most significant users (1.17 billion) and building automation showing the highest growth of 42% [2]. A boost towards this direction could be because of the growth in the number of smart devices and smart objects that build the ecosystem for a smart building [24]. This ecosystem of smart devices and smart objects is enabled by the network technology use, with different protocols (e.g., LoRaWAN, ZigBee, Wi-Fi, Bluetooth) being selected based on the desired goals.

### 2.1 Communication in Smart Buildings

Communication protocols enable the exchange of a massive stream of data between sensors and the network. Factors like range, data load, power demand, and security define which communication protocol would be suitable for a particular set of smart devices. A comparison of the major protocols is given in Table 1.

There has been quite a focus on using LoRa for smart buildings [33, 40] due to its low power consumption, low cost, long-range, bi-directional communication made possible by chirp spread spectrum (CSS) [48], standardization and flexibility of selecting bandwidth, code rate and spreading factor as per need. Unlike the other protocols, Modbus is a wired protocol for industrial automation systems that has recently become popular for building management systems, especially for meters and HVACs.

### 2.2 Sensing in Buildings

Building Energy and Comfort Management (BECM) [55] has been a significant cause of the introduction of sensing in buildings. BECM tries to achieve optimal energy consumption [57, 67] and provide a high level of indoor environment quality (IEC) [11] by using different types of sensor. We divide them into two broad categories:

**Dumb Sensors**. This type includes sensors such as pressure mats, IR sensors, $CO_2$ sensors, temperature sensors, particulate matter sensors. They simply read and transmit readings for values they sense, and can be used to understand occupant behaviour patterns [30, 68] or to optimise IEC parameters to improve thermal comfort, visual comfort, or indoor air quality [27, 47].

**Intelligent Sensors**. This type compromise one or more *dumb sensors* with attached computation that can process the raw data and relay analysed results. Example include smartphones, fingerprint sensors, wearable sensors, smart cameras, and body thermometers. In smart buildings these sensors can be used for traditional purposes such as occupancy detection and managing IEC [66], but can also provide more personalised functions [56].

## 2.3 Smart Building Management and Control

Several works provide a framework for building control to balance energy consumption and maintain occupant comfort. iDorm [32] set up a testbed where multiple embedded sensors were fitted in a dorm room to obtain responsive inputs from the user, which helped to learn user preferences using distributed AI and fuzzy-genetic logic. MASBO [41] is a multi-agent system where data arriving from a Building Management System (BMS) is observed and analysed by a set of agents to provide suitable energy-efficient control for the building without compromising on occupant comfort. Chen et al. [18] propose a hierarchical system architecture that emphasises improving savings over the building life-cycle while addressing stakeholder goals.

Another class of work concentrates more on how sensor data could be collected and managed in a smart building. Choubey et al. [19] set up a localised sensor network in an area and perform localised data processing for this set of sensors. LabVIEW [54] provides a data collection framework to collect humidity, temperature, and light data from sensors in a wireless sensor network in the building. Bashir et al. [15] provide an IoT Big Data Analytics (IBDA) based framework for storage and analysis of real-time data that the IoT sensors in a smart building generate.

Most architectures follow the trend of using a particular approach to collect sensor data locally and then store it on the cloud for further processing. Data collection efficiency is achieved by using different means like adding a programmed data acquisition chip [10], setting up a fog server [29], or using IPv6 over Low power Wireless Personal Area Networks (6LoWPAN) [28]. However, all of these systems rely on stored data to perform any analysis rather than analysing it as and when it arrives.

## 2.4 Necessity of Stream Processing

The existing systems [10, 15, 25, 29, 54], utilize the prevalent store-then-analyze architecture, where data is first stored in a data repository and then analyzed as per the needs of the applications (Figure 2). However, several use-cases of sensing in buildings are linked to time-critical responses. It is very important to minimize delay in safety based use-cases like gas leak [31], water leak [44] or fire detection [42] in buildings. When considered as basic components of a smart-grid network, timely analysis of energy consumption is important to optimize smart-grid management [21, 46]. When the building is used for specialized purposes like hospitals [37] or elderly care homes [12, 61], adhering to timeliness of sensor data becomes crucial. When trying to act on such scenarios, which generate always moving time-critical data, the store-then-analyze approaches fall short. The reliance on a source of data repository would involve several to-and-fro network transactions resulting in crucial time loss. This fails a real-time system's adherence to timeliness [60, 62].

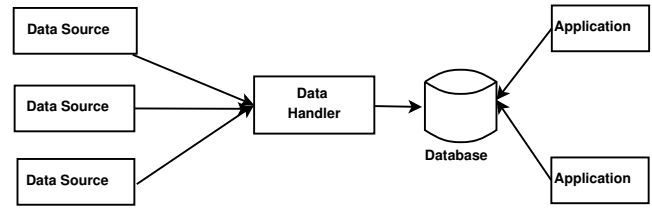

**Figure 2: Store-then-Analyze Model**

Along the line of these time-critical solutions, ScaleOut [53] has pointed out the importance of real-time stream processing [60] for digital twins. ThoughtWire [63] who also work with stream processing based systems for digital twins share a similar notion. CityPulse [64] and Zhou et al. [69] advocate the necessity of stream processing based data analytics for smart city projects.

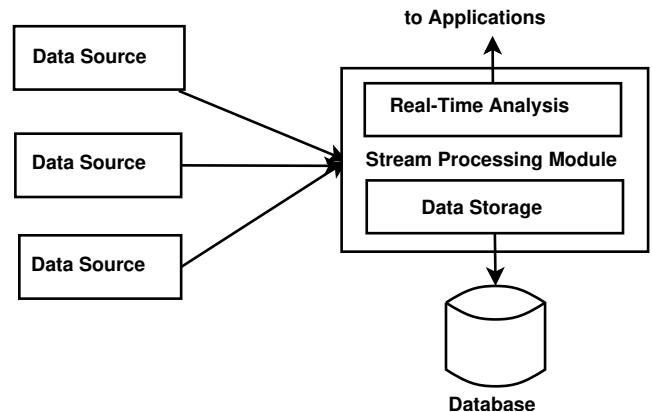

**Figure 3: Stream Processing**

Figure 3 shows the stream processing approach which unlike Figure 2 performs storage and analysis simultaneously. The stream processing module has two (or more) concurrent processors which analyse and store data at the same time. The analysed results are then made available to any applications which have subscribed a priori.

## 3 DESIGN & ARCHITECTURE

We first describe the design principles *SenseRT* is built upon and then its high-level architecture.

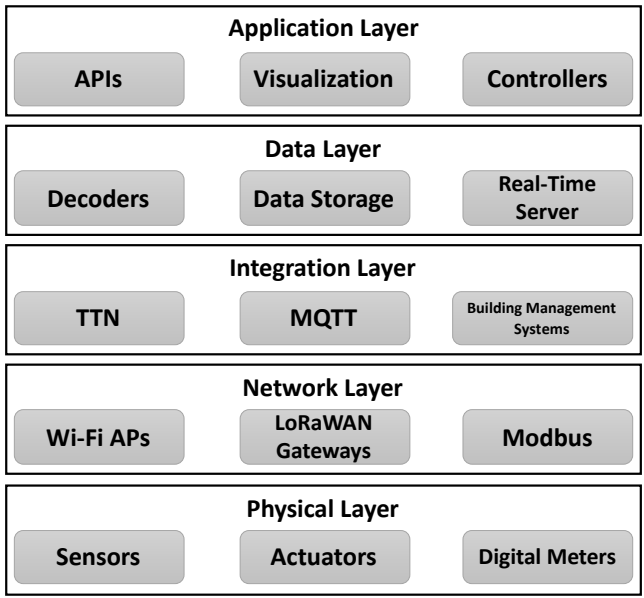

Figure 4: *SenseRT* high level system architecture.

## 3.1 Design Principles

The *SenseRT* architecture must be flexible to changes at any period of deployment and robust enough to handle the continuous stream of real-time data entering the system through the numerous sensors in the building via different channels. To achieve this, we apply the following design principles.

**All data is spatio-temporal**. Over a long enough period, the metadata about any part of the building or sensors is bound to change. For instance, a sensor could be moved to a new room changing its location, or a room be divided into two rooms changing its boundary. Tagging every datum with a timestamp and location information, enables change tracking of deployments.

**Spatial hierarchy of container objects**. Sensors could be deployed in different locations having different usage and access control policies depending on the granularity of location considered (e.g., building, floor, room, desk). This defines each component as an object which could hold multiple other objects, each of which could have its own set of sensors. This ensures physical changes in the building are handled with minimal changes, e.g., changing only the parent when a desk is moved to another room, or reusing the floor model on adding a new floor to the building.

**Stream processing of data**. *SenseRT* uses a publish/subscribe model for data exchange, ensuring no polling is required for the real-time stream of data arriving at any time instance to the system. This reduces delivery latency, essential for real-time data analysis.

**Asynchronous message transfer in a non-blocking framework**. Asynchronous message transfer ensures that any message from any sensor or system module could be analysed in real-time, and multiple modules can work on the same data concurrently. This in turn guarantees a non-blocking data processing framework.

## 3.2 System Architecture

The high-level decomposition of *SenseRT* results in five layers, shown in Figure 4: Physical, Network, Integration, Data, and Application.

**Physical Layer**. Holds all hardware devices like sensors, actuators, and building meters. Devices report data and exchange messages using one or more of their supported protocols and formats. Our deployment uses a wide range of sensors outlined in Table 2.

**Network Layer**. Houses the devices required to support message transfer from the sensors such as Wi-Fi access points, ZigBee or LoRaWAN gateways, or Modbus components. Our deployment uses Zigbee and Wi-Fi networks for short-range connectivity within particular areas, and a LoRaWAN network to provide backhaul interconnection within and between buildings.

**Integration Layer**. Provides services to collate and homogenise data received over different network types from different hardware devices, making it much more straightforward to build applications to analyse and react to data from one or more sources. Consider a room that has smart plugs, LoRa sensors, and an electric meter. The smart plugs could be uploading data over Wi-Fi, LoRa sensors through LoRaWAN via The Things Network (TTN) [65], and the electric meters through a Modbus. Services in this layer ensure that data received over all these protocols are available through a single channel. Furthermore, this layer also ensures that data exchange in the system is through a publish/subscribe method. Our deployment achieves this by making use of the Message Queuing Telemetry Transport (MQTT) [59] and MQTT bridging.

**Data Layer**. Provides services to manage data streams arriving from the integration layer. A set of decoders normalises the data from multiple devices, which is made available to a Real-Time Server (RTS). The RTS holds various crucial modules of the architecture. These modules handle real-time stream processing, routing messages to other similar systems as required, storing data for future usage, and making data available to the external application. In our deployment the RTS is implemented to be asynchronous and non-blocking using Vert.x. The data layer also houses the database, which stores the spatio-temporal metadata of all the sensors and the object-level components in a hierarchical structure.

**Application Layer**. The client-facing layer, providing APIs and user interfaces for those wishing to access sensor data. Our deployment comprises a server that acts as the point-of-contact with *SenseRT* for all client-side applications trying to access sensor data. The details of application architecture are out of scope for this paper as we are primarily concerned with the overall system.

## 4 IMPLEMENTATION

We deployed a prototype of *SenseRT* over two department buildings on our University campus. The two buildings each have multiple floors containing offices, labs, communal areas, and corridors. Figure 5 depicts the overall architecture of the prototype, and we detail each layer next.

### 4.1 Physical Layer

Table 2 lists the sensors used in our deployment, dumb and intelligent.

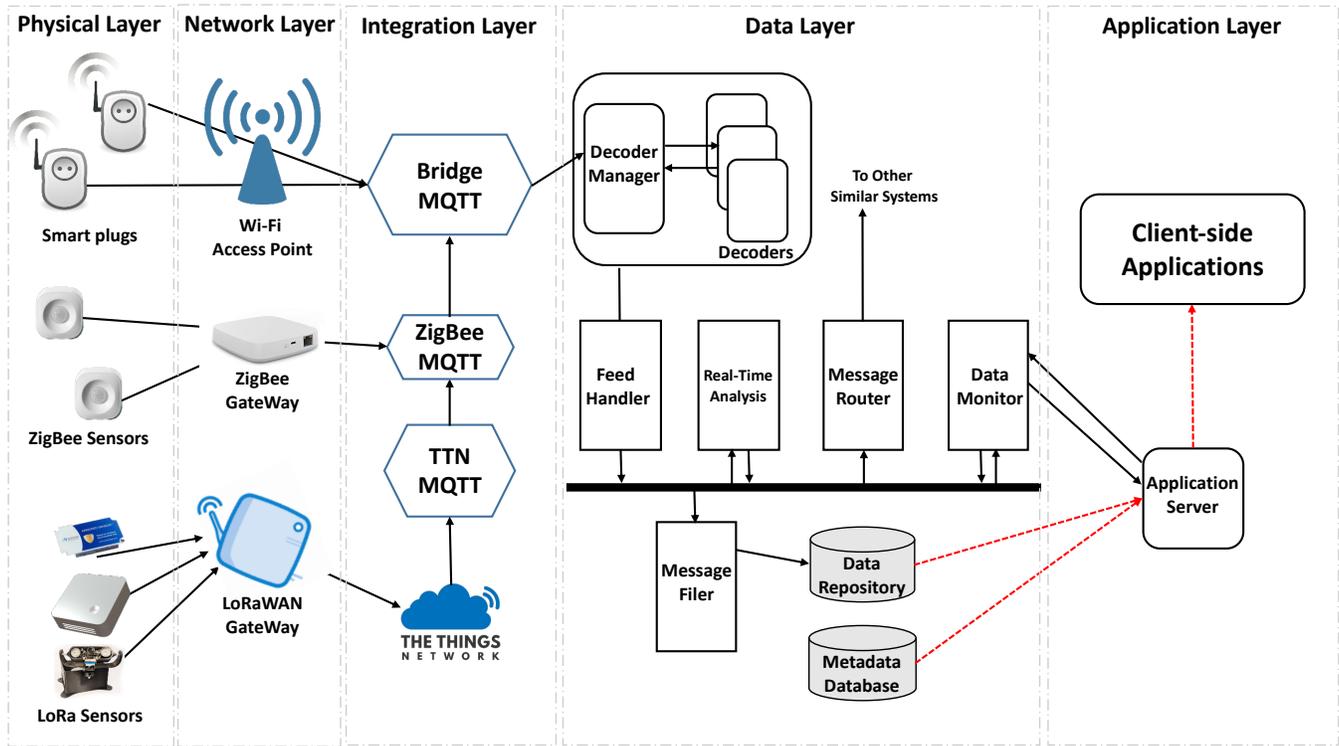

Figure 5: Our prototype implementation of the *SenseRT* architecture. Solid black lines show real-time data flow while dashed red lines show request-response data flow.

| Sensor | Measures | Channel | Cost |
|---|---|---|---|
| Smart plugs | power | Wi-Fi | $45 |
| Infrared Motion | motion | ZigBee | $17 |
| Door/Window | open/close events | ZigBee | $9 |
| $CO_2$ | $CO_2$, humidity, temperature, light, motion | LoRaWAN | $205 |
| Temperature | temperature | LoRaWAN | $129 |
| Tilt | tilt angle | LoRaWAN | $99 |
| Door/Window | open/close events | LoRaWAN | $99 |
| Water Leak | presence of water | LoRaWAN | $119 |
| Occupancy | occupancy, temperature, humidity, light, motion | LoRaWAN | $123 |
| DeepDish | people count | Wi-Fi | $100 |

Table 2: Sensors used in our deployment.

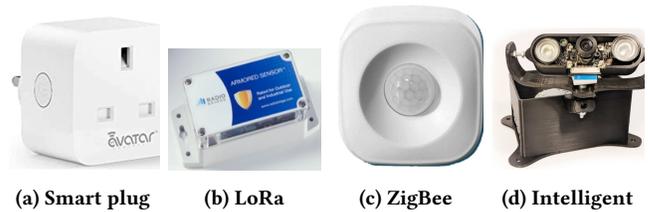

(a) Smart plug  (b) LoRa  (c) ZigBee  (d) Intelligent

Figure 6: Four types of sensors used in our deployment.

*Dumb Sensors.* The three types of dumb sensor we deployed are depicted in Figure 6, and are:

**Smart plugs** (Figure 6a). These are COTS smart plugs from several vendors built around the ESP8266 part [1]. We replaced their default firmware with the Tasmota firmware [3] so we could control where data were sent. The smart plugs were controlled over Wi-Fi using Message Queuing Telemetry Transport (MQTT).

**LoRaWAN Sensors** (Figure 6b). We use different types of COTS LoRaWAN sensors, measuring e.g., $CO_2$, temperature, and occupancy from vendors including Elsys and Radio Bridge. These were managed over LoRaWAN via The Things Network (TTN).

**ZigBee Sensors** (Figure 6c). We use two types of ZigBee sensor: infra-red motion sensors, and door/window open/closed sensors. These sensors were accessed via ZigBee gateways.

*Intelligent Sensors.* The DeepDish [23] (Figure 6d) intelligent sensor counts the number of people in an area. DeepDish uses TensorFlow to identify and track selected objects (e.g., cars, bicycles, people) in the video feed, and supports occupancy counting by counting how many people cross a line in the scene in each direction. Only the cumulative occupancy count is reported, over Wi-Fi; no video data is collected or transmitted.

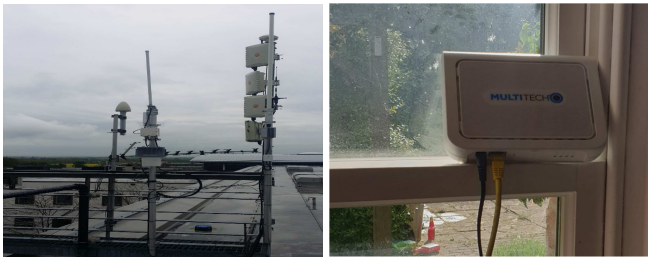

(a) LoRaWAN gateways, on roof and in building

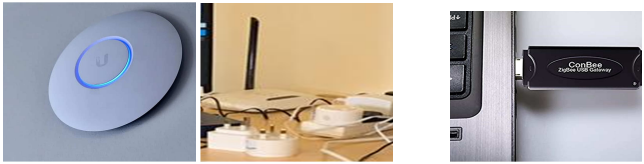

(b) Wi-Fi access points    (c) ZigBee gateway

Figure 7: Deployed network devices.

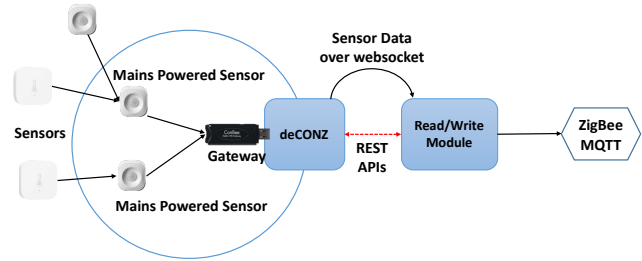

Figure 8: ZigBee Sensor support framework.

## 4.2 Network Layer

The sensors used in our deployment required LoRaWAN, Wi-Fi, and ZigBee for data transfer.

**LoRaWAN**. We deployed LoRaWAN gateways from Semtech and Multitech on the rooftops and inside the buildings (Figure 7a). Network access was made available via The Things Network (TTN). The hardware not only provided long-range connectivity but also low power (sensors target several years lifetime on a single battery) and low bandwidth (51 bytes/message) support. The gateway used Wi-Fi to connect to TTN.

In our implementation the gateways were within 2 km range of all the sensors allowing for any transmission to use the *spreading factor* suggested for real-time monitoring systems, *SF*7 [9]. Such a low spreading factor guarantees low latency and supports around 1500 devices transmitting through the gateway on the same channel with a low packet error rate [38].

**Wi-Fi**. We used two classes of Wi-Fi APs. The first, from tp-link and D-link, follows the IEEE 802.11b/g/n standards and supports a maximum of 32 clients at a time and a maximum rate of 300 Mbps. The second, from Ubiquiti, follows the IEEE 802.11ac standard, supports a maximum of 250 clients, and provides a maximum rate of 450Mbps over the 2.4GHz frequency and 867Mbps over the 5 GHz frequency. In a practical setting based on factors like size, cost, range, one or both APs could be used. Based on the requirement, several of these APs were set up in different parts of the building (Figure 7b).

**ZigBee**. We used USB-based ConBee II gateways (Figure 7c) which do not require Internet access and have a range of up to 30 m inside. For more distant devices we created a ZigBee mesh network (Figure 8) for which any ZigBee device connected to mains power acts as a repeater and routes signals.

## 4.3 Integration Layer

We used the Things Network (TTN) services, deCONZ [7], and MQTT to integrate readings from the sensors. MQTT also served to bridge messages received over multiple channels, implementing the publish/subscribe design principle. We next describe the three phases of integrating the physical layer with the layers above.

*4.3.1 Smart plug Message Exchange.* We used the MQTT protocol for message exchange with the smart plugs. MQTT is a popular lightweight protocol for IoT projects, providing publish/subscribe support for real-time data exchange. MQTT has three primary components; (*i*) *broker*: which is the server handling data exchange, (*ii*) *publisher*: a client which sends a message, and (*iii*) *subscriber*: a client which retrieves messages. Publishing and subscribing is with reference to a unique *topic*, and a client can act both as a publisher or subscriber. We use the *mosquitto* broker [5]. Each smart plug publishes and subscribes to a topic unique to the device id. The broker receives periodic messages from all the sensors.

*4.3.2 ZigBee Sensor Message Exchange.* We set up our framework to obtain ZigBee sensor data through MQTT, as shown in Figure 8. We used deCONZ [7], an application which communicates with the gateway to expose the devices connected to the gateway. It provides websocket support for real-time data exchange and a set of APIs to manage sensors. We introduced the *Read/Write* module, which continuously listens for data from the deCONZ websocket and publishes the same on the ZigBee MQTT broker.

*4.3.3 LoRa Sensor Message Exchange.* On deployment, The LoRa sensors are registered with the TTN using Over The Air Activation (OTAA). Once deployed, these sensors start broadcasting the LoRaWAN messages over the LoRa radio protocol, which is received by the deployed gateways. These gateways forward the LoRaWAN messages to TTN over the Internet, which is made available through the MQTT API. The messages obtained through the MQTT API are linked to a topic to which the sensors and the server subscribe. TTN also ensures that message transfer is secured.

*4.3.4 Bridging.* In our implementation, we had three channels, each having its own MQTT brokers. In order to integrate messages from all channels at a single broker, we utilise the MQTT bridging technique. Effectively, we subscribe the Wi-Fi MQTT broker (say *localMQTT*) to the TTN MQTT broker (say *ttnMQTT*) as well as the ZigBee MQTT broker (say *zigbeeMQTT*) by configuring *localMQTT* appropriately. This configuration is flexible because we could decide if the bridging is required on all the topics or a subset of topics. Once set up, *localMQTT* acts as the sole broker for data exchange between the Physical and other layers. This ensures that the Data layer need not tackle any changes in inclusion or removal of brokers in the architecture.

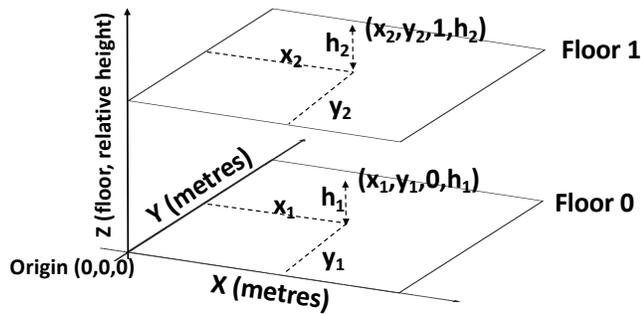

**Figure 9: The in-building coordinate system followed in the implementation. X and Y coordinates are distance from origin in metres and Z coordinate is the combination of floor number and relative height on the floor.**

## 4.4 Data Layer

This handles three major tasks, (*i*) decoding and homogenising the data obtained from different channels via the *localMQTT* broker, (*ii*) storing data for future use, with associated metadata, and (*iii*) making the data available in real-time for processing by client-side applications.

*4.4.1 Decoders.* The messages transmitted by the sensors vary based on factors like the type of sensor, vendor, design, transmission channel. For example, the smart plugs we used include their unique device number in the MQTT topic and not in the message, while most of the LoRa sensors include this information in the message itself. Moreover, although *SenseRT* strives that the time of a reading is generated as far upstream the architecture as possible, preferably at the sensor, it is not always achievable. Several real-world sensors do not include time in the message as they do not have a real-time clock. Simply receiving and storing the message without assigning any time to it would make any further processing a complex task, especially for a real-time system. The decoders take care of these problems and generate a normalised message for each sensor. We implemented a set of decoders for the different classes of messages received from the *localMQTT* broker. The decoder program consists of two main components, (*i*) the decoder set, which includes all the decoders for the available sensors, and (*ii*) the decoder manager, which based on the message decides which decoder to use as well as automatically registering new decoders added to the decoder set. For our prototype, we simply append the timestamp and unique device id to all messages. However, more complex processing could be done to generate a more advanced payload.

*4.4.2 Storage.* SenseRT has two types of storage: a data store containing all the data received from the physical devices in separate JSON files, and a metadata store containing metadata for all devices except network devices.

We used PostgreSQL to implement the metadata store. As different devices have different metadata available, we use a simple two column table, one with the sensor's unique id, and the other a *jsonb* column storing all other information in JSON format. The JSON entry for each device contains at least a timestamp property, guaranteeing that we have all the historical changes of a sensor in the database instead of only having the latest information; and a location property, indicating where in the building the sensor is deployed. The location information also included the coordinates in the XYZ plane, with one corner of the building as the origin to calculate the XY coordinates, and the Z coordinate combines the floor the sensor is on and the height relative to that floor (Figure 9).

*4.4.3 Real-Time Server.* The Real-Time Server (RTS) supports minimal latency processing of real-time data and asynchronous & non-blocking data management. This is guaranteed by following the *Actor Model* [34] using Vert.x [20] to receive and support analysis of data in real-time by multiple Vert.x modules called *verticles*. The *actor model* supports concurrency by enforcing that each *actor* (here implemented as a verticle) only interacts with other actors (verticles) through messages posted to its *message-box*. This model provides the following advantages:

**Asynchronous message passing**. The actor model guarantees that the RTS adheres to an asynchronous message-passing paradigm, providing real-time data handling from whichever source it is received. The *FeedHandler* receives a message arriving at the RTS and publishes on the *EventBus* to be used by other verticles.

**Non-blocking modules**. Using the Vert.x library with the actor model provides support to build a non-blocking framework for the RTS. Unlike the standard actor model, where each actor has its own *message-box*, in our implementation, the *EventBus* acts as the common *message-box* for all verticles. The message exchange is performed using a publish/subscribe approach. Any verticle accesses data by subscribing to the *EventBus* and sends messages by publishing it to the *EventBus*. The model also ensures that any communication between two verticles also happens only through the *EventBus*.

**Modular server**. Each verticle is an independent actor ensuring that the RTS is modular. This guarantees that any number of modules could be added or removed as needed without affecting any existing verticles. As a result, a standard production implementation could have thousands of verticles accessing data concurrently.

As well as following the actor model for concurrency, all verticles in *SenseRT* act as *stream processors*. Unlike most systems, which store the data in a storage unit and then query or perform computation over it, stream processing differs in two key ways:

(1) **Events substitute messages**. Verticles in *SenseRT* react to the incoming stream of events instead of a message or a batch of messages. Many sensors will send periodic updates reporting the status quo – these are typically not of interest to verticles, which are concerned rather with events indicating some change of state. For instance, a verticle controlling lights in a room might only be interested in the events indicating a change from unoccupied to occupied, or vice versa, and not in processing periodic messages of current occupancy which the sensor sends. Working with events also ensures that *SenseRT* handles *timeliness* of the event being processed by a verticle.

(2) **Reversing the norm**. Unlike the store-then-analyse approach, stream processing focuses first on enabling real-time reactive processing of data (events). Upon receiving a relevant event, a stream processing application (a verticle in *SenseRT*) reacts to the event by updating some information, creating another event, or simply storing it. The result is that data can still be archived for historical processing, but this does not negatively affect the performance of real-time processing.

There are four classes of verticles that *SenseRT* requires: (*i*) *Data ingestion verticles*, which receive data from the MQTT broker and publish the same on the *EventBus* (e.g., *FeedHandler*), (*ii*) *Data storage verticles*, which subscribe to the *EventBus* for any new data and store it for future usage (e.g., *MessageFiler*), (*iii*) *Real-time analysis verticles*, which analyse the stream of data in real-time and publish updates on the *EventBus*, and (*iv*) *Outbound verticles* which make the data available to the outside world. As shown in Figure 5, additional verticles include the *MessageRouter* verticle, used to share data to other similar systems; the *Data Monitor*, used to interact with client-side applications; and the *Real-Time Analysis* verticle, comprising one or more verticles and performing tasks like identifying events such as the measured $CO_2$ level crossing a threshold or a power outages, and broadcast the results of such analysis as derived events on the *EventBus*.

We set up three backup servers, which could act as the primary server during any outage. These backup servers subscribed to the *EventBus* to receive any new message from a sensor. This ensured that the backup server had all the data that the primary server had.

## 5 CASE STUDY

We next examine an end-to-end example of a system using *SenseRT* to fuse sensor data to provide a useful application: a modernisation of the *Trojan Room Coffee Pot* [58]. The original deployed one of the first webcams to monitor how full was a research group's coffee pot.

In our modernised version, where an opaque coffee pot renders the webcam approach ineffective, we measure and transmit in realtime the coffee-making and consuming events of the coffee pot in one of our buildings. The system analyses data from a set of sensors deployed at the coffee pot to recognise one of five events; (*i*) `pot-removed`, indicated the pot is not present, (*ii*) `new-pot`, indicating the presence of freshly-made coffee in the pot, (*iii*) `pot-poured`, indicating that coffee has been poured, (*iv*) `pot-empty`, indicating that no coffee remains, and (*v*) `coffee-grinding`, indicating that the coffee bean grinding machine appears to be active.

The coffee pot setup (Figure 10a) is designed as a *sensor node*, a coordinated collection of multiple sensors: weight sensors connected to the Raspberry Pi periodically monitor the weight of the coffee pot, while two smart plugs monitor the power usage of the grinder and the coffee brewing machine respectively. The Raspberry Pi also provides a Wi-Fi gateway to connect the two smart plugs. The sensor node accumulates data from each sensor and transmits a message to the local MQTT broker over Wi-Fi. The message consists of the weight and power readings and the time when the reading was recorded.

After being homogenised by the corresponding decoder, the message is published on the *EventBus* by the *FeedHandler* verticle. The *MessageFiler*, subscribed to the *EventBus* for any new message receives the new message and stores the attached data. A real-time analysis verticle, *RTCoffee*, looks for two types of events in the published data: did the power consumed by either the grinder or the coffee machine cross a threshold (40 W in our case), indicating the grinder or the coffee machine was in use; and has the measured

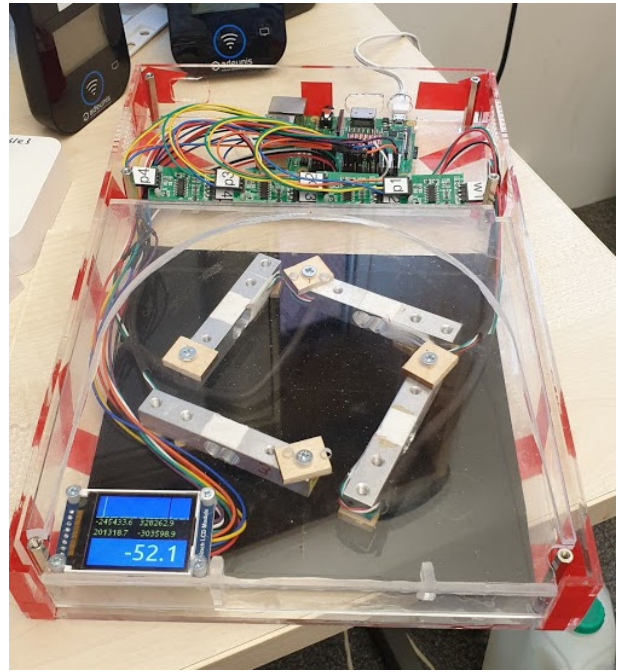

(a) Coffee pot sensor node comprising four weight sensors, a Raspberry Pi and Wi-Fi gateway, with a local display for convenience.

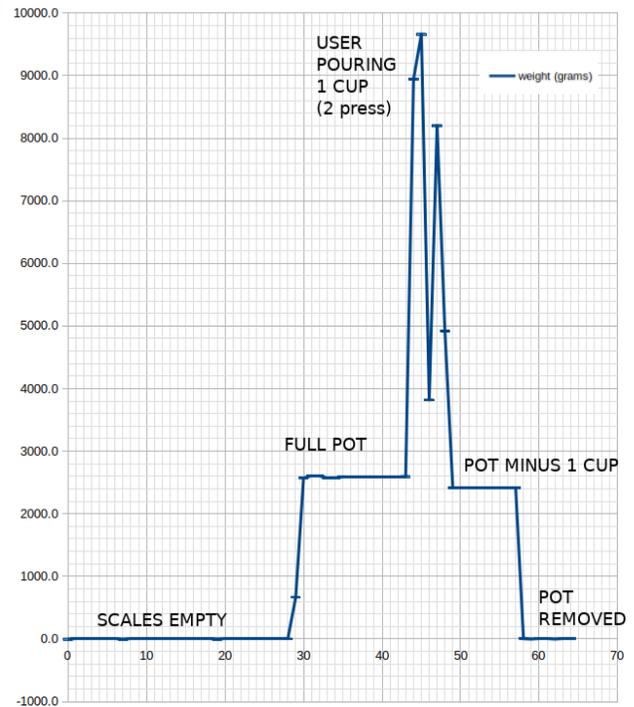

(b) Readings during coffee preparation.

**Figure 10: The Coffee Pot sensor node. The pot weighs ~0.5 kg, and when full holds 2 kg of coffee. Each cup taken is ~0.25 kg.**

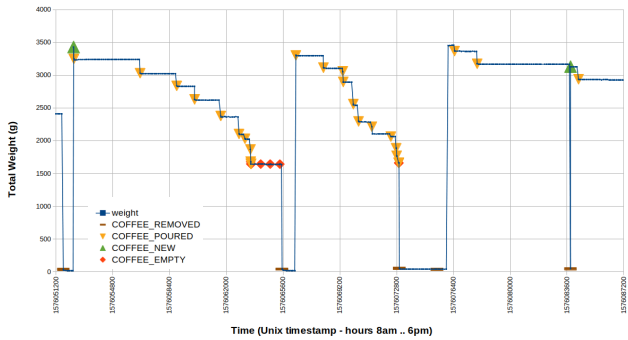

Figure 11: Events observed by RTCoffee.

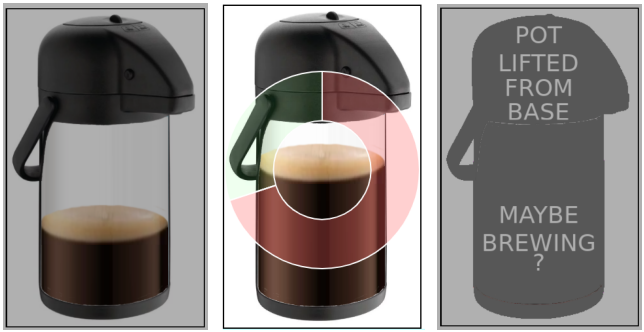

(a) The level of coffee in the pot at a given time.

(b) Level of coffee in the pot after a *coffee-grinding* event with approximate time remaining.

(c) Following a *pot-removed* event.

Figure 12: Web-Client UI showing status of the coffee pot.

weight of the coffee pot changed, indicating one of the five events described above, as depicted in Figure 10b. Figure 11 gives an example of the events observed by *RTCoffee* on a particular day between 8am and 6pm. *RTCoffee* publishes its derived event on the *EventBus* to be consumed by other verticles.

We implemented a simple web client to consume the data received from *SenseRT* and display the status of coffee in the coffee pot. The web client subscribes to a *DataMonitor* verticle for event updates, which in turn subscribes to the *EventBus* and receives any updates provided by *RTCoffee*. Whenever an update occurs, the web client receives the event, and the UI is updated accordingly. Some example UI updates are shown in Figure 12.

## 6 EVALUATION

Our deployment of *SenseRT* has been live since March 2020, and we present measurements taken over seven months, to October 2020. We show that *SenseRT* provides an architecture for minimal latency real-time data processing. Many sensors were added during the experiment period, and we examine how this impacted performance. We then compare against some of the existing alternative systems to highlight the advantage of the design choices made in *SenseRT*.

### 6.1 Real-time Performance

As a real-time architecture, it is important for *SenseRT* to show minimal latency between the generation of data at the sensor and the consumption of data by an application. In order to validate this, we measured the latency at four key points in the *SenseRT* architecture: (*i*) the gateway receiving the first hop message from a sensor (*ii*) the bridge MQTT broker in the Integration layer, (*iii*) the EventBus in the Data layer, and (*iv*) a client-side application similar to that described in (§5).

|  | Gateway | MQTT Broker | Event-Bus | Client Application |
|---|---|---|---|---|
| **Mean** (ms) | 57.15 | 147.86 | 157.86 | 159.55 |
| **Std Dev** (ms) | 10.21 | 63.56 | 2.35 | 0.56 |

Table 3: Latency at key points of data flow in the *SenseRT* architecture. All values are in ms and are calculated from the time of message generation at the sensor.

As is evident from Table 3, at all four points latency is minimal. With as low as 57*ms* average latency at the gateway, an increase is observed at the integration layer (approx 90*ms* in average) owing to the processing involved with the bridging of information from different protocols. The messages are published at the *EventBus* with a latency of just 10*ms*, while the client-side application receives the desired data with almost no latency (2*ms* in average). For all practical purposes, an average latency of 159.55*ms* is a negligible value which validates *SenseRT* as a real-time system.

Furthermore, *SenseRT* is robust enough that addition of new sensors or category of sensor in the system doesn't affect the overall working of the architecture. It is clear from Figure 13a that even on increasing the number of sensors in the system, the overall latency remains close to a similar mean value of 200*ms*. This is true also for when we calculated the mean latency for the different category of sensors used in the implementation of *SenseRT* (Figure 13b). Here, the maximum average latency is observed for *DeepDish* (400*ms*) which is because of 200*ms* processing time involved in computing the number of people from a video frame.

### 6.2 Comparison Against Alternatives

We compare different aspects of *SenseRT* with three similar systems. The first is the work by Al-Ali et al. [10] which sets up a Wireless Sensor Network with each sensor interfaced with a data acquisition system on a chip. The sensors exchange data through a MQTT broker which is then sent to a central server for analysis. Any analysis in this system is request/response based. The second system by Fayyaz et al. [29] does something similar, however uses a fog server as the first hop point for the sensor messages. The third work we compare with is developed by Evangelatos et al [28] which uses IPv6 over Low power Wireless Personal Area Networks (6LoWPAN) to collect sensor information and store it for further processing. However, data is only requested for events such as a person entering a room, so no real-time data flow is required.

We compare systems on two fronts, (*i*) latency at the first hop point for the sensor messages (at the gateway), and (*ii*) latency for a complete data flow transaction from sensor to application. Both results were not reported for all the three competing systems, hence we only include the ones which were reported for each criteria. Also, the results for Fayyaz et. al. [29] are based on simulation.

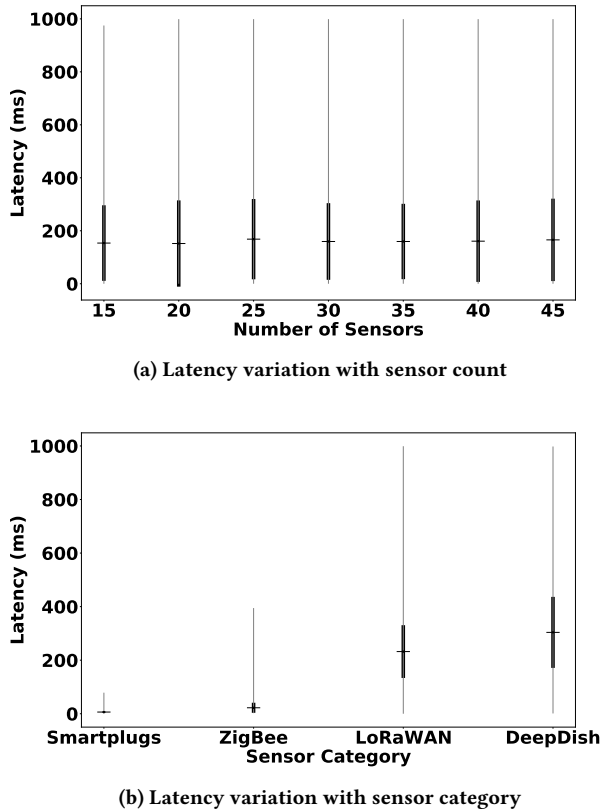

(a) Latency variation with sensor count

(b) Latency variation with sensor category

Figure 13: Observed latency variation.

| System | *SenseRT* | [10] | [29] | [28] |
|---|---|---|---|---|
| Sensors Used | 45 | 10 | 12 | 20 |
| First hop Latency (ms) | 57 | 66 | 60 | - |
| End-to-End Latency (ms) | 160 | - | 6000 | 259 |

Table 4: Comparing state-of-the-art systems with *SenseRT*.

*First hop latency.* We calculated the average first hop latency observed over all category of sensors which we report in Table 4. It is evident that the latency is at par with the state-of-the-art values (and in fact it beats them by a few ms).

*End-to-End latency.* When compared to the fog-based system [29], *SenseRT* performs 37 times better. This is primarily because the application server subscribing to the Data Monitor and hence receiving the messages almost at the same time as the EventBus (2 ms latency, as shown in Table 3). However, that system stores all the data on the cloud and then queries that from every application, increasing the latency. The 6LoWPAN system [28] does achieve almost similar latency as *SenseRT*, but only for the data stored at the server. Additional analysis would have further increased latency.

## 7 DISCUSSION

*SenseRT* provides a robust architecture for in-building real-time data flow but there remain some key elements requiring further work.

### 7.1 Scalability

This is more a feature of our prototype implementation than the *SenseRT* architecture itself. The prototype implemented in this paper covers only two buildings so far, with two more buildings in the deployment pipeline. It will be essential to observe the impact on performance as this number is increased.

As well as extending to cover more buildings, we are extending the types of sensors used in the implementation. We would like to integrate other types of data sources like Modbus or Monnit sensors [45], and understand how to fit them within the architecture.

As the set of sensors being deployed increases, optimising the number of sensors in the building becomes important. For instance, if a ZigBee and a LoRaWAN sensor provide the same readings, only one might be used dependent on the client's needs. As sensors are increasingly integrated, a single sensor might provide multiple readings and so could replace multiple sensors. With more sensors in place, deciding the optimal placement strategy for effective building coverage is essential. Others have investigated this [39, 50], providing directions for finding a strategy for *SenseRT*.

### 7.2 Privacy

Privacy is a key concern with IoT-based systems, and many have looked into how to achieve this in smart buildings and smart cities [35, 36, 43, 51]. Currently, *SenseRT* provides basic privacy support through encryption provided by network protocols and limited data access. However, *SenseRT* is amenable for more fine-grained privacy approaches. Our intended approach has three aspects.

First, we will ensure all messages from any sensor are encrypted. As some sensors do not have the capacity to encrypt data at source, decisions will need to be made about where is the most appropriate point in the architecture to provide encryption, and what scheme should be used.

Second, we need to define mechanisms to support different data access strategies governing who can access what data. For example, a person might have access to all the data generated by sensors in her room, but only specific sensors on the floor. This could involve assigning the different stakeholders, building managers, third-party clients, or building occupants, into groups by which access is controlled. Another option could be to provide unique tokens to each stakeholder, and access is provided based on tokens. This could also include building visitors to whom limited data could be made available, e.g., in a time of COVID-19 with social distancing recommendations, current (but not historical) room occupancy could help keep visitors and occupants safe.

Third, under whatever access control regime is provided, we must still determine how occupants' privacy should be protected. For instance, a building manager receiving power readings from smart plugs in a room every hour could infer when the occupant was in their office, whereas receiving overall power usage for an office over a week might be sufficient for managing energy efficiency in a building. We thus need to examine how client applications can receive sensor data so as to ensure privacy while still meeting the differing goals of users of the system.

## 8 CONCLUSION

As the world moves towards more and more smart buildings, we anticipate considerable (perhaps exponential) increase in the number of sensors and consequently volumes of real-time data generated. Building Management Systems (BMSs) need to be re-architected to better support both data management and reliable real-time information dissemination. *SenseRT* provides a robust architecture satisfying these goals by considering several key aspects: (*i*) spatio-temporal aspects of real-time sensor data, (*ii*) improving *timeliness* and maintaining low *latency* throughout, (*iii*) ensuring a homogeneous data flow through the architecture notwithstanding the wide range of sensor types and capabilities, and (*iv*) performing efficient real-time analysis of sensor data with minimal latency.

Our implementation of a prototype and the experiments carried on for seven months show that *SenseRT* does provide an efficient network architecture for in-building real-time data flow and real-time data analysis. There do exist key aspects of scaling and privacy, which would improve *SenseRT* further. However, as it stands, *SenseRT* is a first step towards providing a robust network architecture that could tackle the incoming challenges BMS faces regarding the increasing volume of sensors and the real-time data being generated by these sensors every day.

# *RACER*: Real-Time Automated Complex Event Recognition in Smart Environments

Rohit Verma, Justas Brazauskas, Vadim Safronov, Matthew Danish, Ian Lewis, Richard Mortier
University of Cambridge
Cambridge, United Kingdom
{rv355,jb2328,vs451,mrd45,ijl20,rmm1002}@cam.ac.uk

## ABSTRACT

As smart environments become laden with more and more sensors, there has been a need to develop systems that could derive useful information from these sensors and make the smart environments smarter. Complex Event Processing (CEP) has emerged as a popular strategy to identify crucial events from sensor data. However, the existing CEP strategies overlook the relationship with other sensors in the spatial vicinity and understate the temporal variation of sensor data. In this paper, we develop *RACER*, which is an end-to-end complex event processing system that takes into consideration both the spatial location of the sensor in observation and the varying impact of temporal changes in the sensor data. Experiments performed for a duration of five months over both collected and live streaming data shows that *RACER* fares well compared to the other state-of-the-art approaches.

## CCS CONCEPTS

• **Human-centered computing** → **Ubiquitous and mobile computing**; • **Information systems** → **Spatial-temporal systems**.

## KEYWORDS

real-time spatio-temporal data, complex event processing, smart environments

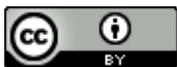

## 1 INTRODUCTION

The Smart Environment –a physical world where connected devices like sensors, actuators, and computational elements work together to make users' lives more comfortable– has seen several advances in recent years. Several applications are designed to achieve the goal of user comfort like remote health monitoring, air quality monitoring, power-grid management, building automation, etc. The key components of these aforementioned applications are the events that either bring in opportunities (lowering the power usage in vacant rooms) or threats (gas leaks or fires). Effective and fast identification of these events is crucial for any such application. These events could be simple (atomic) events like a patient missing an appointment or complex like power failure in a complete building unit.

Recently, Complex Event Processing (CEP) [5] has been of interest in the field of smart environment applications [4, 7, 17]. CEP utilizes the understanding that multiple atomic/simple events are an indication of a much important complex event.

There have been several works that utilize a CEP system to design alerting applications [11, 16]. The approach is to assign area experts who identify a set of atomic events and design rules which could lead to a complex event. These rules are then fed into a CEP engine to be matched later from the incoming event streams.

These state-of-the-art approaches overlook important points. First, the atomic event detection methods only consider a single sensor's data for an atomic event. However, other sensors in the spatial vicinity would also have a relationship with the sensor's reading [12]. For example, in a smart wearable, simply using the accelerometer to detect walking could be erroneous as a person swaying their hand could give similar signatures of walking. Similarly, a high-temperature reading on a sensor in a room would have a relationship with other temperature or humidity sensors in the room. Second, the time of the year/day when the data is being observed is also important. High temperature during summers is normal but a similar high temperature during winters could mean a broken heating system.

In light of this, we develop an end-to-end system, **R**eal-time **A**utomated **C**omplex **E**vent **R**ecognizer (*RACER*), which provides a robust complex event detection mechanism ensuring that the spatial positioning of the sensor being observed and the temporal variation in data is taken into account when doing any computation. We first extract two sets of features (§ 3) from incoming sensor data; relationship with other sensors in the spatial vicinity of the sensor being observed, and the temporal features linked to the sensor data. Following this, we design a model (§ 3.2) which takes these features as input and detects if an atomic event has occurred. Next, we design a CEP engine, which takes as input atomic events and a set of user-generated rules to detect complex events in the incoming data stream (§ 4). We evaluate the end-to-end system over our deployment of a National Digital Twin in the university campus, consisting of five buildings (§ 5).

## 2 RELATED WORK

Complex Event Processing (CEP) [5] is an approach that observes real-time events in data streams and the causal relationship between these events to identify opportunities or threats as they happen. The usual approach is to assign area experts the task to define such rules in a query language [2] and deploy these as operator graphs [8] used to match event patterns in the incoming event streams.

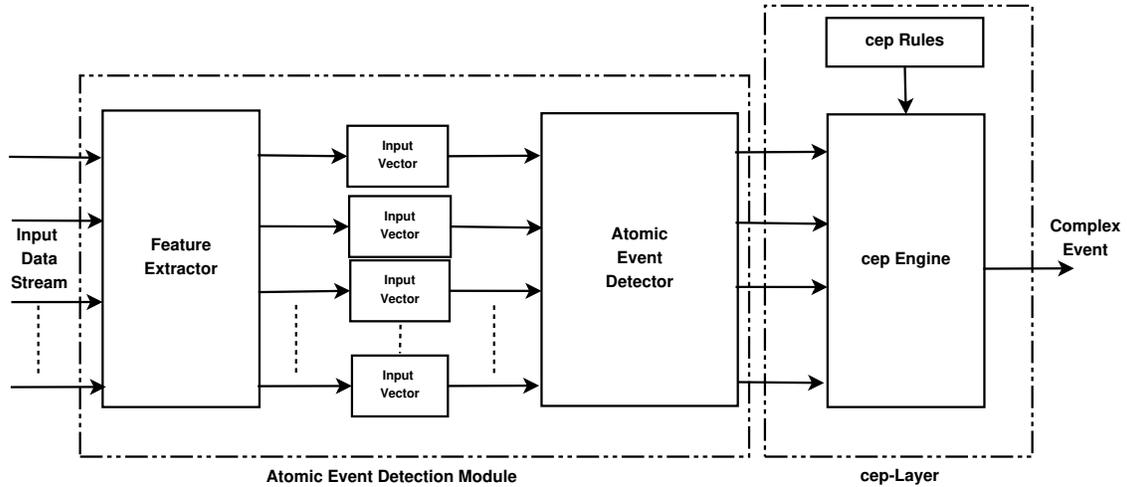

Figure 1: System Architecture

CEP systems have seen wide usage in numerous use-case scenarios like stock market trading [14], predictive maintenance of sensors/equipments in large facilities [15], and real-time marketing [1]. Recently, there has been increasing interest towards applying CEP to autonomous vehicles [6]. CEP is also being widely used in smart environments which we discuss next.

In a smart environment, it is critical to react to any opportunities or threats with low latency. These could be accidents like fire or gas leaks, sudden change in weather conditions, power-grid faults, sudden load variations, etc. A CEP system continuously monitoring event streams from the deployed sensors would serve well to address these scenarios. Many works have employed CEP to such scenarios in the context of smart-grids [4, 10], smart-hospitals [16, 17] or smart-buildings [7, 11].

Although there has been a lot of research in the field of CEP systems, in these systems, the relationship with the sensors in the spatial vicinity is not taken into consideration [7, 10, 11]. However, several works have pointed out that the readings in the same spatial vicinity, be it a room, building or just a sensor box, are correlated to each other [12]. Essentially, the need is to identify atomic events based on the temporal distribution of sensor data over a period and not the instantaneous values or a minor divergence from the data distribution.

## 3 ATOMIC EVENT DETECTION MODULE

As shown in Fig 1, the Atomic Event Detection Module comprises of the Feature Extractor and Atomic Event Detector. We describe these two sub-modules in this section.

### 3.1 Feature Extractor

Any atomic event linked to a single sensor could be defined as a function of the data it generates ($X$), its location in the smart environment ($\mathcal{L}$) and at what time is the data collected ($T$). Hence, the system needs to extract features that cover all three aspects of the sensor data. The *Feature Extractor* thus works with three types of features; (a) correlation with nearby sensors, (b) raw sensor data, and (c) temporal parameters.

*3.1.1 Sensor Correlation Score.* The Sensor Correlation Score ($C$), extracts the impact of the sensor's location. The primary idea being that any other sensor in the spatial vicinity of the sensor being observed would in one way or another impact the data being transmitted.

We define two parameters before computing $C$. First is the *spatial vicinity* ($\mathcal{S}$)– which is the spatial area inside which a sensor ($s_c$) to be correlated with the sensor being observed ($s_o$) lies. $\mathcal{S}$ is a configurable parameter based on application requirements. Second, the time window $\omega$ over which the correlation is to be measured. This is important both for storage purposes (larger $\omega$ would require higher memory consumption)

Once $\mathcal{S}$ and $\omega$ are defined, the Sensor Correlation Score for the sensor $s_o$ with respect to sensor $s_c^i$, represented as $C_i$, is computed as the Pearson Correlation Coefficient [9] between the readings of $s_o$ and $s_c$ in the time window $\omega$. For $n$ sensors in the vicinity, we obtain a vector $C_{i=1}^n$

*3.1.2 Temporal Parameters.* The timestamp of sensor data collection would impact the predictions linked to any event. For example, low readings on a temperature sensor in winter is obvious, but similar reading in summer could be an HVAC issue. Hence, to capture the temporal characteristics of the sensor readings we extract three temporal parameters, *month* ($m$), *day* ($d$), and *hour* ($h$).

These set of features, along with the sensor readings in $\omega$ time window, and the type of sensor compose the input vector at time $t$, represented as,

$$\mathcal{I}_t = \{(\mathcal{X} = \{x_i | i \in [t - \omega, t]\}), C_{i=1}^n, m, d, h, s\}$$

## 3.2 Atomic Event Detector

The *Atomic Event Detector* takes the input vector $\mathcal{I}_t$, obtained from the feature extractor, and predicts if it signifies any atomic event in $\omega$ time window. We design the *Atomic Event Detector* using a Convolution Neural Network (CNN), and train for each event in isolation. The model is trained using sparse categorical cross-entropy loss function with $60 - 20 - 20\%$ data split for training, validation, and testing.

The identified atomic event, if any, are then fed into the CEP Module for further processing. It should be noted that at the same time instance multiple atomic events could be detected which are fed into the CEP module.

## 4 cep-LAYER: COMPLEX EVENT PROCESSING

The *cep-layer* takes care of identifying a complex event from the incoming atomic, based on human-generated rules. In this section, we will first describe how the rules are generated and their structure and following that describe how the *cep-Engine* works.

### 4.1 CEP Rules

Given a set of $n$ atomic events, $\{E_a^i\}_{i \in (1,n)}$, a complex event can be defined as a function of one or more of these atomic events, represented as;

$$E_c^i = f(E_a^1(e^1, t^1, v^1, s^1), E_a^2(e^2, t^2, v^2, s^2), ....) \quad (1)$$

where, $e$, $t$, $v$, and $s$ are the atomic event id, timestamp, the sensor data reading, and the sensor associated with the complex event.

As we observe, from the above example, a complex event rule depends on the type of the event, when the event occurred, what values were linked to the atomic event, and the location of the sensor which provided those readings. Hence, we define a template for any complex event as;

$$\begin{aligned}
E_c^i(e^i, t^i, [sensor\_ids]^i) \\
<= E_a^x(e^x, t^x, v^x, s^x) \cap E_a^y(e^y, t^y, v^y, s^y) \\
\cap \ t^x < t^y \\
\cap \ v^x \leq \mathcal{V}_1 \\
\cap \ v^y \geq \mathcal{V}_2 \\
\cap \ distance(s^x, s^y) < \Delta
\end{aligned} \quad (2)$$

The above rule states that, the complex event $E_c^i$ is implied when both atomic events $E_a^x$ and $E_a^y$ have been detected and $E_a^x$ occurred before $E_a^y$. Additionally, the sensor data values at the time of detection were related to the thresholds $\mathcal{V}_1$ for $E_a^x$ and $\mathcal{V}_2$ for $E_a^y$ and both sensors were within $\Delta$ distance of each other. The rule outputs the detected complex event id, the detection time and the sensors which were used to identify the complex event.

### 4.2 CEP Engine

Owing to its roots in formal logic, we use Prolog to design the CEP engine. The incoming atomic events and the fed back complex events are stored as a set of Prolog *facts*. However, this set of *facts* is kept of a finite size $\mathcal{K}$, and the arrival of a new atomic/complex event removes the oldest entry. This is done to keep only the events that are relevant at the current time. The parameter $\mathcal{K}$ is configurable and depending on the requirements could be increased or decreased. The CEP rules are loaded as Prolog *rules* in the environment.

Once a pattern match is obtained by the CEP engine, it outputs the detected complex event, the timestamp when it is detected and the sensor's responsible for the event.

## 5 EVALUATION

We performed the experiments on *RACER* with 185 LoRaWAN sensors, the data of which arrived from five different buildings that included two offices and three residential buildings. The data used in this section was collected in experiments run from January 2021 to May 2021, for a period of five months.

| Total | None | $CO_2$ Rise | $CO_2$ Fall | $CO_2$ DC | Temp DC |
|---|---|---|---|---|---|
| 3553027 | 3494477 | 1032 | 3864 | 114 | 59 |
| Lux Drop | Lux Rise | Water Low | Occ Mod. | Occ High | Motion |
| 793 | 9764 | 350 | 685 | 5 | 41884 |

Table 1: Number of training data points per atomic event. (DC: At Discomfort level, Mod.: Moderate)

| Corridor Walk (C1) | Group Meet (C2) | Cook (C3) | Office Vacated (C4) | Window Opened (C5) |
|---|---|---|---|---|
| 50 | 10 | 100 | 140 | 80 |

Table 2: Number of data points per complex event labelled in the data. The label in brackets are what these complex events will be referred to as henceforth in the text.

Table 1 lists the number of atomic events used in training the CNN model. The number of complex events identified in the live data stream and the existing training data is listed in Table 2.

### 5.1 Comparison with Competing Systems

We compare *RACER* with three existing systems which also perform complex event processing for smart environments.

*5.1.1 Raj et al. [11].* This work develops a system that performs threshold-based atomic event detection and then feed these detected atomic events to the CEP rule engine to identify complex events in smart buildings.

*5.1.2 Sun et al. [13].* This work relies solely on anomaly detection to identify any complex events in the building. The system uses an Isolation Forest (IFO) [3] algorithm for anomaly detection in the incoming data, however, only detects complex events that could be observed from one sensor at a time.

*5.1.3 Mongiello et al. [7].* This system is built to alert for fire and danger management but is easily extensible to other complex event detection. Instead of a single threshold based atomic event detection, the system uses a threshold range.

We ran the four systems along with *RACER* for detection of the five complex events as listed in Table 2. In our implementation, we used sensors that were in the same room or the same sensor box as being in the vicinity of each other. We set the configurable parameter $\mathcal{K}$, which limited the number of facts with which the CEP engine works as 100 based on empirical results.

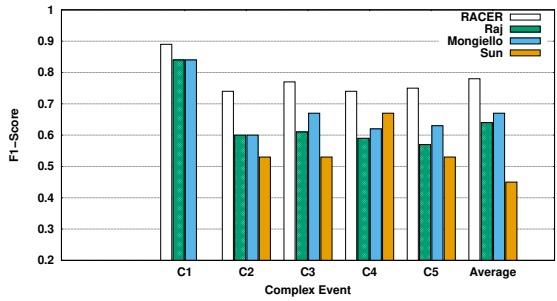

Figure 2: Comparing F1-Score of Complex Event Detection with competing systems for all five complex events.

Figure 2 shows the F1-Score for the complex events and the average F1-score for all four systems. *RACER* fairs better compared to all other systems, and there are few important points to be highlighted regarding this result. As Sun et al. [13] only works with single sensor data, hence, it wasn't able to detect C1. Moreover, for other events too, we had to select one sensor which gave the best result for this system. Although the result for other systems seems similar for C1, it should be noted that the CEP rule had to be different for Mongiello et al. [7] and Raj et al. [11].

## 5.2 Real-time Analysis of *RACER*

|  | Feature Extraction | Atomic Event Detection | cep-layer |
|---|---|---|---|
| Latency (ms) | 1.21 | 35.57 | 37.35 |

Table 3: Mean latency at key points of the work flow in *RACER* from message arrival time.

As a real-time system, *RACER* should complete computations with minimal latency. As we don't have any control over the latency of the sensor data originating at the sensor and reaching the system, we compute the latency from the time when the sensor data arrives at the system. We consider three key points in the *RACER* workflow: (i) feature generation, (ii) atomic event detection, (iii) cep-layer.

Table 3 shows that at all three points the latency is minimal. Owing to the neural networks processing, the atomic event detector takes 34.36*ms*. However, the other feature extractor and cep-layer show negligible latency of 1.21*ms* and 1.78*ms* respectively. For all practical purposes, an average latency of 37.35*ms* is negligible validating that *RACER* could operate as a real-time system.

## 6 CONCLUSION

As smart environments turn smarter, detection of complex events becomes a crucial requirement. *RACER* provides a robust end-to-end system that detects such complex events using sensor data received from these smart environments in real-time. *RACER* achieves this goal by considering several key aspects: (i) the impact of other sensors in the spatial vicinity of the sensor being observed, (ii) the temporal variations in the data that arrives from the sensors.

Experiments run for five months show that *RACER* can detect complex events with high accuracy. There do exist key aspects of automatic CEP rule generation, and timeliness of data, which require further analysis to improve *RACER*.

## ACKNOWLEDGMENTS

This research forms part of Centre for Digital Built Britain's work within the Construction Innovation Hub. The funding was provided through the Government's modern industrial strategy by Innovate UK, part of UK Research and Innovation.

# Real-Time Data Visualisation on the Adaptive City Platform

Justas Brazauskas, Rohit Verma, Vadim Safronov, Matthew Danish, Ian Lewis, Richard Mortier
University of Cambridge
Cambridge, United Kingdom
{jb2328,rv355,vs451,mrd45,ijl20,rmm1002}@cam.ac.uk

## ABSTRACT

In smart buildings research, the integration of Building Information Models (BIM), Building Management Systems (BMS), and Internet of Things (IoT) is of paramount importance. However, such integration often overlooks real-time building data visualisation. In this demo, we examine challenges related to spatiotemporal data representation and novel visualisation methods in smart environments. Following this, we present the front-end design of our Adaptive City Platform (ACP), a system for collecting, processing and visualising building information and sensor data in real-time.

## CCS CONCEPTS

• **Human-centered computing** → **Heat maps**; *Information visualization*; • **Computer systems organization** → Real-time systems.

## 1 INTRODUCTION

The next generation of smart environments will be driven by BIM, BMS and IoT fusion. Such fusion will give rise to digital twins with real-time asset monitoring capabilities. In such environments, spatiotemporal real-time data flow (or *flux*) from IoT sensors and BMS will combine with contextual BIM data to facilitate energy monitoring, accident prevention, increase comfort and resilience.

While creating such integrated smart building management systems comes with challenges like crafting efficient APIs and flexible ontology schemas [5], contextually meaningful spatiotemporal data visualisation often recedes to the background despite being a key interaction aspect that enables assessment of the created platforms.

To tackle this, we have crafted the Adaptive City Platform, a real-time building monitoring system capable of asynchronously handling spatiotemporal data from BIM and IoT sources with low latency and high throughput. Here, we showcase the ACP's front-end interface and describe how data visualisations enable capturing spatiotemporally contextual data flux from BIM-IoT fusion[1].

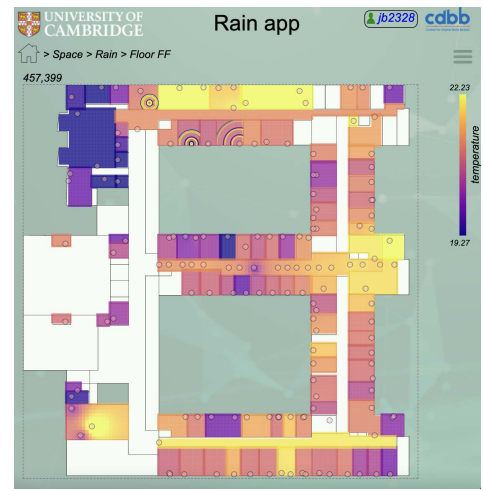

**Figure 1: An example application used to visualise current temperature levels in every office. The radiating circles, called *ripples*, show the real-time data coming in.**

### 1.1 Challenges

Related literature [1–4] highlights three major issues with smart building platforms and their spatiotemporal data representation.

**Facilitating building expandability:** Any smart building platform should have expansion capabilities so that the system could either be ported to other buildings or have new buildings be added to it. Such capabilities imply that the platform should use an easily accessible and modifiable building database containing contextual and spatial information, such as floorplans.

**Capturing high granularity data flux:** In contrast to ageing BMS, a high number of deployed IoT sensor devices create the need for high-bandwidth, low-latency data management platforms. With sensors being set up to either send messages periodically or after interrupts, capturing the incoming data flux is crucial for any real-time building monitoring system.

**Hierarchical data representation layers:** A hierarchical template structure with an increasing granularity of data allows scale-specific information to be displayed at necessary moments. In return, this leads to contextually relevant applications for every spatial scale, as well as privacy-preserving measures.

---

[1]Supplementary video showcase of the platform: www.vimeo.com/593360345

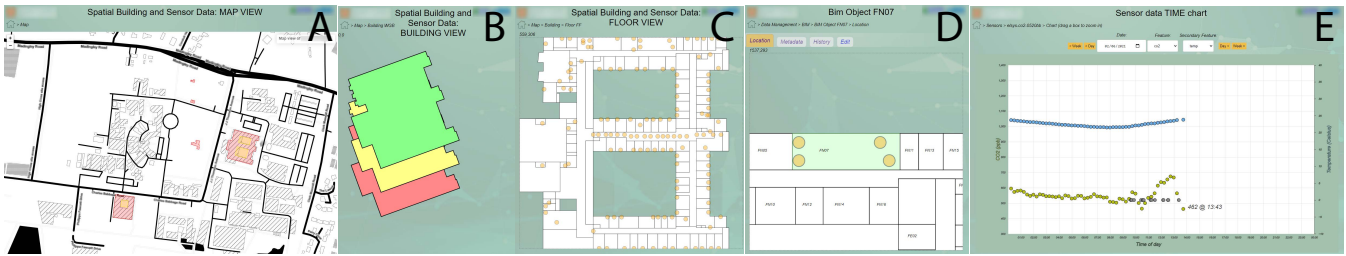

Figure 2: The five templates used to show BIM and IoT data on different scales: (A) Site, (B) Building, (C) Floor, (D) Floorspace, and (E) Sensor. In (E) scatter plots feature two zoomable Y-axes, allowing comparison of $CO_2$ and temperature readings.

## 2 DESIGN

For the ACP, we first create different hierarchical templates, each capturing spatiotemporal building data at different granularities. These templates are presented in Figure 2, where moving from a high-level overview of the site, we eventually enter floor and room level views, ultimately reaching scatter plots for individual sensors.

The platform is designed to accommodate multiple buildings easily. Rather than relying on manually crafted SVGs for floorplan representation, our platform either parses existing BIM files or generates SVG files from a list of coordinate data. We then manipulate the SVG on screen as well as draw sensor icons using the visualisation library *D3.js*.

### 2.1 Sample Applications

To illustrate how high granularity and low latency data can be effectively utilised in a smart building context, we develop two sample applications. One is a heatmap overlay over a floorplan called *Rain* (Figure 1), the second, called *Splash*, is used as a maintenance tool to monitor the state of the deployed sensors.

Our primary goal was to meaningfully and contextually demonstrate the flux of spatiotemporal data. To do so, we use a rain metaphor to facilitate the design process. Next, we aim to portray the real-time high-granularity data using room-bound heatmaps that show intricate differences even in room-scale environments.

**Data as Water Droplets:** We envision the data flux to be similar to a torrent of rain pouring on a puddle surface, where the sensors' messages were akin to splashes of water on the floorplan.

The analogy enables us to both make use of the size of the ripple effect over the floorplan or individual rooms to illustrate the magnitude of the reading, as well as give a clear visual indication to the observer that new data has arrived. Furthermore, as multiple ripples appear in the same space, the observer can perceive more complex events happening, such as a meeting.

**Room-bound heatmaps:** Existing smart building platforms use floorplan-wide heatmaps or choropleth maps [1, 2, 4] to colour parts of the floorplan or rooms with single colour values to represent the sensor readings. These solutions are not ideal in our research context due to the inherent inability to maintain high granularity sensor data. To tackle this, we generate room-constrained heatmaps with a dense resolution for every floor.

We first draw a dense cell network in our heatmap to show fine differences in sensor values. For example, in long corridors, sensor readings might be affected by the sensor being positioned close to windows or hidden away in corners. ACP is capable of displaying such differences with sub-metre accuracy, as shown in Figure 1.

Second, our solution prevents data leaking from one room to another, e.g. server rooms might have significantly different temperatures than surrounding offices. The lack of hard boundaries between rooms with different temperatures renders such floor-scale heatmaps unusable. We create individual heatmaps for every room on the floorplan with surrounding walls. Hence, we can display room-level differences without data *spilling* to surrounding spaces.

Finally, low-latency and high-throughput enable instantaneous building monitoring, where we can detect people walking across a corridor in real-time, as the ripples propagate through the visualisation, immediately catching an observer's attention.

## 3 CONCLUSION

In this paper, we argue the importance of efficient and clear data visualisation methods for such platforms. The presented Adaptive City Platform showcases our novel visualisation tools such as room-bound heatmaps, and through the use of a splash metaphor, can capture the arriving data flux in an immediately obvious and spatiotemporally meaningful way.

## ACKNOWLEDGMENTS

This research forms part of Centre for Digital Built Britain's work within the Construction Innovation Hub. The funding was provided through the Government's modern industrial strategy by Innovate UK, part of UK Research and Innovation.